# Tactile Perception through Fluid–Solid Interaction

Arman Goshtasbi[1], Minke Berghuis[2†], Aida Parvaresh[1†], Saravana Prashanth Murali Babu[1], Robert W. Style[3], Ahmad Rafsanjani[1*]

[1] SDU Soft Robotics, Biorobotics Section, The Maersk McKinney Moller Institute, University of Southern Denmark, Odense 5230, Denmark

[2] Department of Mechanics of Solids, Surfaces, and System, University of Twente, Enschede 7522 NB, The Netherlands

[3] Department of Materials, ETH Zürich, 8093 Zürich, Switzerland

† These authors contributed equally.

* Corresponding author ahra@sdu.dk

**ABSTRACT**

Soft tactile sensors elevate robotic touch through enhanced flexibility and adaptability, yet most existing designs depend on embedded electronics that are susceptible to interference and environmental limitations. In this work, we leverage fluid–solid interactions to develop a class of soft tactile sensors that operate entirely without electronics at the sensing site. The sensor comprises a fluid-filled elastomeric channel connected to only two external pressure sensors. Touching different regions of the elastomeric surface displaces the viscous fluid, producing distinct pressure patterns that encode both touch position and force. These signals are decoded through a machine learning framework that integrates feature extraction, soft clustering, and adaptive neuro-fuzzy inference to achieve accurate localization and force estimation. We validate this concept through single-point touch localization and force estimation in a linear (1D) sensor and extend the same sensing principle to 2D tactile mapping by routing the channel across the surface using space-filling curves, while maintaining the same minimal hardware setup. This simple approach remains effective in environments where conventional electronic sensors often fail, such as underwater or in the presence of magnetic interference.

## INTRODUCTION

Robots transitioning from structured factory floors to unstructured environments—such as homes, healthcare settings, and collaborative workspaces—require the ability to perceive and physically interact with their surroundings[1-7]. While vision has traditionally dominated robotic perception[8-10], visual sensing can be unreliable under occlusions, poor illumination, or reflective and cluttered scenes[9,11]. In these contexts, tactile perception becomes crucial for applications that involve direct physical interaction with humans or the environment[1,12,13].

Embedding tactile sensors with high spatial and temporal resolution, however, remains challenging due to fabrication complexity, integration constraints, and environmental sensitivity. Many established tactile sensing technologies rely on embedded electronic components, including capacitive[14,15], resistive[16,17], and optical[18], as well as hybrid architectures. While effective, these approaches can face limitations when integrated into soft or deformable elastomeric substrates, where repeated strain may lead to debonding, conductor fatigue, or signal drift[19].

Motivated by these challenges, soft tactile sensors have emerged as a promising alternative, leveraging compliant materials to better conform to complex surfaces and interactions[20,21]. A wide range of transduction mechanisms has been explored, including magnetic[22,23], barometric[6,24-27], resistive[28,29], capacitive[30-32], optical[33-38], and vision-based methods[39-42]. Beyond material compliance, several studies have investigated strategies to relocate or minimize electronics at the contact interface[43-46], motivated by considerations of mechanical compliance, electromagnetic interference[47], and system integration.

Removing electronics and electrical components from the sensing site is one of several strategies investigated to improve robustness. One approach includes sensing based on acoustic wave propagation[48-52], in which a speaker emits a sound wave into a channel embedded within a soft membrane. Deformation of the channel under an external load alters the pressure wave, enabling touch detection. In addition, wave propagation can also be detected through the solid interface via Lamb wave in a thin elastomeric layer[48]. In another approach, Soter et al.[53] proposed a vision-based tactile sensor that utilizes a network of fluid channels embedded in an elastomer. By observing fluid displacement using an externally mounted camera, the system can infer deformation and applied load.

Given the importance of fluidic systems in soft robotics, particularly pneumatics, many studies have focused on developing soft pneumatic tactile sensors[54]. One technique for touch detection involves using the concept of an electrical touchpad, where arrays of pneumatic lines are arranged in a row and column formation. For instance, Zha et al.[55] fabricated a 3×3 grid of pneumatic pouches, placing pressure sensors at the end of each row and column, enabling the detection of nine contact areas using only six pressure sensors. Similarly, Choi et al.[56] introduced a 3×3 configuration of pneumatic channels embedded in an elastomer, using a multi-layered design to simultaneously actuate and sense the deformation for haptic interactions. To further expand the sensing area and enhance spatial resolution, Lampinen et al.[19] developed a 16×16 microchannel array, where 32 pressure sensors enabled dense touch detection over a larger surface area. Truby et al. reconstructed the shape of a cubic lattice prism with 3×3 axially embedded, air-filled channels and qualitatively detected the touch location by processing the corresponding nine pressure readings[54].

In this work, we propose a soft tactile sensing approach that exploits fluid–solid interactions in viscous fluid-filled channels combined with machine learning, enabling single-point touch localization and force estimation using only two external pressure sensors. Pressure sensors are positioned away from the contact interface, while touch-induced deformation generates characteristic differential pressure signatures that encode contact information. Unlike prior fluidic tactile systems that scale resolution through sensor count, our approach leverages data-driven inference to extract spatial information from minimal hardware. We validate this approach for single-point touch localization and force estimation, first in a linear (1D) configuration and then extended to two-dimensional tactile mapping by routing channels along space-filling curves, enabling 2D tactile mapping with the same simple hardware setup. While improved mechanical resilience is a motivating consideration of this architecture, damage tolerance and long-term durability are not quantitatively evaluated in this study. Instead, the focus is on perceiving touch position and strength with minimal sensing hardware, supported by experimental characterization, simplified modeling, and computational simulations, including operation under environmental constraints and in human–robot interaction scenarios.

## RESULTS

### Sensor design and working mechanism

Our sensor consists of a fluid-filled channel embedded within an elastomeric structure and connected to two solid-state pressure sensors positioned at both ends of the channel. When the elastomer surface above the channel is touched, a small volume of fluid is displaced toward the ends, creating a pressure difference between the left and right sides, $\Delta P = P_L - P_R$, detected by the pressure sensors. In its simplest form, the sensor features a linear channel within a rectangular elastomer block (Fig. 1A). During a touch event, interactions between the viscous fluid and the soft elastomer generate distinct pressure responses that depend on the fluid medium and touch position (Fig. 1B, Supplementary Movie 1), as well as on the applied touch force (Fig. 1C). By processing these pressure profiles with a machine learning algorithm, we can accurately estimate both the touch position and force.

This simple design can be extended to more complex surfaces by covering them with a meandering channel using a space-filling curve. In this work, we used the Hilbert curve[57], which covers a 2D surface with a single continuous non-overlapping path (Fig. 1D). Therefore, we can apply the same principle used for the 1D linear sensor with only two sensor readings at both extremes. Fig. 1E shows recorded pressures along the dimensionless channel length $\tilde{s}$ at four selected locations within each quadrant of the sensing area. While increasing the length of the channel may reduce the amplitude of the pressure signal at distant points (Fig. 1F, top), the viscous effect, reflected in the rise time, $t_r$, in pressure differentials (Fig. 1F, bottom), provides an extra feature that further enables the machine learning algorithm to detect touches.

To translate pressure readings into touch position and force estimates, we implemented the machine learning pipeline shown in Fig. 1G. Pressure time series were recorded from multiple touch locations for both 1D and 2D sensors. From the recorded signals, we extracted key features by calculating the maximum pressure at each end for the 1D sensor and both the maximum pressure and rise time $t_r$ at each end for the 2D sensor. This approach reduced the data dimensionality and enhanced computational efficiency compared to analyzing the full time series.

We then applied soft clustering[58] by partitioning the feature space into overlapping regions so each touch event can belong partially to multiple clusters. This divides the variable dataset into smaller, more homogeneous subsets, enabling assigning separate models to each cluster and reducing overfitting. The resulting membership values, together with the extracted features, were fed into an adaptive neuro-fuzzy inference system[59] (ANFIS) to model the nonlinear mapping between sensor outputs and touch parameters. This method achieved high-accuracy, simultaneous prediction of touch location and applied force using only two pressure sensors for both 1D and 2D sensor configurations.

**Tactile response of the 1D sensor**

We began our analysis by performing comprehensive tests on a 1D linear sensor to thoroughly characterize its response to touches applied both on-channel (directly above the channel) and off-channel (laterally offset from the channel), across channels of varying dimensions and filled with different fluids.

**On-channel touch**. Fig. 2A reports the difference between maximum left and right pressure ($\Delta P = P_L - P_R$) and measured normal forces $F$ as a function of indentation depth $\Delta z$ when the sensor is touched on-channel at $x/l = 0.1$ for three channel diameters ($ø = 1, 1.5, 2$ mm) filled with air, water, or oil ($\mu = 100$ cSt). The touch response exhibits three distinct phases. At low $\Delta z$, the equilibrium pressure difference $\Delta P$ is nearly zero, making touch interpretation unfeasible (Supplementary Materials S3). This behavior results from rapid backflow within the channel, which equilibrates $P_L$ and $P_R$. As $\Delta z$ increases, the channel begins to collapse, restricting the backflow. Consequently, the pressure difference becomes more pronounced with deeper indentations, leading to a larger $\Delta P$ values (Supplementary Movie 2). This middle region defines the sensor's effective working range, where accurate touch detection is possible. Beyond a certain indentation depth, the channel becomes fully occluded, preventing further fluid displacement. In this regime, $\Delta P$ saturates and ceases to vary with increasing $\Delta z$, rendering touch detection unreliable (Supplementary Materials S3). The measured force $F$ is greater for smaller channels with more elastomer volume, offering structural support, resisting indentation. Moreover, due to a slight diffusion of silicone oil into the silicone rubber, we observed that sensors filled with silicone oil exhibited slightly lower forces at high indentation depths.

Across all fluids, increasing the channel diameter ø, amplifies $\Delta P$ at the same $\Delta z$, because more fluid volume is displaced during touch, but at the same time, the sensitivity of the sensor at a small $\Delta z$ diminishes. Therefore, the channel diameter should be chosen based on the expected range of touch intensities. Among the three media, air—being compressible—produces the smallest pressure variations. In contrast, water and oil, which are nearly incompressible, produce notably larger $\Delta P$. The oil pressure is much higher than the water pressure due to its higher viscosity, which more strongly restricts flow between the two sides of the touch point.

In the following, we investigate the effect of the off-channel touches, the rise time in pressure histories, and the channel length in 1D sensors as a transition from 1D to 2D sensor design.

**Off-channel touch**. In Fig. 2B, we report $P_L$ for off-channel touches at $x/l = 0.1$ and varying depths $\Delta z$ when the fluid in the channel is oil ($\mu = 100$ cSt). The results indicate that as the touch points move farther from the channel, the pressure decreases, with sharper changes (higher sensitivity) observed for channels with larger diameters ø. Sensitivity to off-channel touches diminishes significantly at positions farther from the channel ($y \approx \pm 5\ \mathrm{mm}$), where the pressure readings converge to similar values across all diameters. For larger diameters, a greater area of the indenter comes into contact with the sensor at higher touch depths, further enhancing the pressure readings over a wider range. The off-channel response is relevant for practical scenarios as well as 2D sensor design since every touch has a finite radius of influence which is not necessarily always on-channel.

**Time response**. We investigated the effects of different media on the pressure history under tactile stimulation. The left panel in Fig. 2C shows $P_L$ and $P_R$ over time during touching, holding ($t = 2\ \mathrm{s}$), and removal of the indenter at $x/l = 0.1$, highlighting the 100% *rise time* ($t_L^r$) and *fall time* ($t_R^r$) of the pressure signal. In this plot, the touch depth reached $\Delta z = 2.5\ \mathrm{mm}$, and the displacement rate of the indenter for both touching and removal period was 100 mm/s. In this plot, it is evident that $P_R$ requires more time than $P_L$ to reach its maximum value. This behavior reflects the way pressure propagates through the oil: the sensor furthest from the touch point experiences a longer rise time, whereas the sensor closer to the touch point shows a shorter rise time ($t_R^r > t_L^r$). A similar trend in the fall time was observed during the removal of the indenter ($t_R^f > t_L^f$). Next, we extracted the rise time ($t_L^r$ and $t_R^r$) as a function of the normalized position ($x/l$) for different media. For oil-filled channels, as the touch moves further from the left side towards the right, $t_L^r$ increases while $t_R^r$ decreases, demonstrating a consistent pattern of delayed pressure rise as the distance from the touch point increases. This phenomenon could provide valuable additional data for more precise training of machine learning models. However, while oils with very high viscosity display more pronounced rise times, this also introduces a significant drawback: a slower response of the sensor. The extended rise time results in a delay in registering touch inputs, making the sensor less suitable for applications that require fast, real-time responses. For instance, in the case of the oil with $\mu = 200\ \mathrm{cSt}$, the combined rise and fall time for a moderate touch ($\Delta z = 2.5\ \mathrm{mm}$) is approximately 2.5 seconds, whereas for the oil with $\mu = 1000\ \mathrm{cSt}$, this value reaches nearly 10 seconds. Therefore, to balance fast response and information-rich pressure signals, we selected the oil with $\mu = 100\ \mathrm{cSt}$ for the rest of this study.

For air- and water-filled channels, the rise time and fall time are very similar to each other, and compared to oil, they occur quicker and with a reversed trend: $t_L^r$ decreases and $t_R^r$ increases as the touch moves further from the left probe, but the changes are very minimal. This reversal can be attributed to the lower viscosity of water and the high compressibility of air, leading to a faster pressure saturation. For air and water, the rate of pressure change is the same for left and right probes, whereas in oil-filled channels, the viscosity determines the rise time.

**Channel length**. We studied the effect of channel length ($l = 50\ \mathrm{mm}, 100\ \mathrm{mm}, 150\ \mathrm{mm}$) on the pressure response of the sensor for three fluid media (air, water, and oil) for a fixed channel diameter $ø = 1.5\ \mathrm{mm}$ (Fig. 2D). Across all media, increasing the channel length reduced $P_L$ for indentation along the channel, with this effect being more pronounced in oil-filled channels due to their higher viscosity and greater resistance to fluid flow. In the shortest channel ($l = 50\ \mathrm{mm}$), the pressure $P_L$ decreased more gradually with distance from the touch point, while in the longest channel ($l = 150\ \mathrm{mm}$), particularly in oil, pressure decayed more quickly. For instance, between $x/l = 0.1$ and $x/l = 0.5$, the pressure decreases nearly 52% in the shortest channel, whereas it drops by about 70% in the longest channel.

The response also varied by medium: oil, with its high viscosity, maintained higher pressure readings over a greater portion of the channel, enhancing sensitivity to touch along its length. In contrast, air-filled channels exhibited the fastest pressure decay, followed by water, highlighting the impact of fluid properties on pressure propagation. These findings suggest that shorter channels have a higher sensitivity, i.e. instantaneous slope of the $P$-$x/l$ curve, at the channel ends, hence, are better suited for applications requiring a faster pressure response, while longer oil-filled channels provide superior pressure resolution for detecting touches at greater distances.

**Understanding the sensing mechanism using theory and simulation**

To better understand the rich tactile response of the 1D linear sensor presented in the previous section, we developed a simplified theoretical model. We fitted this model to our experimental data to explain various observed phenomena. Additionally, we performed numerical simulations using COMSOL Multiphysics, incorporating fluid-solid interaction (FSI) to provide a more realistic representation of the sensor's behavior.

**Theoretical model**. We assume that touching the 1D sensor at a distance $x$ ideally displaces a fluid volume $\Delta V$, which moves equally to the left and right sides of the indenter without any flowback. Half of this volume, $\frac{1}{2}\Delta V = \Delta V_t^i + \Delta V_e^i, i = (L, R)$, will deform both the tube and the channel embedded in the elastomer on each side, pressurizing the left side with $P_L = K_t\, \Delta V_t^L / V_t^L = K_e\, \Delta V_e^L / V_e^L$ and right side with $P_R = K_t\, \Delta V_t^R / V_t^R = K_e\, \Delta V_e^R / V_e^R$ where $K_t$ and $K_e$ are the effective bulk moduli of the tube system and the elastomer, respectively. Since the connecting tubes have the same length, $V_t^L = V_t^R = V_t$ while $V_e^L = Ax$ and $V_e^R = A(l - x)$ where $A$ is the cross-sectional area of the channel. When the sensor is touched at the midpoint ($x/l = 0.5$), $P_L = P_R = P_M$. By defining $\kappa_t = K_t l / V_t$ and $\kappa_e = K_e / A$ and combining these equations, simple relationships for $P_L$ and $P_R$ as a function of the normalized bulk modulus $\kappa = \kappa_e / \kappa_t$ and $P_M$ can be derived (Supplementary Materials S7):

$$P_L = P_M \frac{\kappa + 1/2}{\kappa + x/l}\,, \qquad P_R = P_M \frac{\kappa + 1/2}{\kappa + 1 - x/l} \tag{1}$$

We fitted these equations to our experimental results by identifying $\kappa$ as a function of the indentation depth $\Delta z$. At each $\Delta z$, the experiments consist of 80 pressure points at $x/l \in [0.1, 0.9]$ where $P_M$ measured at $x/l = 0.5$. Fig. 3A shows that the model fits well with the experimental results for an oil medium ($\mu = 100$ cSt) with $ø\ = 1.5$ mm and $\Delta z = 3$ mm. When fitted to experiments for oil medium across different channel diameters $ø$, the normalized bulk modulus $\kappa$ decreases as the channel diameter increases, since the effective bulk modulus of the sensor structure decreases (Fig. 3B). Furthermore, we observed a sharp decrease in $\kappa$ values around $\Delta z \approx ø$ (fully collapsed channel) after which $\kappa$ stabilizes. In theory, we expect a constant $\kappa$ across different indentation depths $\Delta z$. However, flowback during indentation, where fluid moves from high-pressure to low-pressure regions before channel closure, violates the assumption of evenly distributed volume on both sides of the indenter. This discrepancy becomes more pronounced when fitting the model to other fluid media. Water, with its low viscosity, facilitates flowback, while air compressibility results in volume changes, both leading to higher $\kappa$ values as demonstrated in Fig. 3C.

**Numerical simulations**. We conducted a series of fluid-solid interaction (FSI) simulations in COMSOL Multiphysics to support our hypothesis on the effect of backflow on the measured pressure difference at two ends of the channel. The solid domain consists of silicone rubber (Ecoflex 00-30), modeled as a nearly incompressible Neo-Hookean hyperelastic isotropic material with a density of $\rho_s = 1070\,\mathrm{kg/m^3}$ and a shear modulus of $G = 27\,\mathrm{kPa}$. The fluid domain inside the channel is treated as a single-phase, laminar Newtonian flow under isothermal conditions, characterized by fluid density $\rho_f$ and dynamic viscosity $\mu_f$. The evolution of pressure over time was analyzed by solving the Navier-Stokes equations in a transient framework. Upon indentation, the viscous, laminar flow in the partially open channel can be compared to Poiseuille flow through a pipe, with the volumetric flow rate given by $Q = ø^3 \Delta P / 8 \mu L$.

Fig. 3D shows snapshots of the FSI simulations when the sensor is indented at three regions ($x/l = 0.1, 0.5, 0.9$), demonstrating the pressure difference on both sides of the indenter. The calculated pressure curves presented in Fig. 3E exhibit trends like those observed in experiments (Supplementary Movie 3). For asymmetric indentations ($x/l = 0.1, 0.9$), a pressure drop was observed immediately after indentation, highlighting the effect of backflow.

To further investigate this effect, we systematically increased the indentation depth by increasing the applied force (Supplementary Materials S8). Fig. 3F shows the pressure history at two applied pressure levels corresponding to moderate ($\Delta z = 1.1\,\mathrm{mm}$) and strong ($\Delta z = 1.8\,\mathrm{mm}$) touches for channels filled with oils with three different viscosities. At moderate indentations, the pressure drop was largest for the oil with the lowest viscosity and decreased as viscosity increased. For strong touches, since the channel was nearly blocked under the indenter pressure, backflow significantly decreased, and the pressure drop diminished for all viscosities. We quantified the flux across the channel for these two cases in Fig. 3G, confirming that less viscous fluids and lighter touches result in larger backflow. Fig. 3H shows snapshots of the transient pressure contours for the low-viscosity oil under the two indentation depths considered above. These results show the pressure difference fades out for the lighter touch due to the flowback, but it persists for the stronger touch (Supplementary Movie 4).

Finally, we measured the pressure difference $\Delta P$ once the maximum indentation load is attained under two assumptions at different indentation depths. First, we considered an extreme case with a virtual interior wall that fully closed the channel, dividing it into two separate fluid domains, similar to the simplified theoretical model. Second, we considered an initially open channel that allows backflow. In this scenario, the channel is considered closed once the channel walls come into contact, with a small offset to ensure numerical convergence. We performed simulations for air and different oils with varying viscosities, as presented in Fig. 3I. The results show that under the channel closure assumption, $\Delta P$ is the same for all oil viscosities and slightly higher than in a compressible air medium. When backflow is allowed, the pressure quickly equilibrates for air, whereas for oil, higher viscosity leads to a larger $\Delta P$. As expected, stronger touches always result in higher pressure differences. The nearly linear behavior of $\Delta P$ versus $\Delta z$ under the closed-channel assumption or for high-viscosity fluids confirms the validity of our simplified theoretical model in these limiting cases.

**Tactile response of the 2D Sensor**

Next, we characterized the tactile response of our proposed 2D sensor, which is an extension of the 1D design where the channel follows a meandering path to cover a 2D surface. Fig. 4A shows a 2D sensor designed with a 4th-order Hilbert curve. To better visualize the indentation response, we divided the sensor into four quadrants ($Q_1, Q_2, Q_3, Q_4$). The position along the channel was normalized by its total length as $\tilde{s} = s/l$. In the following, we examine the response of the 2D sensor to on-channel touches.

We measured the pressures at both ends of the 2D channel while indenting the sensor at 16 locations, defined as $\tilde{s} = (2i - 1)/32$ (where $i \in [1,16]$), and at 16 indentation depths $\Delta z \in [1\ \mathrm{mm}, 4\ \mathrm{mm}]$ for three fluid media (air, water, oil). The results are shown in Fig. 4B for oil ($\mu = 100\ \mathrm{cSt}$) and in Supplementary Materials S9 for air and water. The findings confirm the trend observed in the 1D sensor. $P_L$ was higher for indentations closer to the left probe ($\tilde{s} = 0$) and lower when indenting near the right probe ($\tilde{s} = 1$), while $P_R$ followed the opposite trend. As in the 1D configuration, the pressure response saturated beyond a certain indentation depth, corresponding to channel collapse and the cessation of fluid displacement.

For air and water, the 2D sensor is not sensitive to slight and moderate touches due to the rapid flowback. However, in the case of oil, the combination of higher viscosity, the tortuous Hilbert path, and the longer channel minimizes flowback, enhancing both sensitivity and spatial resolution for detecting slight touches. Fig. 4C zooms in on a selected indentation depth ($\Delta z = 3\ \mathrm{mm}$), showing the evolution of the pressure and force across different locations and quadrants (Supplementary Movie 5). Fig. 4D looks closer at one indentation event (#6), revealing a delay ($\Delta t \approx 0.85\ s$) between pressure peaks of $P_L$ and $P_R$. This finding suggests that, in addition to $\Delta P$, the rise time can be leveraged as an additional feature in the learning process to improve touch position and force estimation.

**Learning touch position and force**

We evaluated several learning algorithms to estimate position and force in the 1D sensor and selected Fuzzy C-Mean (FCM) clustering as the most effective model based on performance (Supplementary Materials S12). We then applied FCM to the 2D data. Unlike traditional clustering methods that assign each data point to a single cluster, FCM allows data points to belong to multiple clusters simultaneously, with varying degrees of membership. This soft assignment helps capture the nonlinearity in sensor readings and reflects the continuous nature of the sensor, thereby improving the accuracy of position and force estimation. Final estimates are computed as weighted sums of cluster outputs based on membership values.

First, we collected pressure data by indenting a 1D oil-filled sensor ($\mu = 100\ \mathrm{cSt}$, $\o = 1.5\ \mathrm{mm}$) at coarse spatial intervals. We then visualized the feature space as a function of touch position $x$ and indentation depths $\Delta z$ (Fig. 5A and Fig. 5B) to identify the sensor's effective operating range. Three zones were identified: (1) a *position-insensitive zone* where $P_L \approx P_R$, (2) a *force-insensitive zone* where the channel collapses and the pressure saturates, further increasing in $\Delta z$ or applied force $F$, do not result in additional fluid displacement or pressure change; and (3) a *working range* between the two, where both position and force can be reliably distinguished. Next, we collected a learning dataset at high spatial resolution in the working range and segmented the position and force data with FCM, as illustrated in the insets of Fig. 5A and Fig. 5B. We optimized the number of clusters by evaluating RMSE and $R^2$ scores for both position and force estimation (Supplementary Materials S12), with 20% of the collected data reserved for validation.

Fig. 5C shows the position estimation results, color-coded with $\Delta z$. The predictions closely match the true values overall, with a slight decrease in performance at shallow indentation depths, where flowback within the channel reduces the pressure difference between the two ends. In contrast, force estimation is more challenging, particularly at deeper indentations where pressure variations become more subtle (Fig. 5D). Furthermore, due to the complex, non-linear relationship between force and pressure, simpler models such as Linear Regression, Polynomial Regression, and Multi-Layer Perceptron (MLP) failed to achieve the same level of accuracy as FCM (supplementary materials S12).

To better understand the spatial structure of the extracted feature used for learning, we visualized the distribution of the maximum left pressure ($P_L$) and the corresponding rise time ($t_r^L$) across the full 2D surface at representative indentation depths ($\Delta z = 2$ mm and $3$ mm). The pressure magnitude (Fig. 6A) exhibits a clear spatial gradient along the Hilbert curve, with stronger signals near regions closer to the sensing outlet and attenuated responses in areas farther away. In contrast, the rise time feature (Fig. 6B) varies more gradually along the channel path and provides complementary spatial information that is less sensitive to absolute pressure attenuation. The combination of these two features enhances separability in the feature space and improves robustness of position estimation, especially in regions where pressure amplitude alone becomes weak. The detailed results of the extracted feature is presented in Supplementary S13.

For the 2D sensor, we identified the working range by analyzing the corresponding feature space (Supplementary Materials S13). Due to the tortuous channel geometry, the flowback was significantly reduced in the 2D configuration, resulting in an expanded working range compared to the 1D sensor. Once this range was established, we acquired a comprehensive dataset for model training. To ensure consistent spatial sampling, we focused initially on “on-channel” indentations by selecting 256 uniformly distributed points along a 4th-order Hilbert curve at each indentation depth, where $\tilde{s} \in [0,1]$ denotes the normalized arc length along the curve.

Fig. 6C shows the estimated normalized position $\tilde{s}$. The model achieves high predictive accuracy ($R^2 = 99.18\%, RMSE = 0.026$), comparable to the 1D sensor. Minor deviations occurred primarily at shallow indentation depths, especially in the second and third quadrants ($Q_2, Q_3$), where the pressure signal is weaker due to a longer distance from the sensing elements. Despite these localized limitations, the trained FCM model reliably maps touch locations along the curve. Real-world Cartesian positions can be recovered from $\tilde{s}$ using the conversion code provided in GitHub[60].

Fig. 6D shows the force estimation performance for on-channel indentations, with strong results across most of the range ($R^2 = 84.74\%, RMSE = 0.66$ N). The model effectively captures force trends at low to moderate indentation levels. However, at higher forces, prediction variability increases slightly, likely due to a combination of pressure saturation and fabrication tolerances related to the 2D sensor’s more intricate geometry.

To better reflect real-world scenarios, where touch interactions are not constrained to predefined trajectories, we extended the dataset to include off-channel indentations across the entire 2D surface. At each depth, we sampled a grid of 31 × 31 indentation points uniformly distributed across the area. Fig. 6E presents a heatmap of the resulting position estimation errors, showing that the model maintains low spatial error across most of the sensor area. A few larger errors were observed, primarily localized in regions where either the left or right pressure readings are weak, reducing the reliability of position estimation in those zones.

The force estimation performance for the full dataset, including both on- and off-channel indentations, is shown in Fig. 6F. While the model performs reasonably well ($R^2 = 82.04\%, RMSE = 0.71$ N), a slight reduction in accuracy is observed compared to the on-channel case. This decline is partly due to the heterogeneous nature of the sensor: off-channel touches often engage stiffer structural regions, resulting in higher absolute force values. At the same time, these touches tend to displace less fluid, producing weaker pressure signals and causing the model to slightly underestimate the applied force. Moreover, for off-channel touches in 2D sensors, the probe may engage multiple locations along the meandering Hilbert curve (corresponding to different $\tilde{s}$), leading to additional localization error.

**Adaptability to Environmental Constraints**

To validate the versatility of our design, we conducted a series of demonstrations under diverse conditions that typically challenge conventional tactile sensors, such as underwater operation and exposure to magnetic interference. We also demonstrated the effectiveness of the sensing architecture and performed real-time touch estimation with indenters of varying shapes and sizes, including on a curved substrate. Collectively, these experiments highlight the effectiveness of the proposed sensor concept and its adaptability to different environments.

**Underwater sensing.** With no electronics present at the sensing site, we were able to operate underwater without the risk of short circuits or the need for insulation (Fig. 7A). The same external pressure sensors used in the standard air-based setup were employed, with the tubing simply extended outside the water. Indentation tests were performed at a depth of $\Delta z = 3$ mm across nine contact points, both under dry condition and with three different water levels above the sensor ($h = 10$ mm, $30$ mm, $50$ mm). We observed that hydrostatic pressure caused a baseline shift proportional to the water level; however, once this offset was accounted for, the pressure response to touch remained effectively unchanged across all scenarios. The differences between the dry case and all underwater conditions were minimal (< 5 Pa), confirming that the sensor maintained its sensitivity regardless of the submersion (Supplementary Movie 6). The same behavior was observed in the 2D scenario, as presented in Supplementary Materials S14.

**Sensing under magnetic field.** We investigated the impact of magnetic fields on sensor performance, since many tactile sensing modalities are prone to interference (Fig. 7B). As a representative case, we tested a commercial barometric sensor, which only responded to indentations directly above or near its surface, leaving many points undetected. With a magnetic indenter, performance degraded further, pressure readings became noisy, and communication soon failed, underscoring poor robustness in magnetically active environments. In contrast, our fluidic tactile sensor remained unaffected. Using the same indentation protocol ($\Delta z = 2$ mm, with $\Delta x = 2.5$ mm), with both plastic and magnetic indenters, we obtained nearly identical pressure patterns, a robustness also confirmed in the 2D sensor (Supplementary Materials S14).

**Effect of sensor size.** We studied sensors of various dimensions through geometric modification (Fig. 7C). Two smaller sensors were fabricated—one half-sized (30 mm × 30 mm, $4^{th}$-order Hilbert curve, channel diameter $\emptyset/2$) and one quarter-sized (15 mm × 15 mm, $3^{rd}$-order Hilbert curve, channel diameter $\emptyset/2$)—without altering the external electronics. Indentation tests were conducted every 2 mm in both axes while recording pressures and applied force, using the same training and evaluation pipeline as for the full sensor (60 × 60 mm). For data collection, we used a 4 mm indenter and the results in both raw and normalized units showed consistent performance across scales results in both raw and normalized units showed consistent performance across scales. The mean spatial errors remained around 5% and 7% in x and y directions, and force estimation was comparable for halved and full sensors. The 15 × 15 mm sensor showed slightly reduced accuracy, likely due to its lower Hilbert curve order. Overall, the results confirm that the fluidic sensing concept allows simple geometric scaling of size and resolution with minimal hardware modification.

**Effect of indenter shape and size.** We evaluated the robustness of the 1D sensor under varying contact conditions by testing indenters of different shapes and diameters, although the system had been trained only with an 8 mm indenter (highlighted in Fig. 7D). Data were acquired and processed in real time using an oil-filled sensor with 1.5 mm channels, without pre-collection or batch processing. Ten random indentation points were selected across the x- (position) and z- (depth) directions to simulate live estimation scenarios. Three 8 mm indenter shapes—flat, half-oblate spheroid (HOS), and hemisphere (HS)—were tested (Fig. 7D left). The mean positional error remained below 5 mm (≈5%) in all cases, confirming reliable performance beyond the geometry of the indenter used for training. The flat indenter yielded slightly higher errors due to less effective channel sealing compared to the curved HOS and HS tips, which enhanced sensing by reducing fluid backflow. Varying the HOS indenter diameter from 4 to 10 mm resulted in mean errors below 2.5 mm (Fig. 7D right). Notably, reducing the indenter size leads to more localized deformation and increased flowback—particularly at shallow depths—thereby weakening pressure signals and resulting in larger deviations. These results demonstrate that the sensor adapts to diverse contact conditions while maintaining low error, even under online operation. The live data collection and processing code is provided in our GitHub repository[60].

**Conformity to curved substrates.** Finally, we evaluated the sensor's performance when mounted on a non-planar substrate composed of three regions with distinct radii of curvature (Fig. 7E). Although the sensor was trained only in a flat configuration, it successfully generalized to curved geometries and accurately detected touches across regions located on both convex and concave surfaces (Supplementary Movie 7).

**Human-Robot Interaction**

We demonstrated the real-time performance of our 1D and 2D tactile sensors by integrating them with a collaborative robotic arm (UR5e, Universal Robots) to perform various human–robot interaction (HRI) tasks (Fig. 8).

For the 1D sensor, the robot's position along the X-direction was controlled using the estimated position obtained either from the sensor mounted on the table (Fig. 8A) or directly on the robot arm (Fig. 8B) (Supplementary Movie 8). Additionally, we performed position estimation for the 1D sensor mounted on the robot using different processing windows (Supplementary Movie 11). To further validate force estimation, we performed a controlled injection task by dividing the sensor's active area (10–90 mm) into five spatial segments (16 mm intervals) and three force ranges (0–5 N, 5–10 N, and >10 N). Each touch was mapped to a corresponding 5 × 3 grid cell representing the estimated contact position and applied force (Fig. 8C). The mapping accuracy remained high across nearly all cases, with only a slight underestimation observed at the highest force level (3rd touch), consistent with the modest reduction in high-force sensitivity previously noted in Fig. 8D (Supplementary Movie 9).

For the 2D sensor, we mapped the sensor's XY coordinates to the robot's XZ workspace and controlled the UR5e arm position accordingly. Similar to the 1D case, we conducted tests with the sensor mounted both on the table and directly on the robot's curved surface (Fig. 8D–E). The sensor was touched in each quadrant, and the robot successfully moved to the corresponding positions (Supplementary Movie 10). When mounted on the robot arm, the sensor's reference frame moved together with the robot, resulting in baseline shifts in the pressure readings even in the absence of physical contact, due to hydrostatic pressure, which posed a challenge for sensing accuracy. To address this issue, we implemented a baseline correction procedure that compensated for the motion-induced drift, allowing accurate estimation of touch positions. Notably, in this configuration, the sensor conformed to the robot's curved surface despite being trained only on flat samples, yet it continued to provide reliable position estimation across all tested points.

## DISCUSSION

In this work, we presented a soft tactile sensor that harnesses fluid–solid interactions to transduce both position and force without requiring electronics at the sensing site, relying solely on two external pressure sensors. By increasing the fluid viscosity—from air-filled to water- and silicone oil-filled channels—we observed stronger coupling between the fluid and solid domains, thereby enhancing sensitivity to external touch. The simple 1D design can be extended to higher dimensions through a space-filling architecture, such as a Hilbert curve, enabling scalable mapping with minimal hardware and achieving accurate localization (Mean Average Error (MAE): 4 $\mathrm{mm}$, <7% error) and robust force estimation across both 1D and 2D configurations. We expect this strategy to be adopted for 3D surfaces as well. To infer tactile information from pressure readings, we implemented fuzzy c-means clustering combined with an adaptive neuro-fuzzy inference system (ANFIS), which partitions the complex nonlinear feature space into soft clusters. This approach improved robustness, outperformed standard regression methods, and better captured the rich dynamics of tactile responses.

Table 1 benchmarks our proposed 1D and 2D soft tactile sensors against representative state-of-the-art tactile sensing technologies that rely on interpreting pressure or acoustic profiles to infer touch position and force. Compared to acoustic-based approaches, which require 1–2 sensors to achieve moderate accuracy over limited ranges, our sensors achieve comparable or better positional and force resolution using only two external pressure sensors. In the 1D configuration, our sensor reaches a position accuracy of 1.8 $\mathrm{mm}$ and reliably estimates forces in the 1–10 $\mathrm{N}$ range, matching or exceeding the performance of pressure sensor arrays with far more elements. Extending the design into a 2D Hilbert curve layout enabled spatial mapping over a 60×60 $\mathrm{mm}^2$ area, achieving positional errors of 3.3–4.2 $\mathrm{mm}$ and force estimation with 0.7 $\mathrm{N}$ accuracy, while dramatically reducing the number of required sensing elements, such as 7[56], 25[26], 32[19], to 2 physical sensors. Furthermore, the absence of electronics at the sensing site allows reliable operation in environments that are challenging for conventional tactile sensors, such as underwater or in the presence of magnetic interference. While underwater and deep-see tactile sensing has been previously demonstrated using a range of established technologies[61-65], our results show that the proposed sensor remains functional when submerged and therefore represents an additional sensing modality applicable to underwater and harsh environments.

We conducted comprehensive experimental characterizations of the 1D sensors, complemented by theoretical modeling and fluid–solid interaction (FSI) simulations to elucidate the sensing mechanism. First, we examined the effect of fluid viscosity by comparing silicone oils of different viscosities with water-filled and air-filled channels. The results revealed that incompressibility and higher viscosity significantly enhance the pressure response, yielding greater sensitivity and an extended working range. Next, we developed a simplified physical model linking the left and right pressures to the touch position along the channel. The model captured the main steady-state trends observed experimentally, providing an analytical description of the tactile response. However, it could not fully reproduce the system's behavior at low indentation depths, where transient effects dominate and flowback alters the dynamics. Therefore, we resorted to FSI simulations, which addressed this limitation by revealing that pressure equilibrium arises from rapid fluid flowback during the initial indentation stage. These simulations further confirmed that this effect is suppressed at higher viscosities, supporting our experimental observations.

A practical limitation encountered during experimentation arose from the chemical compatibility between the fluid and the elastomer. Due to PDMS's high permeability, many silicone oils gradually diffused into the elastomeric walls, leading to a progressive pressure drop over time[66]. This effect restricted our usable fluid selection to a specific silicone oil formulation exhibiting reduced reactivity with PDMS. While adequate for stable measurements, this constraint prevented systematic exploration of intermediate viscosities. Future work could benefit from water-based solutions (e.g., polyvinylpyrrolidone, see Supplementary Materials S5) to mitigate diffusion and enable a broader viscosity range (1–100 $cSt$). Such fluids may also improve response time in 2D configurations, where pressure losses are more pronounced.

Three sources of sensor calibration drift were identified, including diffusion-induced baseline changes, variations in the sensor frame due to three-dimensional motion, and shifts in ambient pressure. Diffusion effects become negligible once the system reaches steady state after fabrication and are effectively compensated by baseline adjustment, as confirmed by long-term measurements (see Supplementary Materials S4, and Supplementary S10). Motion-induced frame changes and ambient pressure variations introduce hydrostatic pressure offsets; however, in all cases, subtracting the baseline improves position estimation and yields consistent pressure responses. Overall, this approach robustly resolves calibration drift and eliminates the need for repeated model retraining, enabling reliable performance across experiments conducted months apart using a single initially trained model.

Increasing the sensor size inevitably lengthens the fluidic channels, thereby increasing pressure dissipation. This effect is particularly evident in regions $Q_2$ and $Q_3$ and near the middle of the Hilbert curve—areas farthest from the reading points—where pressure decay and a reduced signal-to-noise ratio result in higher localization errors. Such behavior is intrinsic to space-filling patterns, as their total path length increases with higher Hilbert orders or larger sampled domains. Nevertheless, as shown in Fig. 6B, the rise time associated with viscous effects remains distinguishable and exhibits a more gradual spatial variation along the Hilbert curve, providing a complementary cue that is less sensitive to signal attenuation. Accordingly, fine-tuning the fluid viscosity to minimize pressure loss, together with more efficient learning algorithms, is expected to further improve localization accuracy.

Furthermore, the 2D sensor exhibits non-homogeneous pressure–force behavior between adjacent channels, with intermediate regions generating larger forces for smaller pressure changes. Increasing the resolution of the space-filling curve or reducing channel spacing could yield a denser and more uniform pressure–force response across the surface. Alternatively, adding a small number of additional pressure readings—still far fewer than in conventional pressure arrays—could mitigate this effect, although this approach is not ideal. Employing more advanced fabrication methods[67] for the 2D sensor could minimize force non-uniformity within each layer, thereby improving the accuracy of force estimation. Finally, one approach is to augment continuous space-filling curves using advanced fabrication techniques, such as freeform injection molding[68], to add an array of fluidic reservoirs ('taxels') that locally amplify pressure signals without increasing the number of transducers. As a more radical alternative, one could employ a maze-generation algorithm to construct a connected maze with two entrances, such that every point in the maze lies within a prescribed maximum distance from at least one outlet of the sensor, thereby preserving signal strength.

From a processing standpoint, we deliberately adopted a minimal feature extraction strategy, relying on the maximum left and right pressures for 1D sensors and, for 2D sensors, the corresponding rise times. This approach ensured high computational efficiency and enabled real-time operation with limited data bandwidth. Despite its simplicity, the system achieved satisfactory localization and force estimation. Nonetheless, performance could be enhanced by incorporating richer temporal features or employing time-series-based learning models (e.g., LSTM) to better capture dynamic behaviors, particularly in 2D cases.

Overall, this study establishes a baseline demonstrating how viscous fluids can enhance tactile sensing relative to air-based systems. Viscosity suppresses flowback, increases sensitivity, and broadens the detectable range, all while maintaining minimal hardware and computational requirements. Although challenges such as material compatibility, non-uniform pressure distribution, and slow response time persist, the findings provide a foundation for a simple, adaptable, and computationally efficient tactile sensing in soft robotics.

Future work should investigate various touch modalities—such as multi-point contact, sliding interactions[69], and responses to shear forces and moments—since real-world interactions often involve multiple or dynamic surface engagements. Current experiments focused on flat surfaces, but investigating angular and curved geometries and benchmarking their performance against flat configurations will be critical for applications such as collaborative robots, where contact surfaces are rarely planar and often in motion. Moreover, to account for environmental effects such as temperature-dependent changes in fluid viscosity and related properties, more sophisticated recalibration algorithms such as dynamic time warping[70] can be used in addition to baseline correction, rather than resorting to full model retraining. Finally, the developed sensor exhibits a response time on the order of seconds, partly due to the high viscosity of the selected fluid and chosen signal features. Systematic dynamic characterization highlights loading-rate–dependent behavior in this proof-of-concept (see Supplementary note S6), pointing to clear opportunities for performance optimization through viscosity tuning and advanced signal-processing approaches to enable more responsive tactile estimation for human–robot interaction, where immediate feedback is essential.

## METHODS

**Fabrication**: We fabricated the 1D sensor from silicone rubber (Ecoflex 00-30, Smooth-On) following a one-step molding process. First, we 3D printed a two-part mold using a stereolithography (SLA) resin printer (Form3, FormLabs) with two alignment holes on either side for positioning the central rod (Supplementary Materials S1). The sensor's dimensions were $w \times t = 20\ \mathrm{mm} \times 5\ \mathrm{mm}$. To investigate the effects of channel length and diameter, we created molds with channel lengths of $l = 50, 100,$ and $150\ \mathrm{mm}$, and diameters of $ø = 1, 1.5,$ and $2\ \mathrm{mm}$. The silicone prepolymer was degassed at -1 bar, poured into the mold, and cured in an oven at 60°C for 20 minutes. After curing, the sensor was demolded, and the central rod was removed. Two flexible tubes (GESSULTRA-C-040-0F, Sani Tech) were then attached to both ends using Sil-Poxy (Smooth-On). Finally, the channels were filled with different fluids, including air, water ($\mu = 1\ \mathrm{cSt}$), and different silicone oils (Sigma Aldrich) with the viscosity of $\mu = 100, 200, 1000\ \mathrm{cSt}$ to evaluate the effect of fluid compressibility and viscosity (Supplementary Materials S1).

For the 2D sensor, we used a two-stage process. First, a mold featuring the Hilbert curve was 3D printed using SLA. The dimension of the sensing interface was $l_x \times l_y \times t = 60\ \mathrm{mm} \times 60\ \mathrm{mm} \times 5\ \mathrm{mm}$ comprised of a continuous channel in the form of a 4th-order Hilbert curve with square cross-section, and the side edge of $a = 1.8\ \mathrm{mm}$, maintaining the same volume as a circular channel with $ø = 2\ \mathrm{mm}$. After pouring and curing the silicone rubber, similar to the process for the 1D sensor, the Hilbert curve section was demolded, and the surface was brushed with uncured silicone rubber to ensure that the channels remained unfilled. A separate 1.5 mm-thick sheet of Ecoflex 00-30 was then applied over the Hilbert curve, ensuring full contact without obstructing the channels. The entire assembly was cured again, resulting in a uniform 2D sensor with encapsulated channels within the elastomer. Finally, flexible tubes were attached to the sensor, and the channels were filled with silicone oil (AP100, Sigma-Aldrich). Supplementary Materials S1.

**Experimental Setup**: The measurement setup consists of two liquid-compatible differential pressure sensors (Honeywell, SSCDRRV100MDAA5) connected to a data acquisition card (NI-DAQ USB-6009) that transmits recorded signals to a personal computer. To perform comprehensive testing and generate large datasets for training the machine learning algorithm, we employed a robotic arm (UR5, Universal Robots) with an attached indenter featuring a rounded tip (8mm diameter, CAD files in Supplementary Materials S15). A miniature load cell (5 lbs, Futek) was mounted to accurately measure the applied force during touch events (Supplementary Materials S2).

**Numerical simulations:** The mechanical response of the sensor was modeled in COMSOL Multiphysics as a 2D fluid-solid interaction problem. The computational domain ($l = 100\ \mathrm{mm}$, $t = 5\ \mathrm{mm}$) consists of a $2\ \mathrm{mm}$ fluid channel flanked by a solid material. Since the plane strain assumption would imply that the fluid domain extends out of the plane, this cannot sufficiently constrain the in-plane deformation, we opted for an axisymmetric formulation where the rotation axis was placed at the lower edge of the sensor ($r = 0$), Supplementary Materials S8. In this case, any in-plane displacement of the solid material results in circumferential stress due to the nonzero material's Poisson's ratio, thus effectively providing a natural spring boundary condition. For $r \to \infty$, the plane strain condition is recovered. The solid domain consists of silicone rubber (Ecoflex 00-30), modeled as a nearly incompressible Neo-Hookean hyperelastic isotropic material model with a density of $\rho_s = 1070\ \mathrm{kg/m^3}$ and a shear modulus of $G = 27\ \mathrm{kPa}$. The fluid domain inside the channel is treated as a single-phase, laminar Newtonian flow under isothermal conditions, characterized by a weakly compressible fluid density $\rho_{f,ref} = 1000\,\mathrm{kg/m^3}$ and dynamic viscosity $\mu_f$. The evolution of pressure over time was analyzed by solving the Navier-Stokes equations in a transient framework. Gravity effects are neglected. The structural problem is governed by the nonlinear elasticity equations, assuming large-strain behavior, geometric nonlinearity, and isotropic material properties. The solid is discretized using quadratic Lagrange elements in a free tetrahedral mesh, which is refined along the contact boundaries. The accuracy of the flow field along the radial channel cross-section is not considered relevant, as we only want to resolve the pressure profile along the main channel axis. The fluid field and boundary layer were therefore coarsely discretized using standard piecewise linear interpolation elements (type P1 + P1). See Supplementary Materials S8 for the mesh realization in the indenter region. All channel walls are modelled with no-slip fluid flow conditions, while the deformable structure obeys a dynamic Neumann condition based on fluid-structure interaction forces.

The indentation of the soft sensor is implemented as a force-controlled experiment. A Gaussian pressure distribution with peak magnitude $F_{max}$ at the indenter centerline $\mu = x$ and width $6\sigma = d_{indenter}$ is ramped up over time with a constant velocity of $v = 120\,\mathrm{N/s}$, see Supplementary Materials S8. The contact between the sensor's top and bottom walls is modeled using a dynamic penalty formulation with a sufficiently small contact offset, ensuring that the fluid domain topology remains unchanged upon contact. Supplementary Materials S8.

**Data collection:** We first collected 1D data using position-controlled indentation at 9 x-axis locations ($x = [10, 90]$ mm with $\Delta x = 10$ mm) and 41 depths ($\mathrm{z} = [0,4]$ mm with $\Delta z = 0.1$ mm). Each dataset contained $P_L$, $P_R$ and $F$. The lower bound of the working depth was defined between the first detectable pressure difference ($\Delta p = 0.05$ kPa at $x = 10$ and $90$ mm). The upper bound of the working depth where both $\Delta P$ and the average pressure ($\bar{P}$) changed by <1% relative to the previous step. Within this range, data were collected at 41 x-locations ($\Delta z = 0.1$ mm) For each indentation, maximum $P_L$, $P_R$ and $F$ were extracted, paired with position, and stored for training.

For the 2D sensor, preliminary data were obtained from 16 locations along the normalized Hilbert curve($\tilde{s} = (2i - 1)/32$, $i \in [1,16]$), and 21 depths ($z = [0,4]$ mm with $\Delta z = 0.2$ mm). The working depth range was defined as in the 1D case, with the added condition of a minimum pressure threshold ($P > 0.01$ kPa), due to weak responses at shallow depths. Two datasets were then acquired: (1) on-channel indentations at 256 points along a 4th-order Hilbert curve ($\tilde{s} = [0,1]$ $\Delta\tilde{s} = 1/256$, $16 \times 16$ resolution), and (2) combined on/off-channel indentations over a $31 \times 31$ uniform grid ($x, y \in [0, 60]$ mm, $\Delta x = \Delta y = 2$ mm). For all indentations, the maximum $P_L, P_R$, rise times ($t_L, t_R$), and $F$ were extracted, paired with spatial coordinates and depths, and stored for training.

**Machine learning:** We developed a hybrid framework combining fuzzy C-means (FCM) clustering and an Adaptive Neuro-Fuzzy Inference System (ANFIS) to predict indentation position and force from pressure response. For the 1D case, inputs were $P_L$ and $P_R$; for 2D, rise times ($t_L, t_R$) were also included. Output labels were the indentation position $X$ and force $F$. Data were shuffled (random state = 7) and split into 80% training and 20% validation sets.

First, FCM clustering reduced dataset dimensionality and improved learning by assigning membership degrees across clusters, with overlap controlled by the fuzzification coefficient to capture pressure-signal nonlinearities at different depths. The optimal number of clusters was chosen using RMSE and mean error. MATLAB's Fuzzy Logic Toolbox *fcm* function performed clustering, and *genfis3* function generated a fuzzy inference system, where each cluster defined one fuzzy rule as the basis of the ANFIS model.

Next, the clustered dataset was trained to an ANFIS model (Takagi–Sugeno fuzzy system with ANN learning). The five-layer architecture performed fuzzification, rule activation, normalization, first-order Sugeno consequents (optimized by least squares), and output aggregation for $X$ and $F$. Membership functions were tuned via gradient descent, while consequents were refined by a hybrid least-squares update. This approach enabled accurate nonlinear mapping of pressure signals to tactile variables (Supplementary Materials S11).

For comparison, we also tested Linear Regression (LR), Polynomial Regression (PR, second-order), and a Multi-Layer Perceptron (MLP) trained with the LBFGS solver (random state = 1, max iterations = 1500). All were implemented in MATLAB R2024a (Statistics and Machine Learning Toolbox for LR/PR; Deep Learning Toolbox for MLP).

**Human-Robot interaction:** The robot experiments were conducted using a UR5e manipulator integrated with the tactile sensing system through a TCP/IP communication interface between MATLAB and Python. The data collection and estimation processes were performed in MATLAB, where pressure readings were processed in real time to estimate the contact position and applied force. The estimated values were then transmitted to Python to control the robot's motion accordingly. For the 1D experiments, the sensor recorded data over a 5 $s$ interval, and a Nordson Ultimus V high-precision dispenser was used to inject the material during testing. The 2D experiments employed the same communication setup and electronics, but the recording duration was extended to 10 $s$ to accommodate the longer settling time of the fluid pressure after contact. In this case, both the maximum pressure and the pressure rise time were extracted as features for position estimation. The robot control and sensor communication scripts are available in the project's GitHub repository[60].

## DATA AVAILABILITY

The codes supporting the findings are available in our GitHub repository[60].

## CODE AVAILABILITY

The data presented in this article are available in our GitHub repository[60].

**FUNDING**

The research was supported by the Independent Research Fund Denmark through Sapere Aude: DFF-Research Leader Grant 1051-00075B awarded to A.R.

**ACKNOWLEDGMENTS**

We thank Farhad Nezami and Hamed Moradi for insightful discussions on FSI simulations, and Smith Khare for valuable input on machine learning.

**AUTHOR CONTRIBUTIONS STATEMENT**

A.G. and A.R. conceived the research and wrote the manuscript. A.G. designed and performed the experiments, collected, analyzed, and visualized the data. M.B. carried out the FSI simulations. A.P. implemented the machine learning model. A.G. and S.P.M.B. designed the 2D sample and performed the initial tests. R.S. developed the theoretical model. A.R. acquired funding and supervised the project. All authors contributed to the discussion of the results and preparation of the manuscript.

**COMPETING INTERESTS STATEMENT**

The authors declare no competing interests.

## TABLE

**Table 1**. Comparison of state-of-the-art soft tactile sensors that are electronic free at sensing site with our method

| Transduction method | Number of sensors | Working length/area | Estimation method | Position accuracy | Force range | Force accuracy | Ref |
|---|---|---|---|---|---|---|---|
| Acoustic | 1 | 50 (mm) | Classification (20mm tolerance) | 89% | 0-30 N | 95.5% (±1 N tolerance) | 46 |
| Acoustic | 2 | 25×25 ($mm^2$) | Regression (MAE) | 0.7 mm | Up to10 N | 0.1 N (MAE) | 51 |
| Pressure | 32 | 40×40 ($mm^2$) | Regression (RMSE) | X: 2.8 mm Y: 2.2 mm | 1-30 N (0.4 mm-0.6 mm) | N.R. | 19 |
| Pressure | 7 | ≈2000 ($mm^2$) | Classification (grid-class) | 87.3% | 0.9-4.5 N | 0.4 N (RMSE) | 56 |
| Pressure | 4 | 15×15 ($mm^2$) | Regression (RMSE) | X: 1.7 mm Y: 1.8 mm | Up to 4 N | 95% (5 mN tolerance) | 45 |
| Pressure | 2 | 100 (mm) | Regression (RMSE) | 1.8 mm | 1-10 N | 0.4 N (RMSE) | 1D |
| Pressure | 2 | 60×60 ($mm^2$) | Regression (RMSE) | X: 3.3 mm Y: 4.2 mm | 0-6 N | 0.7 N (RMSE) | 2D |

## FIGURES

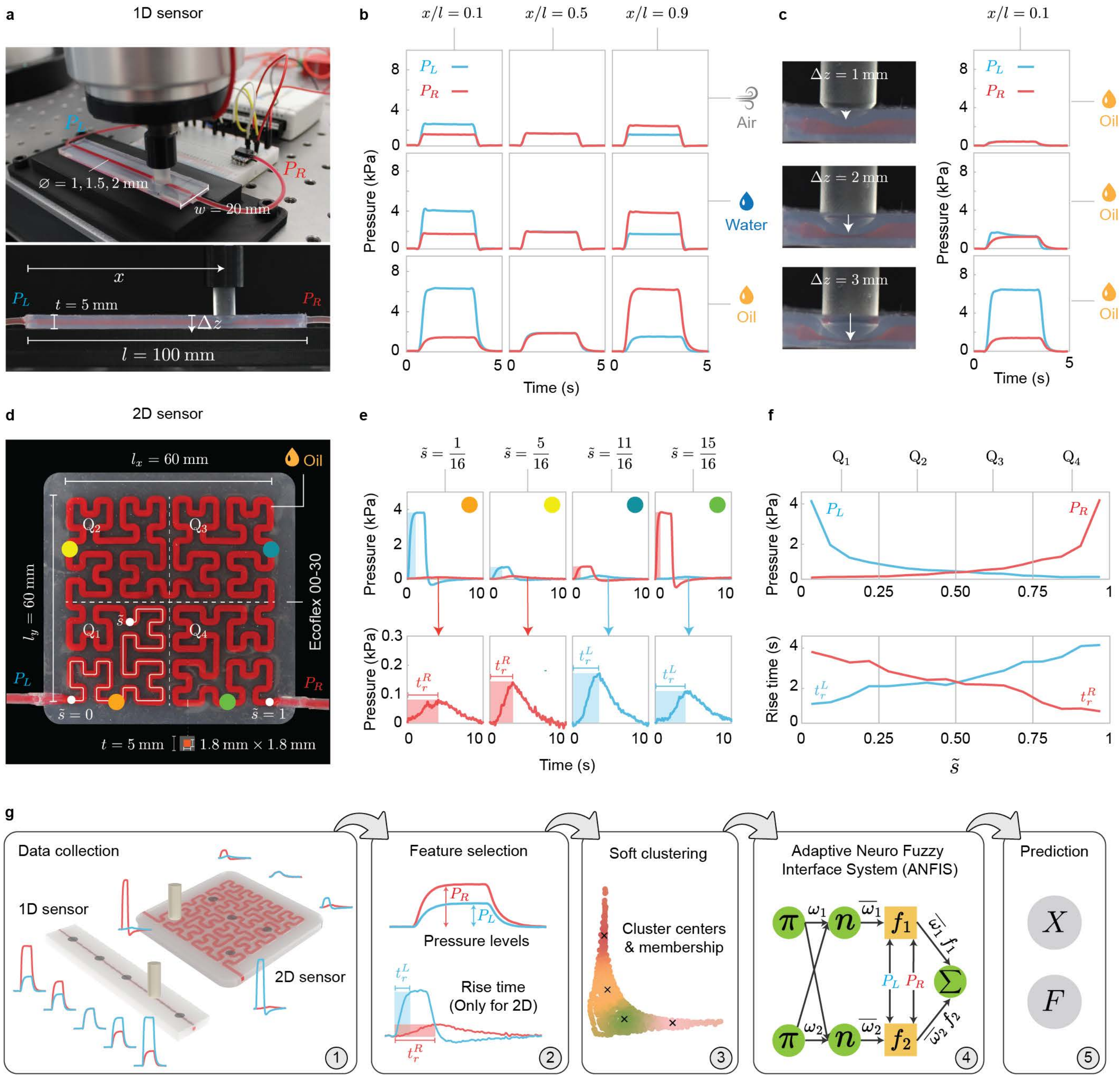

**Figure 1. Fluid-based soft tactile sensor**. **(a)** 1D tactile sensor, featuring a fluid-filled channel embedded in an elastomer, under touch stimulus. **(b)** Selected pressure signals for three different media at three distinct positions, demonstrating the pressure response depends on both position and fluid type (the symbols are made with Adobe Illustrator, Adobe, USA). **(c)** Pressure readings for varying depths of touch, illustrating how deeper touches lead to higher and distinct pressures. **(d)** 2D sensor, utilizing a Hilbert curve to extend the 1D sensor into 2D. **(e)** Selected pressure signals recorded for touches in all four quadrants of the 2D sensor, with a zoom-in plot below highlighting the rise time in the pressure response. **(f)** Top: maximum pressure values recorded at 16 points along the channel, plotted with respect to the touch position along the normalized channel length. Bottom: left and right sensor rise times corresponding to the position of the touch. **(g)** Machine learning pipeline for touch position and force estimation: (1) pressure time series are collected from both 1D and 2D sensors (the renders made using sSolidWorks 2021 SP1, Dassault Systèmes, France); (2) maximum pressure (for both 1D and 2D) and rise time (for 2D) are extracted as features; (3) the feature space undergoes soft clustering, assigning membership values to each data point; (4) an adaptive neuro-fuzzy inference system (ANFIS) models the nonlinear relationships to (5) predict touch position and applied force.

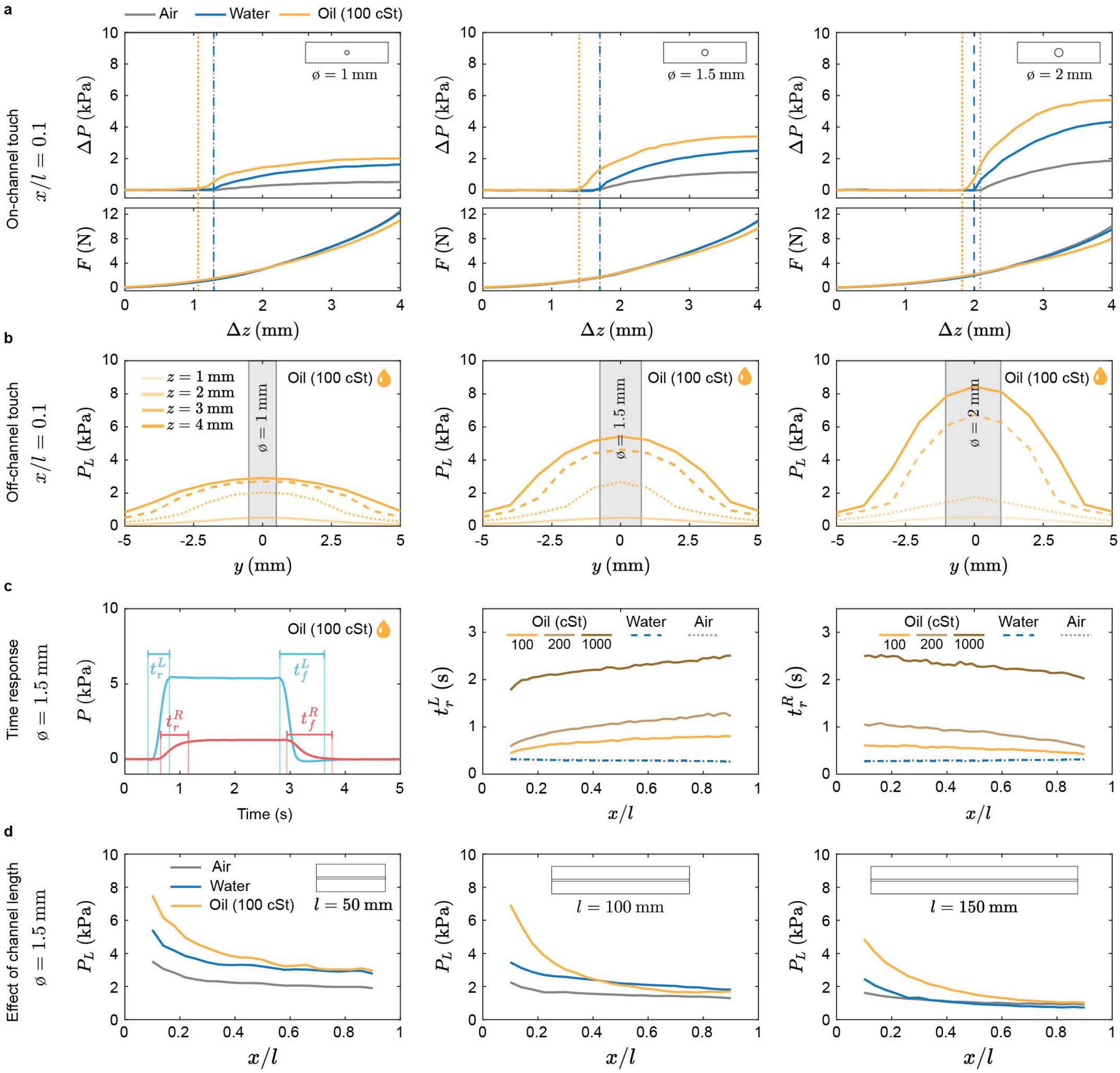

**Figure 2. Response of the 1D sensor. (a)** The pressure difference $\Delta P = P_L - P_R$ and force $F$ for on-channel touches at $x/l = 0.1$ for three different fluid media and channel diameters ø. The start and end of the working range are indicated by dotted lines. **(b)** Maximum $P_L$ during off-channel touches at $x/l = 0.1$ for different channel diameters ø when filled with oil (100 cSt). **(c)** Time response of the pressure readings for ø $= 1.5$ mm and oil (100 cSt) during on-channel touches. **(d)** Maximum $P_L$ for different channel lengths ($l = 50, 100, 150$ mm) and fluid media when ø $= 1.5$ mm.

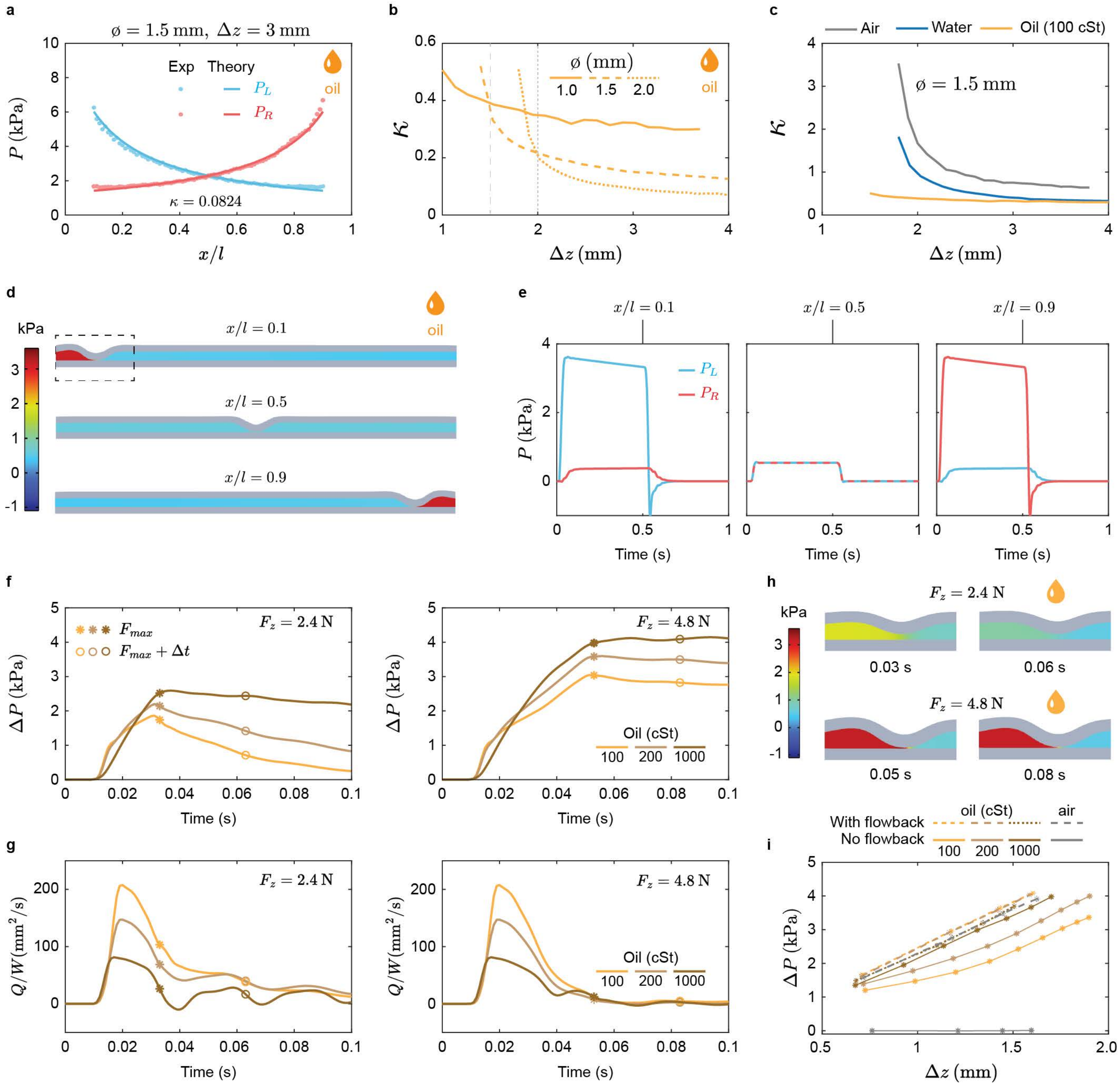

**Figure 3**. **Modeling the pressure response of the 1D sensor**. (**a**) Model fit to experimental pressures ($P_L$ and $P_R$) obtained for oil ($\mu = 100$ cSt), $ø = 1.5$ mm, $\Delta z = 3$ mm ($R_L^2 = 0.976, RMSE_L = 0.21$ kPa, $R_R^2 = 0.983, RMSE_R = 0.19$ kPa). (**b**) Identified normalized bulk modulus $\kappa$ for oil medium ($\mu = 100$ cSt) across different diameters ø. The vertical dashed lines represent the channel diameter for each case (**c**) Identified normalized bulk modulus $\kappa$ for $ø = 1.5$ mm across air, water, and oil ($\mu = 100$ cSt) mediums. (**d**) Simulated pressure distribution along the channel for indentation positions ($x/l = 0.1, 0.5, 0.9$) under $F_{max} = 4.8$ N. (**e**) Corresponding temporal pressure responses. (**f**) Pressure difference $\Delta p$ for indentation at $x/l = 0.1$ for oils with $\mu$ =100, 200, and 1000 cSt under $F_{max} = 2.4$ N (left) and $F_{max} = 4.8$ N (right). (**g**) Flowback rate per unit width ($Q/W$) for the same loading scenarios as **F**, indicating reduced flowback and increased pressure retention with higher viscosity. (**h**) Simulated pressure snapshots for $\mu = 100$ cSt at peak force and $\Delta t = 0.03$ s after, highlighting the reduced pressure dissipation and lower flowback in higher indentation depth. (**i**) Pressure difference versus indentation depth for oils with $\mu = 100$, 200, and 1000 cSt, as well as air, comparing cases with (dashed line) and without (solid line) a separating interior wall inside the channel.

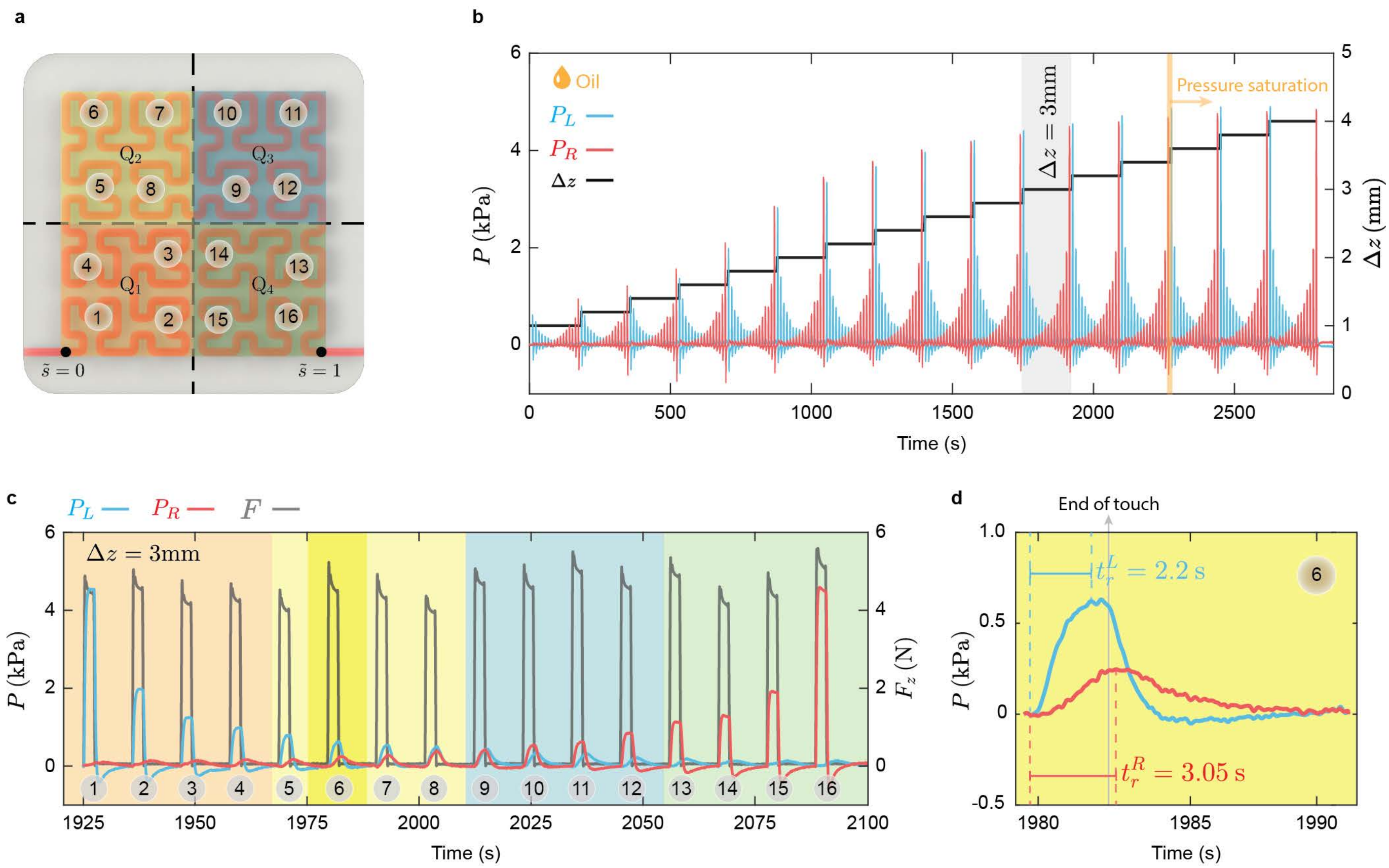

**Figure 4. Characterization of the 2D sensor**. (**a**) The Hilbert curve is divided into four quadrants ($Q_1, Q_2, Q_3, Q_4$), with four representative sensing points selected per quadrant. (**b**) The pressure response for all 16 points as the indentation depth ($\Delta z$) increases from 1 mm to 4 mm, illustrating the periodic response pattern and eventual saturation after ($\Delta z = 3.6$ mm). (**c**) A zoomed-in view of **b** for an indentation depth of 3 mm, showing quadrant-based variation in left and right pressure values over time. (**d**) A detailed view of the pressure response at point 6, displaying the rise time difference between $P_L$ and $P_R$ in addition to the pressure difference, where the right-side pressure exhibits a longer rise time ($t_r^R = 3.05$ s) compared to the left-side pressure ($t_r^L = 2.2$ s), characteristic of points located in the quadrant $Q_2$.

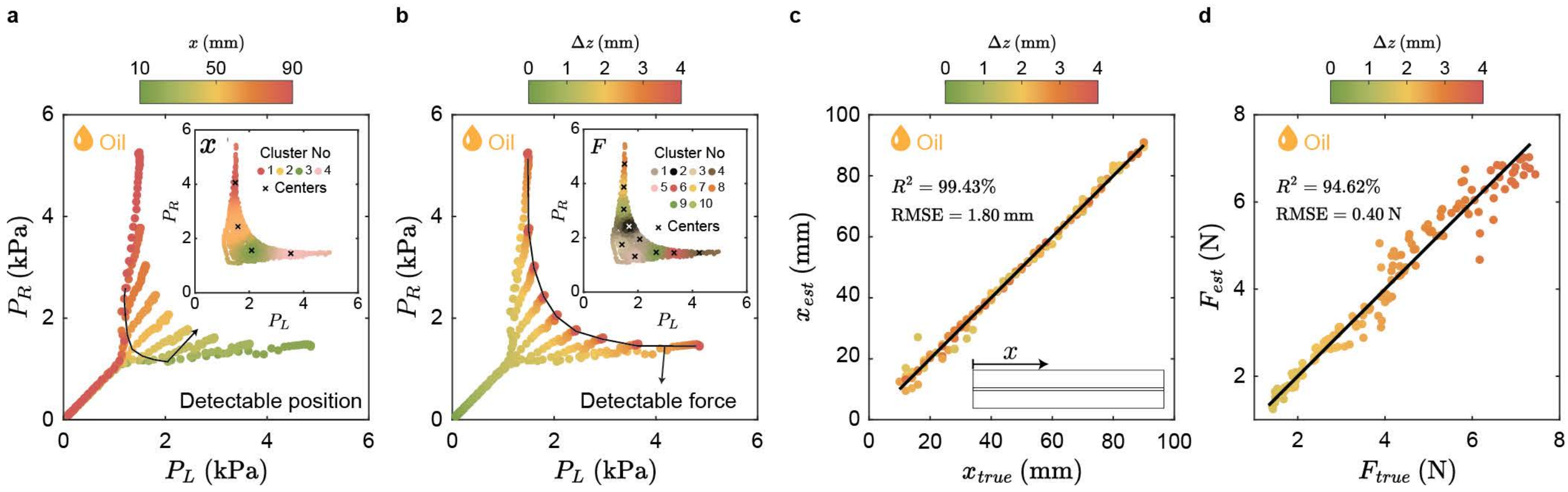

**Figure 5**. **Learning tactile response for 1D sensor**. (**a**) The feature space for oil 100 cSt with the initial dataset was used to determine the sensor's detectable position working range. The inset shows clusters used for position estimation in feature space using an extended dataset. (**b**) Feature space mapped against indentation depth, revealing regions where contact force becomes detectable. (**c**) Position estimation results, and (**d**) force estimation results for oil 100 cSt and $\emptyset = 1.5$ mm.

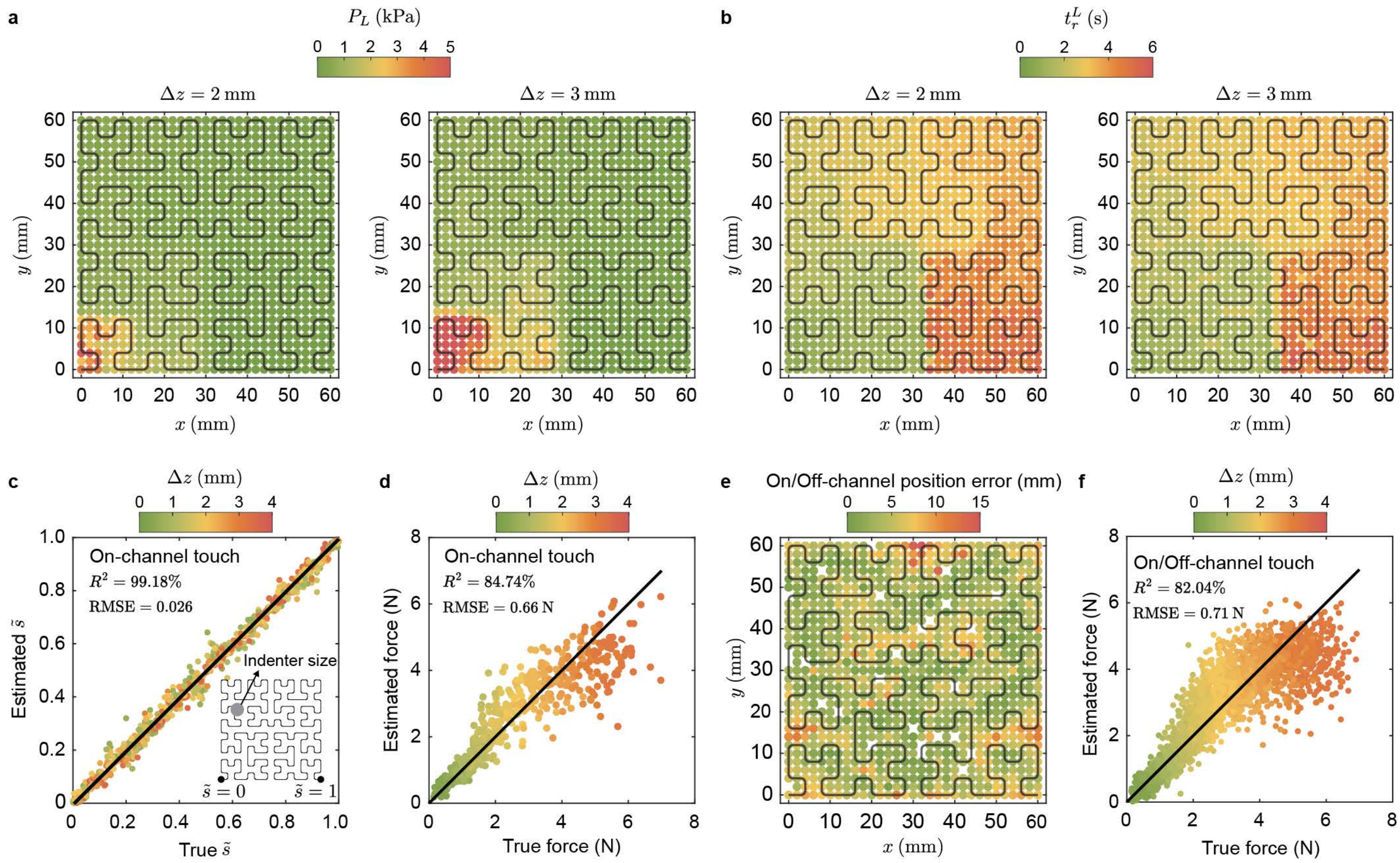

**Figure 6. Learning tactile response for 2D sensor.** (**a**) Spatial distribution of the left pressure magnitude feature $P_L$ $(kPa)$ over the full dataset at indentation depths of $\Delta z = 2\,mm$ and $\Delta z = 3\,mm$. (**b**) Spatial distribution of the left pressure rise time feature $t_r^L$ $(s)$ at $\Delta z = 2\,mm$ and $\Delta z = 3\,mm$. (**c**) On-channel 2D touch localization based on normalized position $\tilde{s}$ along the Hilbert curve. (**d**) Corresponding force estimation for on-channel touch data. (**e**) Heatmap of positional error for both on-channel and off-channel touches across the 2D sensor area. (**f**) Force estimation across the full dataset (on- and off-channel).

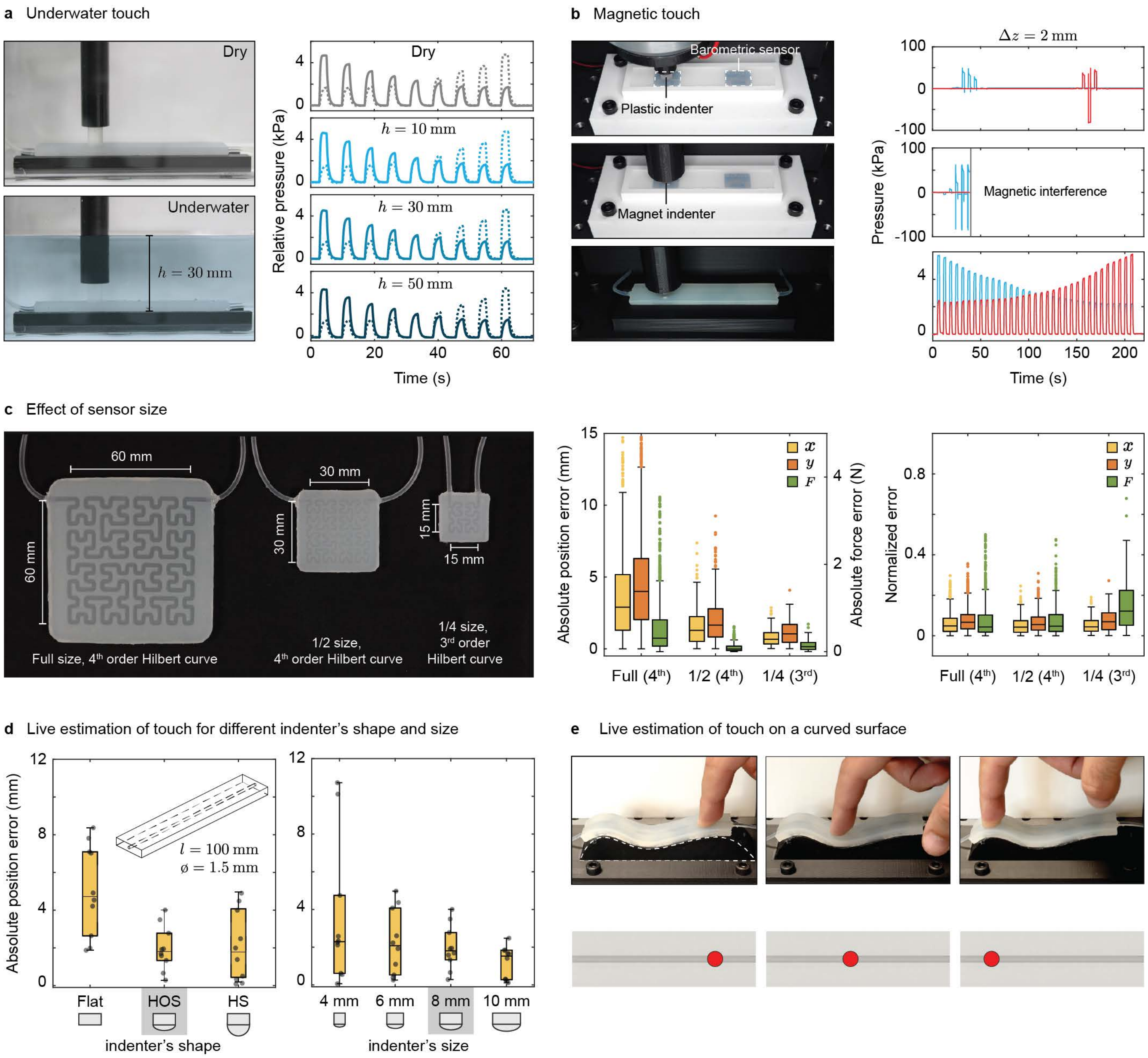

**Figure 7**. **Adaptability of the sensor to diverse environmental conditions**. (**a**) The 1D sensor maintains nearly identical differential pressure response across different water depths as it does in air. (**b**) The 1D sensor reliably responds to magnetic indentation, whereas the barometric sensor fails to detect most indentation points with a normal indenter and experiences communication failure under magnetic indentation. (**c**) The scalability of the 2D sensor across three sizes: 60 mm× 60 mm (Full size) and 30 mm× 30 mm (Half size) with the same 4th-order Hilbert curve, and 15 mm × 15 mm (quarter size, 3rd-order Hilbert curve). Plots show absolute and normalized errors for position and force, with box plots showing data grouped by quadrants and dots representing outliers. The normalized errors remain consistent across dimensions, confirming reliable performance. (**d**) The 1D sensor achieves accurate position estimation (<3 mm error in nearly all cases) across indenters of varying shapes and sizes, despite being trained only with the highlighted reference indenter. Live testing was conducted by randomly indenting the sensor (n=10) at different locations and depths using a robotic arm with different indenters, demonstrating robust real-time performance; box plot bars represent data grouped by quadrant, and all data points are shown in the box plot. (**e**) The sensor performs live position estimation on a curved surface.

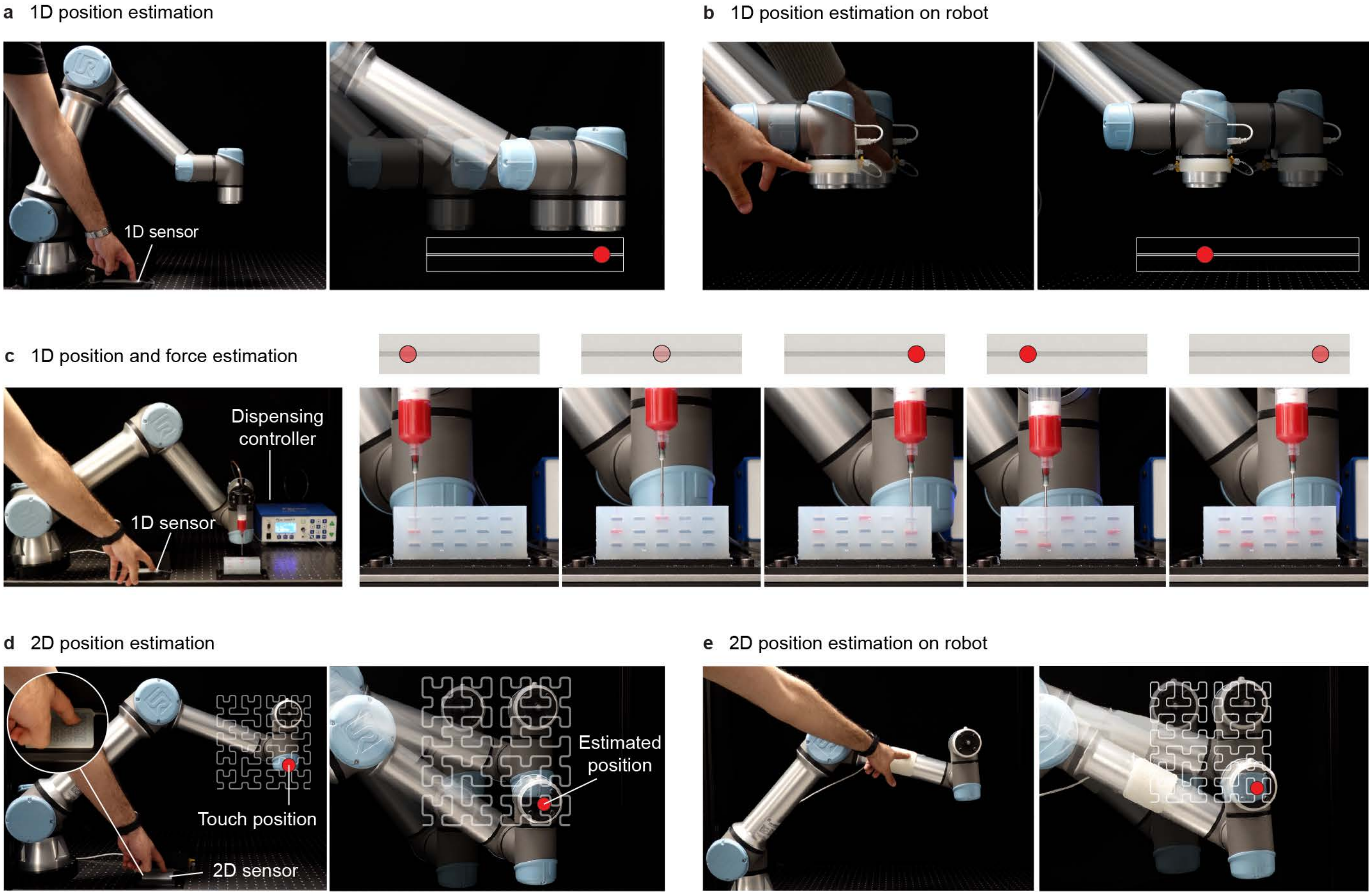

**Figure 8**. **Human-robot interaction demonstrations using 1D and 2D sensors.** (**a**) Controlling the robot arm's X-axis position using the 1D sensor mounted on the table surface (**b**) or directly on the robot's arm. (**c**) Position and Force estimation results for the 1D sensor. Each touch was mapped to a 5 × 3 grid representing the estimated position and force (**d**) Control of the robot arm's XZ position using the 2D sensor mounted on the table (**e**) or on the robot arm's surface where touches in each quadrant moved the robot to the corresponding positions.

*Supplementary information for*

# Tactile Perception through Fluid–Solid Interaction

Arman Goshtasbi[1], Minke Berghuis[2†], Aida Parvaresh[1†], Saravana Prashanth Murali Babu[1], Robert W. Style[3], Ahmad Rafsanjani[1*]

[1] SDU Soft Robotics, Biorobotics Section, The Maersk McKinney Moller Institute, University of Southern Denmark, Odense 5230, Denmark

[2] Department of Mechanics of Solids, Surfaces, and System, University of Twente, Enschede 7522 NB, The Netherlands

[3] Department of Materials, ETH Zürich, 8093 Zürich, Switzerland

† These authors contributed equally.

* Corresponding author ahra@sdu.dk

## S1. Sensor fabrication

Fig. S1 shows the steps-by-step fabrication process of both 1D and 2D sensors. For the 1D sensor, silicone rubber is injected into a custom-designed mold (A–C), cured, and subsequently connected to tubes and sensor electronics after fluid filling (D). The 2D sensor fabrication begins with pouring silicone rubber into a 2D mold (E–G), followed by curing and the insertion of a 3D-printed Hilbert curve to define the internal channel (H). Silicone is then brushed over the surface (I), and once the mold is set, the Hilbert curve is removed to reveal a clear channel (J). A pre-cured silicone rubber layer is carefully placed over the top while the brushed layer remains tacky to ensure bonding without bubbles (K). Finally, the structure is completed by attaching tubes, filling the channel with oil, and connecting to external sensing components (L).

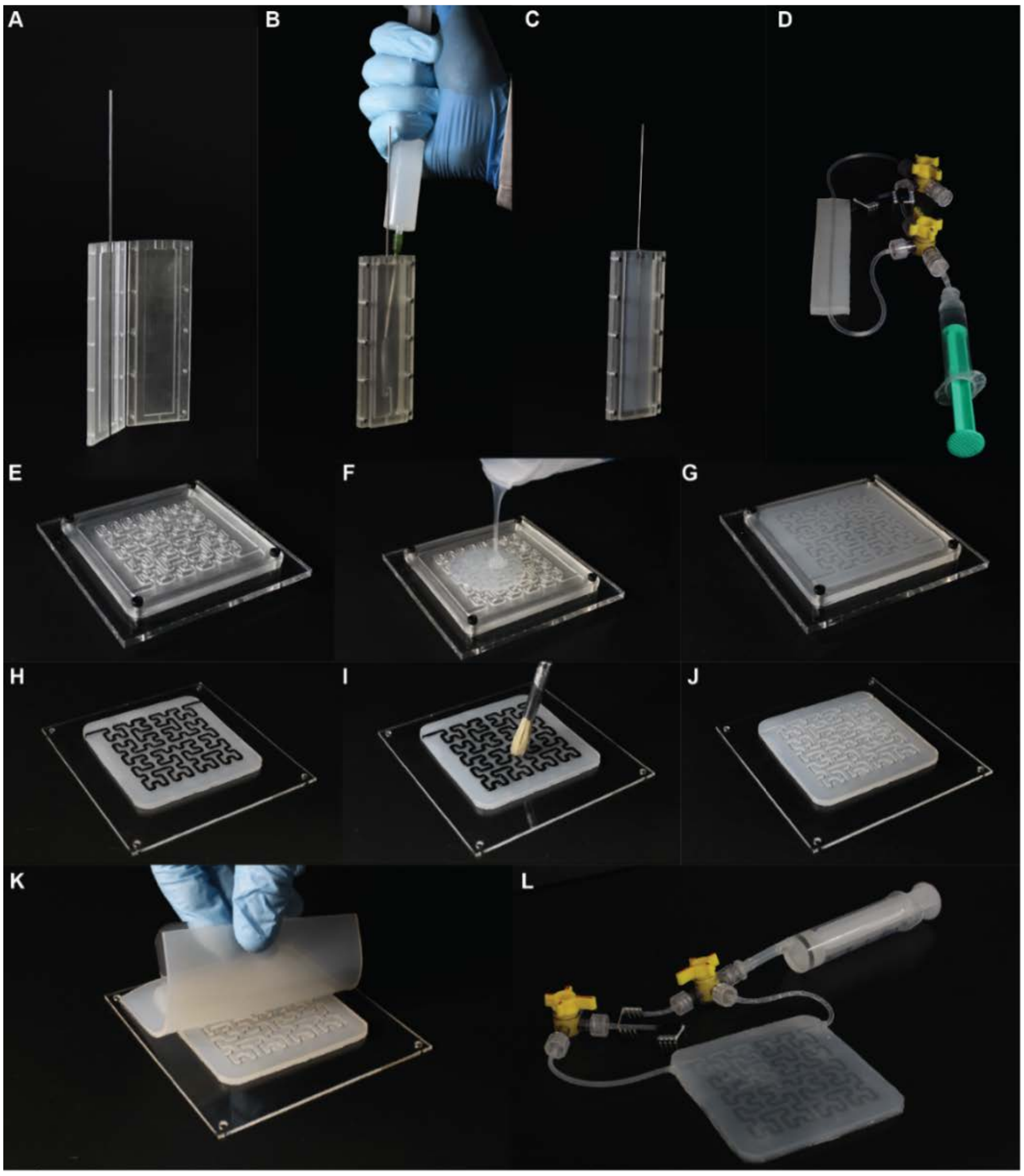

**Figure S1.** Fabrication of 1D and 2D sensors. **(A)** The 1D mold. **(B)** Injection of the silicone rubber into the assembled mold. **(C)** Fully injected mold ready for curing. **(D)** Attachment of tubes, fluid filling, and sensor connections. **(E)** The 2D mold **(F)** Pouring silicone rubber into the 2D mold. **(G)** Fully filled mold awaiting curing. **(H)** Removal of the cured part and insertion of a 3D-printed Hilbert curve into the channel. **(I)** Brushing silicone rubber onto the top surface. **(J)** Removal of the 3D-printed Hilbert curve to ensure the channel is clear. **(K)** Placement of a pre-cured silicone rubber layer on top while the brushed layer is still within its pot life, starting from one side to avoid air bubbles. **(L)** Final attachment of tubes, filling the Hilbert curve channel with oil, and connecting to the sensor.

## S2. Experimental setup

This setup is designed to characterize the response of the soft fluidic sensor under controlled indentation. A Universal Robots UR5 robotic arm is used to move a probe and apply indentation to the sensor in a position-controlled manner. The robot is programmed via a Python script using the UR_RTDE library (available on GitHub[1]). The probe is mounted on a miniature S-beam load cell (LSB200, 5 lbs, FUTEK) to measure the applied force, which is critical since the UR5 does not provide force feedback. The load cell is powered through a DC power supply and connected to a load cell amplifier, with readings acquired through a National Instruments USB-6009 DAQ system and recorded in MATLAB. The soft fluidic sensor is connected to two differential pressure sensors (SSCDRRV100MDAA5, Honeywell) that measure the pressure response during indentation. Pressure data is also collected via the same DAQ at a sampling rate of 100 Hz. The entire system is synchronized to ensure accurate measurement of both pressure and force during each indentation cycle.

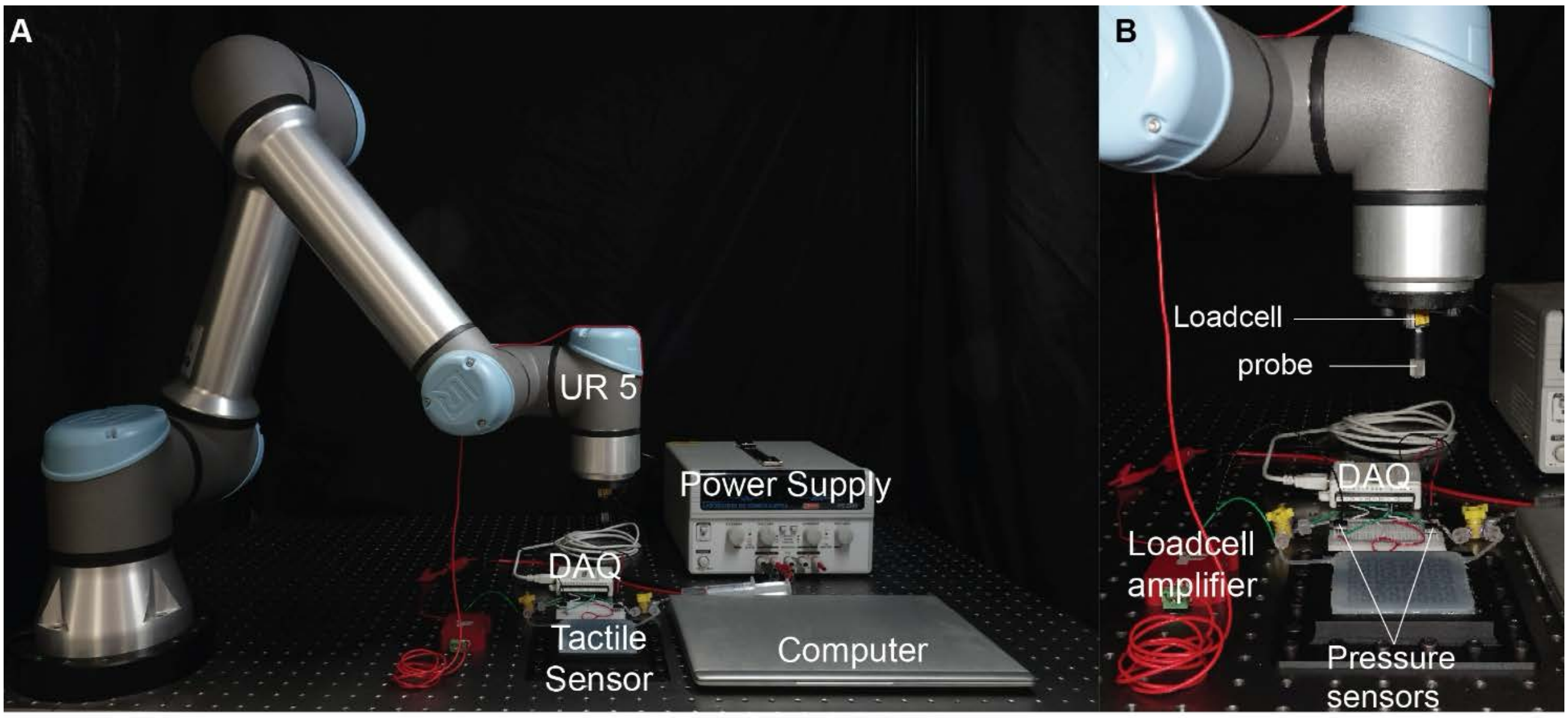

**Figure S2.** Experimental setup for indentation testing of the soft fluidic sensor. **(A)** Overview of the setup, showing the UR5 robotic arm, power supply, data acquisition system (DAQ), fluidic sensor, and computer. **(B)** Close-up view of the probe mounted to a miniature load cell. The soft sensor is connected to two differential pressure sensors, and all signals are recorded via a DAQ. The UR5 is position-controlled using a Python script. The setup enables synchronized measurement of applied force and internal pressure response during robotic indentation.

### S3. Data acquisition for the 1D sensor

We first collected preliminary data for 1D sensor to examine its behavior and the effect of different fluids on the pressure response. Fig. S3 shows the time response of the 1D sensor when filled with different media (air, water, and silicone oil 100 cst, 200 cst, 1000 cst). At low indentation depths, although the sensor registers slight pressure variations, the left and right pressures remain identical, making position estimation impossible in this regime, except for oil 1000 cst, which thanks to the high viscosity the flowback inside the channel is significantly damped and the rise time is more significant. In addition, although the rise time allows us to distinguish the left and right pressure in the oil 200 and 1000, the significant settling time making the sensor not suitable for most applications

Furthermore, when transitioning from air (a compressible medium) to water and oil (incompressible fluids), the overall pressure response increases significantly, reflecting the more substantial confinement effect of incompressible fluids.

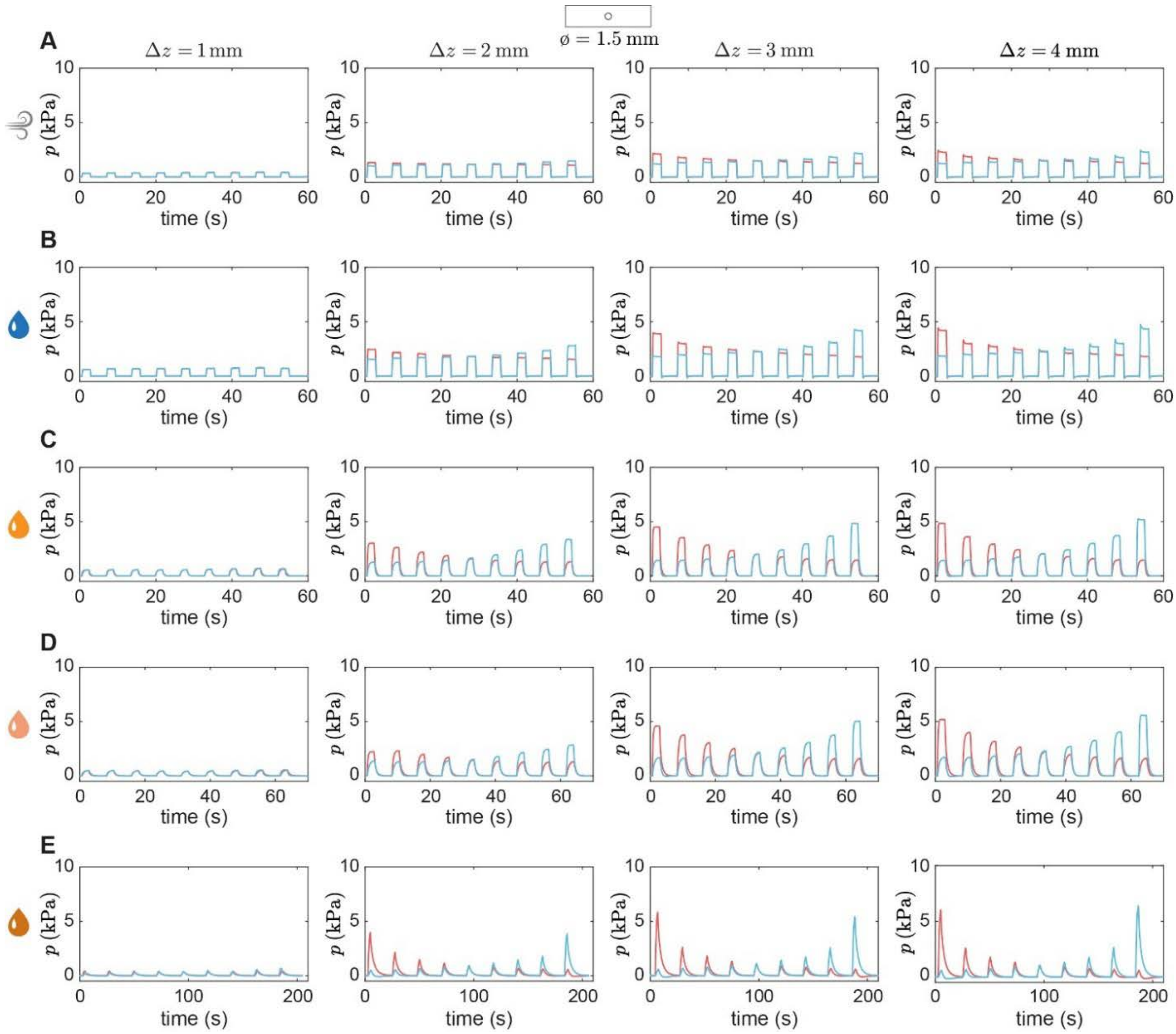

**Figure S3.** The time response of the 1D sensor across different media and at four different indentation depths for (**A**) air (**B**) water (**C**) oil 100 cSt (**D**) oil 200 cSt (**E**) oil 1000 cSt.

Additionally, these preliminary results allowed us to determine the effective working range of the sensor. As noted earlier, at low indentation depths, the pressure difference is negligible, rendering the sensor incapable of providing position estimation. At high indentation depths, however, the pressure difference saturates due to limited fluid displacement. This behavior is illustrated in Fig. S4, where the absolute pressure difference is plotted against indentation position along the 1D channel for different media and channel diameters. For Ø = 1 mm, the detectable pressure difference occurs at shallower indentations compared to larger diameters, but saturation also occurs earlier as the available fluid volume is quickly exhausted. Finally, a small degree of asymmetry can be observed in the measurements, which we attribute to minor fabrication imperfections and setup-related variations.

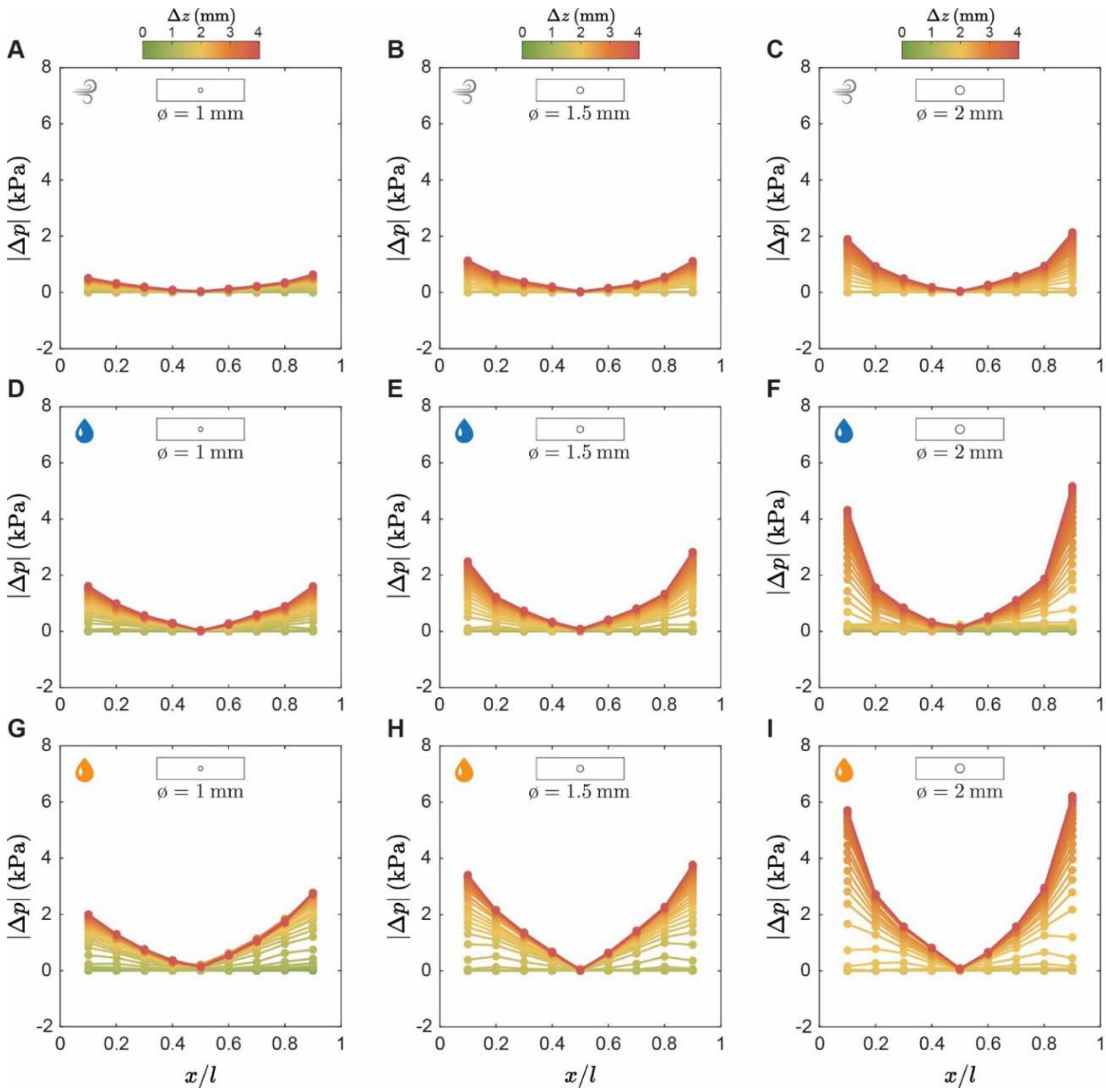

**Figure S4.** Absolute pressure difference $|\Delta P|$ along the 1D channel for different indentation depths ($|\Delta z|$), channel diameters ($ø = 1, 1.5, 2$ mm), and media (air, water, oil 100 cSt). Smaller diameters show earlier onset of measurable response but saturate at lower depths.

## S4. Diffusion of silicone oil into Ecoflex 00-30

To characterize the diffusion of silicone oil into the Ecoflex 00-30 in our sensor, we filled a fresh 1D sensor with silicone oil (viscosity: 100 cst). The pressure drop was monitored daily over a 14-day period, with the channel refilled to its original baseline pressure after each recording. As shown in Fig. S5, the rate of pressure loss decreased significantly after day 6. By the end of the two-week period, the daily drop stabilized at less than 0.1 kPa. This result shows that the diffusion of the silicone oil reaches a plateau. While a negligible pressure loss persists, it can be effectively managed by subtracting the baseline during data processing and periodic oil replacement. The working range of this specific sensor is 2-5 kPa, making the pressure drop negligible compared to the working range.

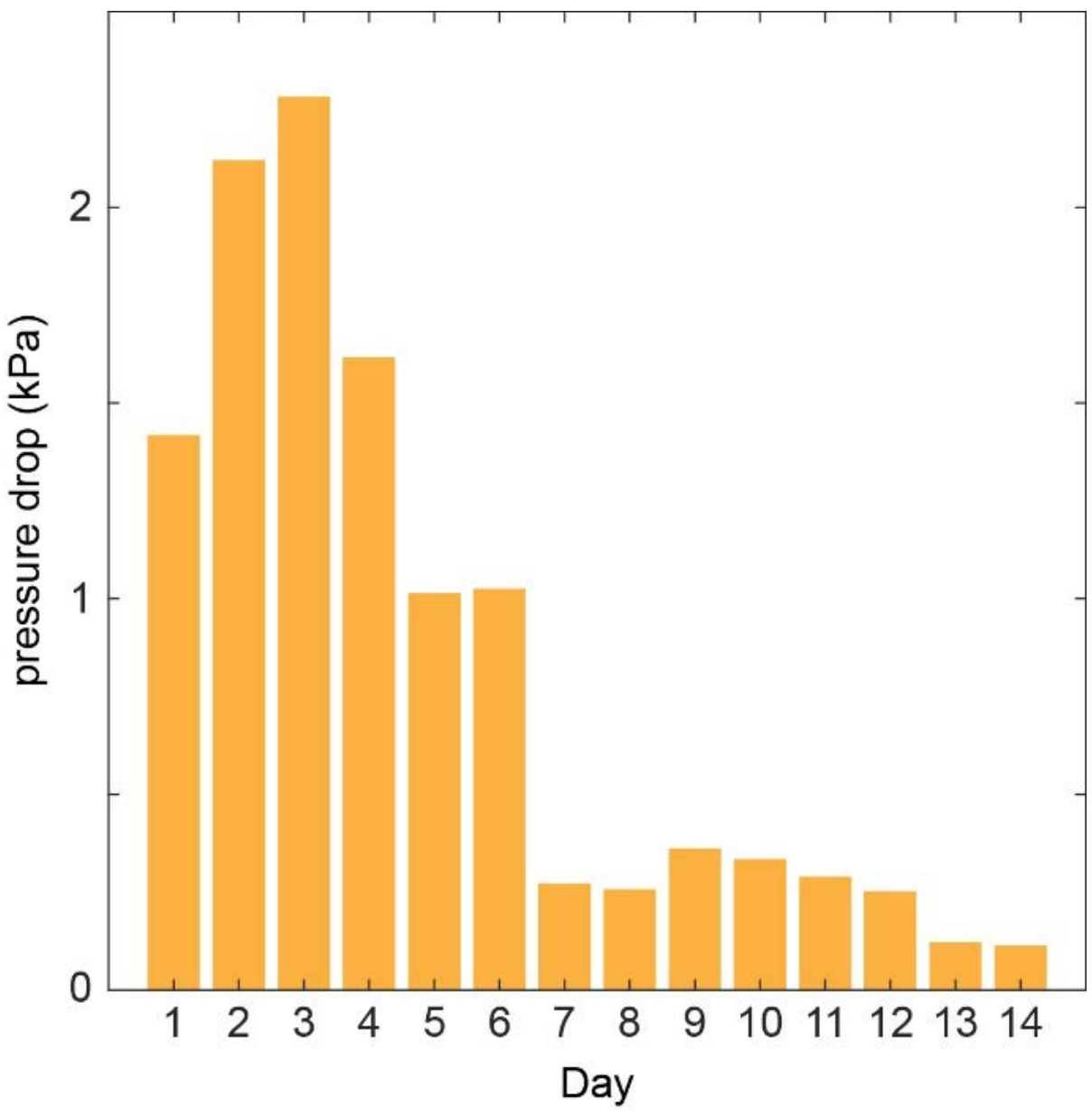

**Figure S5:** Pressure drop over time due to silicone oil diffusion into the Ecoflex 00-30 substrate.

## S5. Effect of working fluid viscosity on sensor response

Silicone oil diffuses into PDMS (Ecoflex 00-30), which limited our experiments to relatively high silicone oil viscosities. To probe intermediate-viscosity regimes while minimizing diffusion effects, we therefore employed water-based polyvinylpyrrolidone (PVP)(K90, Sigma Aldrich) solutions as a model system for viscosity tuning, without introducing additional interfacial or chemical complexity. We varied the PVP concentration to achieve viscosities of approximately 20 $\mathrm{cSt}$ and 50 $\mathrm{cSt}$, that fall between water (≈1 $\mathrm{cSt}$) and our tested silicone oil (100 $\mathrm{cSt}$).

To compare the rise time and pressure value of the PVP solution with silicone oil and water, we performed an experiment analogous to Fig. 2C and 2D at the indentation depth of $\Delta z = 2.5\ \mathrm{mm}$. Fig. S6A shows the pressure distribution along the normalized sensor length ($x/l$) for different fluids in a 100 $\mathrm{mm}$ 1D sensor. As viscosity increases, a larger pressure drop is observed along the channel. Compared to water, the PVP solutions exhibit a noticeably steeper pressure decay, indicating increased viscous resistance to flow. However, the pressure drop remains lower than that of 100 $\mathrm{cSt}$ silicone oil, which shows the most pronounced attenuation along the sensor length.

The corresponding rise-time behavior is shown in Fig. S6B. For water, the rise time decreases with increasing distance from the sensing region, consistent with rapid fluid redistribution in low-viscosity media. In contrast, both PVP solutions display a spatial trend similar to silicone oil, with the rise time increasing along $x/l$. Nevertheless, the absolute rise times of the PVP solutions remain significantly lower than those of silicone oil (≈0.4–0.7 $s$ for PVP versus ≈0.6–0.9 $s$ for silicone oil), reflecting faster pressure propagation due to their lower viscosity.

Fig. S6C shows the pressure difference $\Delta P$ between the left and right channels measured at $x/l = 0.1$ as a function of indentation depth. An increase in $\Delta P$ is observed with increasing fluid viscosity. Water produces the smallest pressure contrast, while the PVP solutions generate intermediate values, and silicone oil yields the largest pressure difference. This trend shows that higher viscosity increases differential pressure, which is directly related to improved sensing sensitivity.

In Fig. S6D, a closed-up view of the pressure difference at smaller indentation depths shows that higher-viscosity fluids extend the sensor's effective working range. Specifically, increased viscosity reduces flowback during indentation, delaying pressure balance and enabling detectable pressure differences at larger $\Delta z$ values. As a result, the onset of measurable $\Delta P$ shifts to higher indentation depths as viscosity increases.

The pressure responses shown in Fig. S6E further illustrate the dynamic trade-offs associated with fluid viscosity. While silicone oil (100 $\mathrm{cSt}$) produces the largest steady-state pressure response, it also exhibits a slow relaxation process, requiring some seconds to return to baseline after unloading. In contrast, the PVP solutions show much faster pressure recovery, with settling times closer to those of water while maintaining notably larger pressure amplitudes. This intermediate behavior highlights the advantage of optimizing the working fluid viscosity to combine enhanced pressure contrast with rapid response and recovery, making it a promising strategy for improving both sensitivity and temporal resolution in real-time sensing applications.

Finally, it is notable that the viscosity of water-based solutions can depend on temperature; therefore, accounting for temperature variations may be beneficial when precise decoupling of thermal effects from pressure measurements is required under fluctuating ambient conditions.

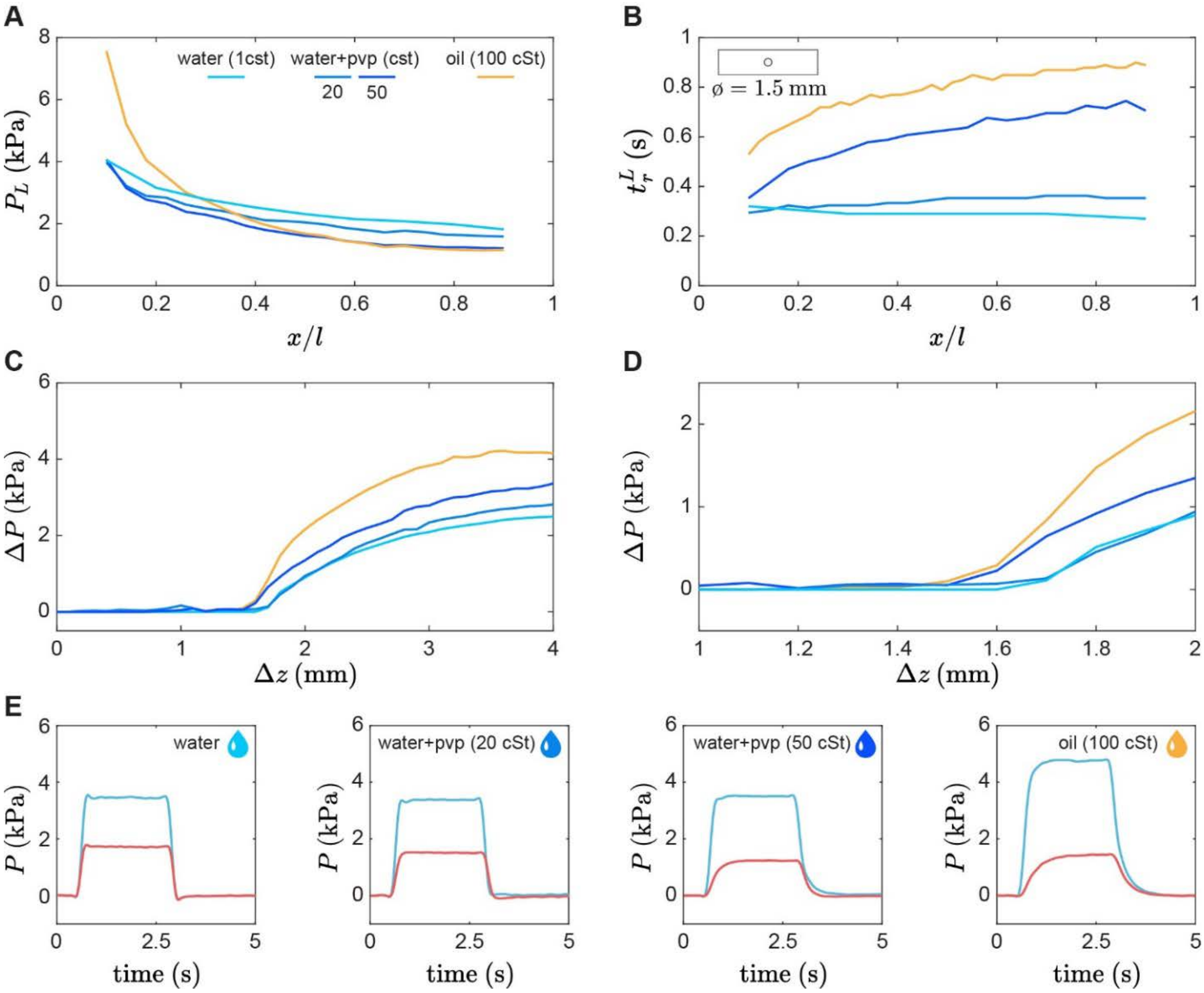

**Figure S6:** PVP solutions response compared to water and silicone oil. (**A**) Pressure distribution along the normalized sensor length ($x/l$) for water, PVP–water solutions (20 and 50 $cSt$), and silicone oil (100 $cSt$) at $\Delta z = 2.5\ mm$, showing increased pressure drop with higher viscosity. (**B**) Corresponding rise time along $x/l$, where the PVP solution shows trends similar to silicone oil but with a shorter rise time. (**C**) Pressure difference $\Delta P = P_L - P_R$ as a function of indentation depth. (**D**) Zoomed view at small $\Delta z$, showing an extended working range at higher viscosity due to reduced flowback. (**E**) Pressure-time responses showing slower pressure settling for silicone oil and faster recovery for PVP solution, while benefiting from the temporal response of viscous fluid.

## S6. Effect of touch frequency on the pressure response of 1D sensors

**Varying touch intervals along the length**: We evaluated the pressure response of the 1D sensor (∅ = 1.5 mm, $\mu = 100$ cSt) by varying the time interval $T$ between successive touches along its length, with a fixed holding time of $t = 2$ s. An effective frequency was defined as $f = 1/T$, and the results are summarized in Fig. S7. Fig. S7A presents representative examples of experiments at two frequencies ($f = 1.06$ Hz and $f = 2.54$ Hz). The corresponding force histories were used to extract $T$, as illustrated in Fig. S7B. Fig. S7C shows exemplary raw data for $P_L$ at these two frequencies, while Fig. S7D compiles the pressure peaks across different frequencies. Overall, the results indicate that the pressure response follows similar trends across frequencies, with slightly higher peak pressures observed at higher $f$ values. In Fig. 7E, the pressure responses were post-processed using baseline correction, and the corrected signals are shown in Fig. S7F. However, the curves do not fully overlap, indicating that further refinement of the learning algorithm is required to better account for frequency-dependent touch estimation.

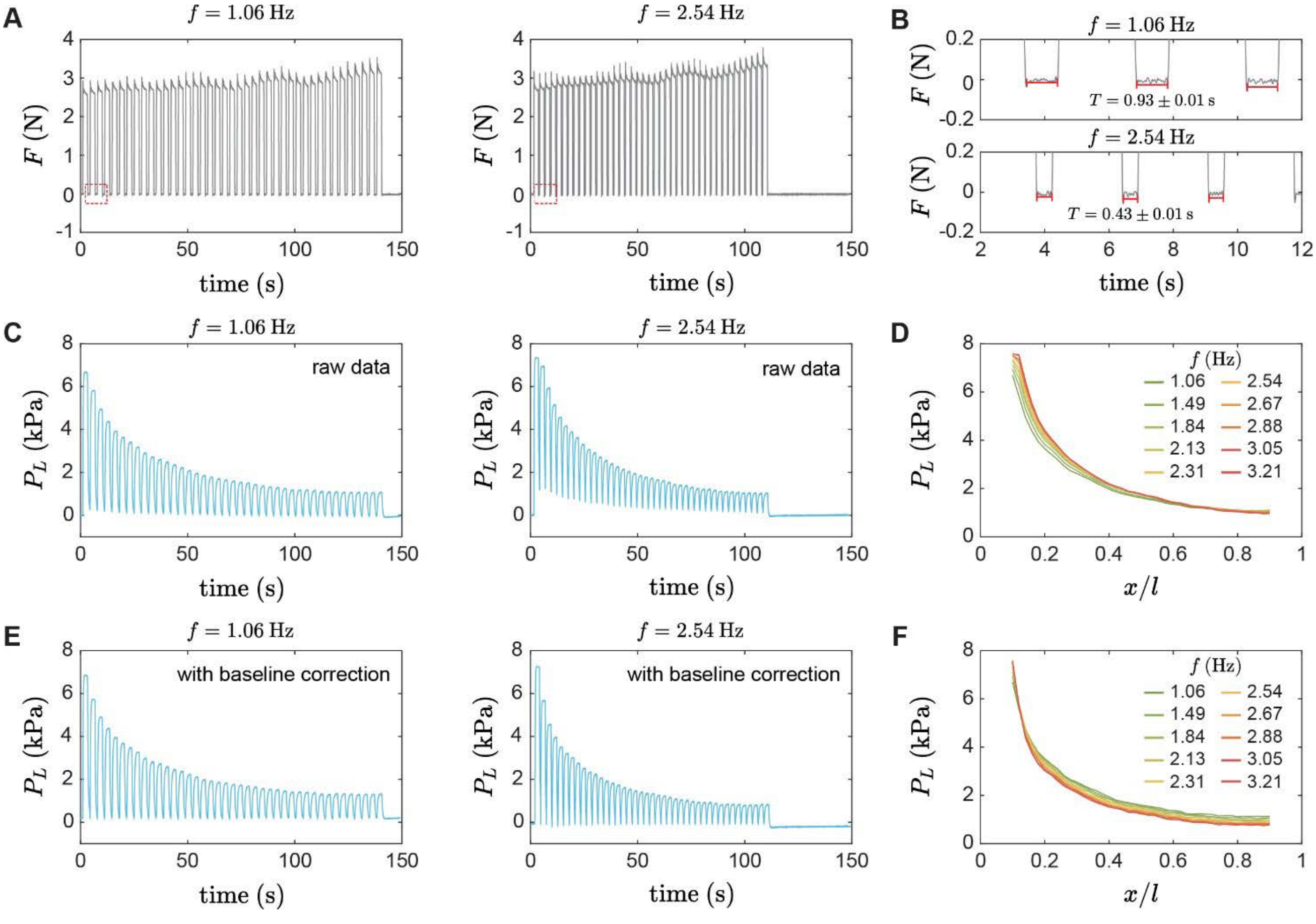

**Figure S7.** Frequency-dependent pressure response of the 1D sensor with ∅ = 1.5 $mm$ and $\mu = 100$ cSt. (**A**) Representative force histories for $f = 1.06$ Hz and $f = 2.54$ Hz. (**B**) Extraction of the time interval $T$ between successive touches from the force histories. (**C**) Exemplary raw pressure signals $P_L$ at the two indicated frequencies. (**D**) Compiled pressure peaks for a range of effective frequencies $f = 1/T$. (**E**) Pressure responses after baseline correction. (**F**) Baseline-corrected peak pressures vs $f$.

**Varying touch intervals at fixed positions:** We performed repetitive touches on 1D sensor by varying the time interval $T$ between successive touches at fixed positions ($x/l = 0.1, 0.3, 0.5, 0.7, 0.9$) with a fixed holding time of $t = 2$ s, as shown in Fig. S8. Peak pressures were compiled for both the raw data and after baseline correction (see the last column of Fig. S8). We observed that at higher frequencies, the pressure notably builds up, particularly at positions close to the external transducer ($x/l = 0.1$), whereas at positions farther from the transducer, the pressure slightly decreases. Applying baseline correction mitigates this effect to some extent; however, for higher frequencies ($f > 2$ Hz), the sensor does not have sufficient time to equilibrate to the ambient pressure due to the high fall time of the oil-based sensor.

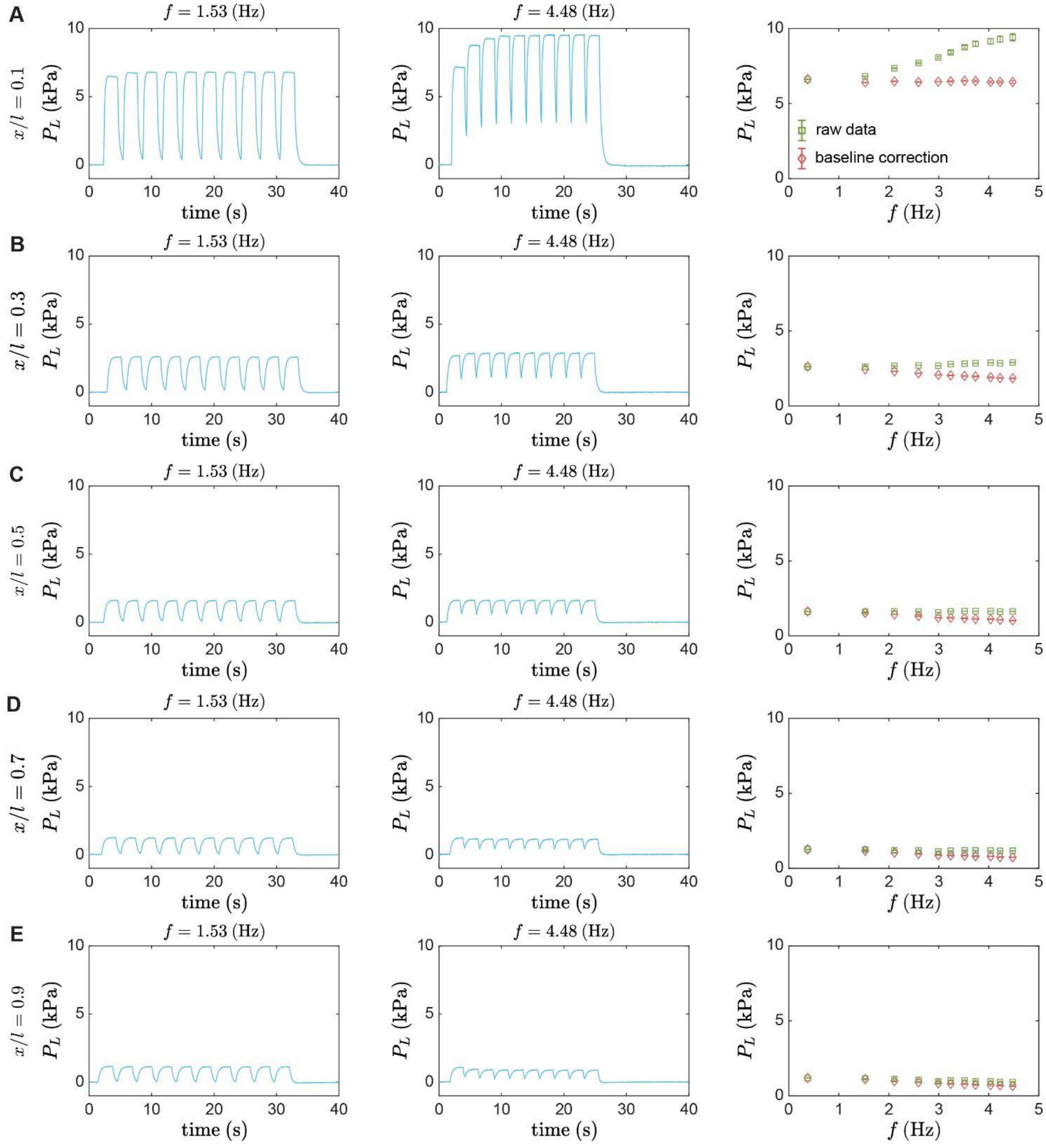

**Figure S8:** Frequency-dependent pressure response of the 1D sensor at fixed positions with Ø = 1.5 $mm$ and $\mu = 100$ cSt. (**A**) $x/l = 0.1$, (**B**) $x/l = 0.3$, (**C**) $x/l = 0.5$, (**D**) $x/l = 0.7$, and (**E**) $x/l = 0.9$. In the first two panels, the raw $P_L$ signals for the lowest and highest indentation frequencies are presented. The rightmost panel shows the maximum pressure recorded from the second touch onward.

**Fixed touch intervals with varying holding times:** In real-world applications, contact duration can vary from case to case, which may affect the pressure signal. To evaluate the influence of the pressure response, we fixed the indentation speed and acceleration to those used in the fastest frequency experiment ($V_{max} = 500$ mm/$s$ and $a = 1000$ mm/$s^2$) and varied the holding time ($t = 0.25, 0.5, 1, 2$ s).

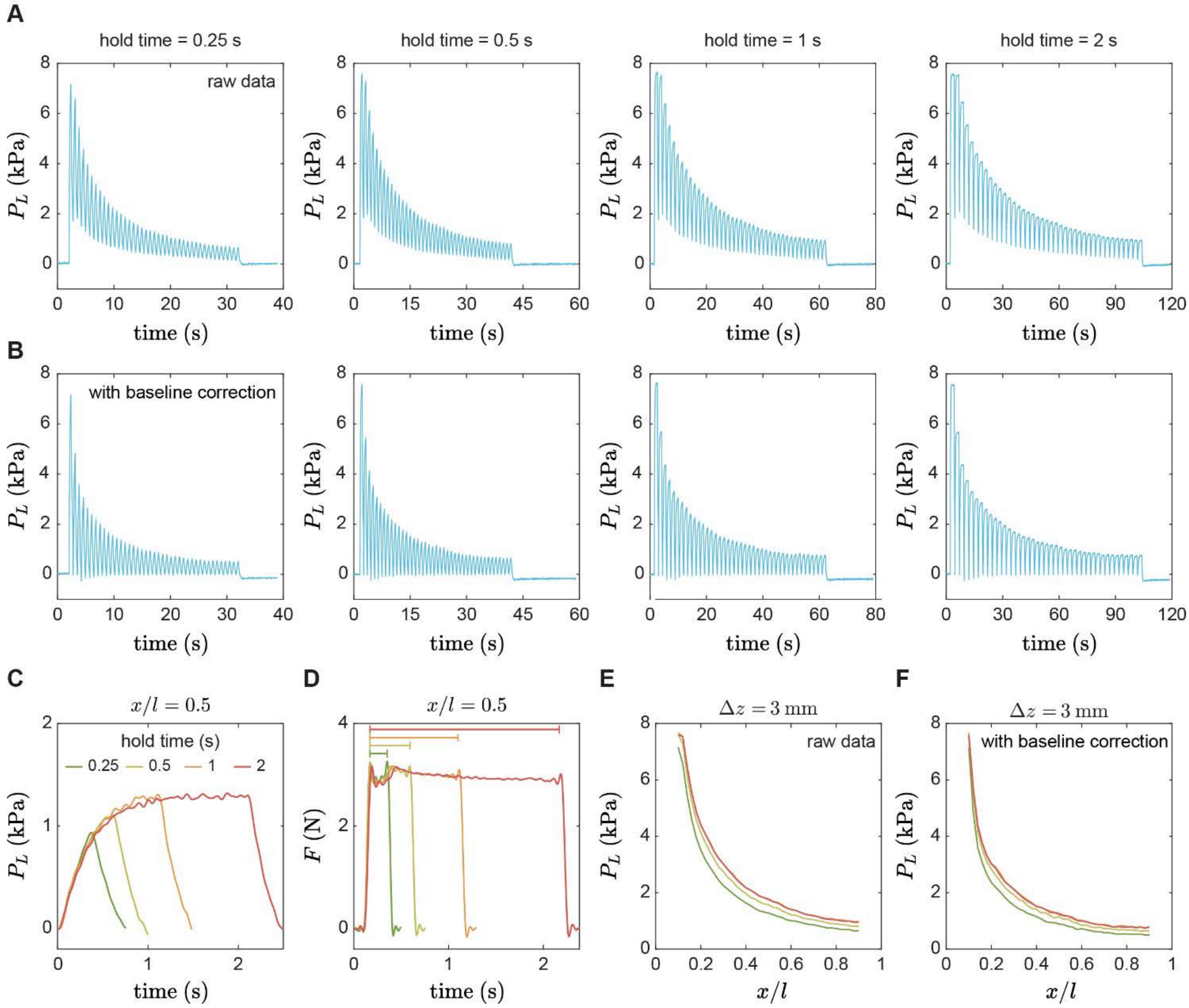

**Figure S9:** Pressure response of 1D sensor with ∅ = 1.5 $mm$ and $\mu = 100$ cSt at fixed touch intervals for varying holding time $t = 0.25, 0.5, 1, 2$ $s$. (A) raw data, (B) with baseline correction, (C) Zoomed-in view for left pressure $P_L$ and (D) force $F$. Peak pressure values for $P_L$ along the sensor length for (E) raw data and (F) with baseline correction.

As shown in Fig. S9A, the overall pressure profiles remain similar across conditions. However, for shorter contacts ($t = 0.25$ s), both the peak pressure and accumulated response are reduced. This behavior arises from the dynamics of the viscous fluid, in which the pressure requires time to reach its steady-state maximum, as shown in Fig. 2C of the main manuscript. When the contact duration is shorter than the characteristic rise time, the signal does not fully develop before unloading. Hence, Baseline correction (Fig. S9B) does not eliminate this effect, confirming that it is not caused by drift but by the sensor's transient response.

A zoomed-in view of the pressure signal at $x/l = 0.5$ (Fig. S9C) further illustrates this effect. For short holding times, the pressure does not reach the same steady-state level observed for $t = 1$ s and $t = 2$ s even though the applied force profile (Fig. S9D) reaches comparable maximum values. This confirms that the reduced pressure amplitude is not due to weaker indentation, but rather due to insufficient time for fluid redistribution and pressure equilibration within the channel.

Importantly, although the peak pressure is reduced for short contacts, the overall temporal evolution of the signal follows the same characteristic pressure profile across all holding times. The difference lies in the moment the touch is released: when the contact ends earlier, the pressure decays before reaching its steady-state maximum. Therefore, the unloading phase and the transient pressure drop still contain valuable information about the touch event. Since our current feature extraction strategy primarily relies on maximum pressure (and rise time), fast touches whose duration is shorter than the characteristic rise time may be underestimated. However, because the pressure signal preserves its dynamic shape and encodes information in both the rising and falling phases, future work could incorporate richer temporal features, such as partial-rise characteristics, slope information, or decay dynamics, to improve detection across a wider range of contact durations.

### S7. Theoretical model

We consider an isothermal chamber with elastic walls filled with a fluid. We assume that touching the 1D sensor at a distance $x$ divides the chamber into two domains and displaces a fluid volume $\Delta V$, half of which moves to the left (L) and the other half to the right (R) of the indenter.

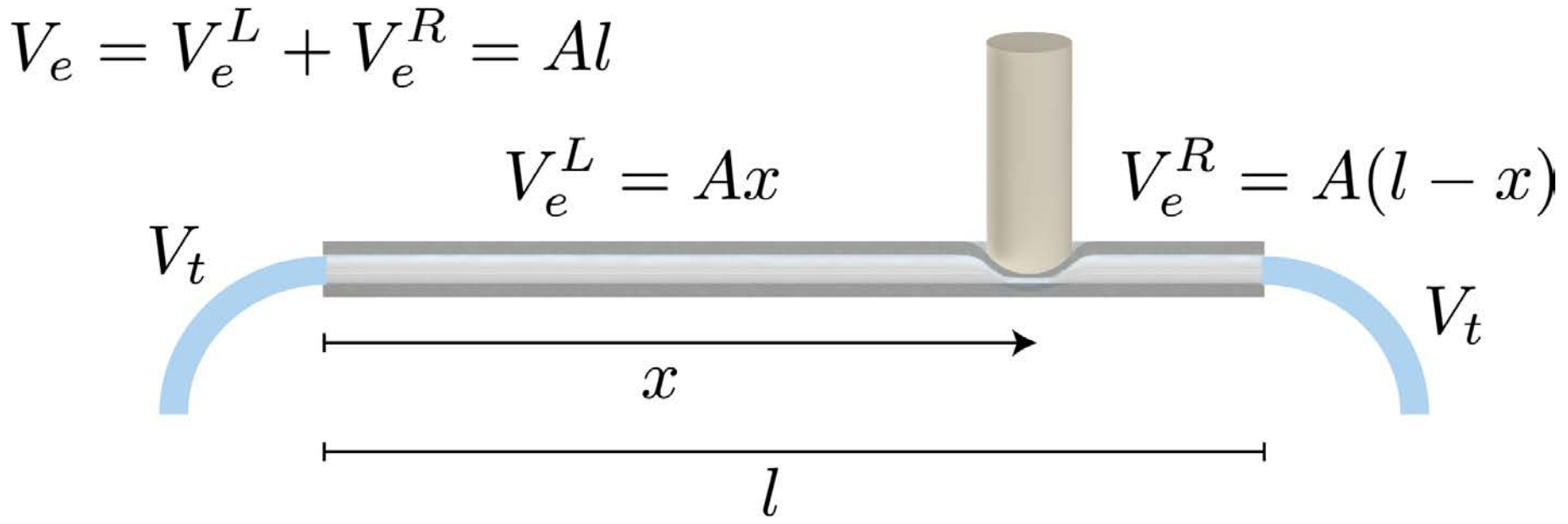

**Figure S10.** The schematic of the theoretical model.

This displacement is assumed to occur without any flowback, thereby deforming both the tube and the embedded channel in the elastomer on each side.

$$\frac{1}{2}\Delta V = \Delta V_t^i + \Delta V_e^i, \quad i = L, R \tag{S7.1}$$

Where $\Delta V_t^i$ and $\Delta V_e^i$ are the added volume to the tubing and the channel, respectively. This deformation pressurizes the left side as

$$P_L = K_t \frac{\Delta V_t^L}{V_t^L} = K_e \frac{\Delta V_e^L}{V_e^L} \tag{S7.2}$$

and the right side as

$$P_R = K_t \frac{\Delta V_t^R}{V_t^R} = K_e \frac{\Delta V_e^R}{V_e^R} \tag{S7.3}$$

where $K_t$ and $K_e$ are the effective bulk moduli of the tube system and the elastomer, respectively. Since the connecting tubes have the same length,

$$V_t^L = V_t^R = V_t \tag{S7.4}$$

Moreover, the volumes of left and right domains are given by

$$V_e^L = Ax, V_e^R = A(l - x) \tag{S7.5}$$

where $A$ is the cross-sectional area of the channel and $l$ is the total length. From Equations (S7.2)–(S7.5), we can calculate $\Delta V_t^L$ and $\Delta V_t^R$ as:

$$\Delta V_t^L = \frac{V_t}{K_t}\frac{K_e}{A}\frac{\Delta V_e^L}{x} = \kappa\frac{\Delta V_e^L}{x/l} \tag{S7.6}$$

$$\Delta V_t^R = \frac{V_t}{K_t}\frac{K_e}{A}\frac{\Delta V_e^R}{l - x} = \kappa\frac{\Delta V_e^R}{1 - x/l} \tag{S7.7}$$

By defining the normalized bulk modulus $\kappa = \kappa_e/\kappa_t$ with $\kappa_t = K_t l/V_t$ and $\kappa_e = K_e/A$, we obtain:

$$\frac{1}{2}\Delta V = \Delta V_e^L\left(1 + \frac{\kappa}{x/l}\right) = \Delta V_e^R\left(1 + \frac{\kappa}{1 - x/l}\right) \tag{S7.8}$$

From which the added volume to each side of the channel are derived as:

$$\Delta V_e^L = \frac{\Delta V}{2\left(1 + \frac{\kappa}{x/l}\right)} \tag{S7.9}$$

$$\Delta V_e^R = \frac{\Delta V}{2\left(1 + \frac{\kappa}{1 - x/l}\right)} \tag{S7.10}$$

Using Equations (S7.9) and (S7.10), the pressures $P_L$ and $P_R$ as a function of $\Delta V$ are obtained as:

$$P_L = \frac{\kappa_e}{x} \frac{\Delta V}{2\left(1 + \frac{\kappa}{x/l}\right)} \tag{S7.11}$$

$$P_R = \frac{\kappa_e}{l - x} \frac{\Delta V}{2\left(1 + \frac{\kappa}{1 - x/l}\right)} \tag{S7.12}$$

When the sensor is touched at the midpoint ($x = l/2$), the pressures at both ends are equal:

$$P_L = P_R = P_M \tag{S7.13}$$

Therefore, using either Equation (S7.11) or (S7.12), $\Delta V$ can be expressed as a function of $P_M$:

$$\Delta V = P_M \frac{l(1 + 2\kappa)}{\kappa_e} \tag{S7.14}$$

From which simple relationships for $P_L$ and $P_R$ can be derived as a function of the normalized bulk modulus $\kappa$ and $P_M$

$$P_L = P_M \frac{\kappa + 1/2}{\kappa + x/l} \tag{S7.15}$$

$$P_R = P_M \frac{\kappa + 1/2}{\kappa + 1 - x/l} \tag{S7.16}$$

## S8. Numerical simulations

Fig. S11 illustrates the axisymmetric boundary condition applied in modeling the 1D sensor. Due to the nonzero Poisson ratio, in-plane displacements are coupled to circumferential hoop stresses, which act as a restoring force. Relevant dimensions are defined in (A), including sensor length, mean indenter position and the virtual radius parameter controlling the strength of the coupling. For a small virtual radius (B), the coupling is strong, resulting in minimal in-plane displacement. As the radius increases (C), the stabilizing effect diminishes, and in the limit of $r_{virtual} = \infty$ (D), the model converges to the plane strain condition.

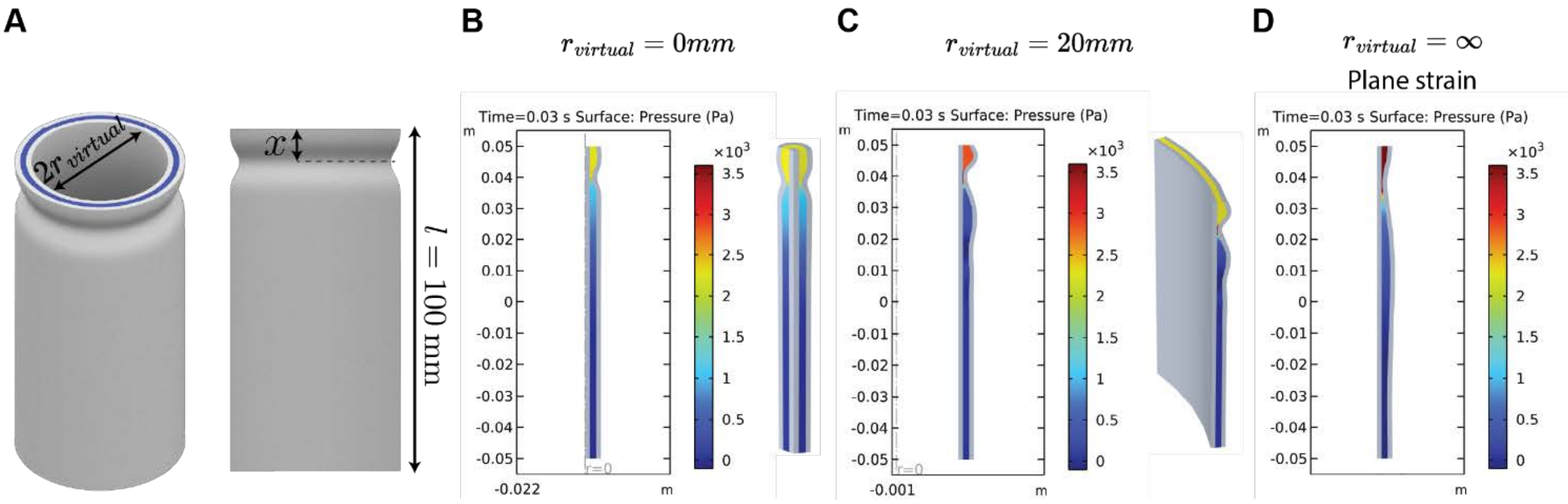

**Figure S11.** Axisymmetric boundary condition for modeling the 1D sensor. A finite Poisson ratio couples in-plane displacement to circumferential hoop stress, acting as a restoring force. (**A**) Key dimensions: sensor length, mean indenter position, and virtual radius. A small virtual radius (**B**) yields strong coupling and limited in-plane displacement, while a larger radius (**C**) reduces the effect, with $r_{virtual} = \infty$ (**D**) recovering the plane strain condition.

Fig. S12 presents finite element simulations of the 1D sensor under indentation. The model includes fluid (yellow, middle) and solid (gray, top/bottom) domains, with the gap defined as the shortest distance between the channel's top and bottom walls (A). Indenter displacement $\Delta z$ corresponds to the change in the top surface's radial position. The applied load follows a Gaussian peak profile (B), with the maximum force marked by an asterisk and a constant holding time $\Delta t$ indicated by an open circle. Because the solid deforms, the indenter displacement (C) does not directly match the gap response (E). The initial channel diameter is denoted $\phi = 2\,\mathrm{mm}$ and closure is defined when the gap reaches the numerical offset used to avoid domain topology changes. Viscosity strongly affects the dynamics: while the indenter displacement is nearly identical across viscosities for the same load (D), higher fluid viscosity slows the response and results in a larger residual gap (F) due to restricted flow from the indentation region.

### Gravity effects

Gravity effects are neglected in both the fluid and structural domains due to their negligible influence at the scale of a 2 mm horizontal channel. Hydrostatic pressure variations (20 Pa for a water-filled channel) and self-weight-induced deformation are orders of magnitude smaller than the dominant pressure-driven and elastic stresses in the system.

### Fluid Domain

The channel is horizontally oriented, so gravity acts perpendicular to the main flow direction. Our model focuses on the axial pressure difference between the $P_L$ and $P_R$ ends. Vertical hydrostatic pressure variations within each cross-section are negligible due to the small channel height. For a $2\,\mathrm{mm}$ diameter water-filled channel ($\rho = 1000\ \mathrm{kg/m^3}$), the hydrostatic pressure difference is approximately ($\Delta P_{hyd} = \rho g h\ \approx 20\ \mathrm{Pa}$), which is insignificant compared to the typical driving pressures in the kilopascal range. Gravity effects in the fluid domain are therefore neglected.

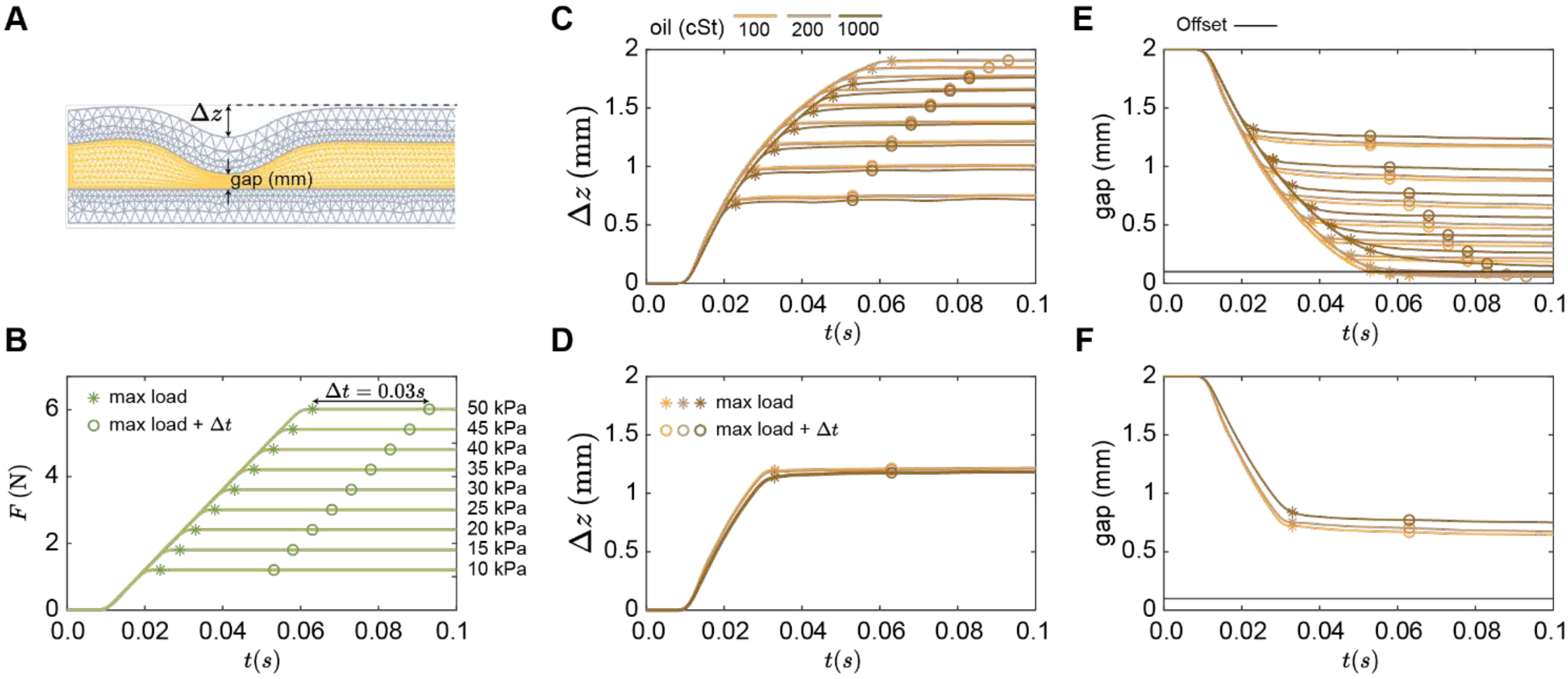

**Figure S12.** Finite element modeling of the 1D sensor indentation response. (**A**) Mesh of fluid (yellow) and solid (gray) domains, with gap defined as the shortest distance between channel walls. Indenter displacement $\Delta z$ is the change in radial position of the top surface. (**B**) Gaussian load profile with peak (asterisk) and holding time $\Delta t$ (open circle). (**C, E**) Indenter displacement and gap differ due to solid deformation, with closure defined when the gap is smaller than or equal to the numerical offset. (**D, F**) Higher fluid viscosity delays peak response and maintains a larger gap under the same load.

### Structural Domain

Gravity is also negligible in the structural response of the Ecoflex material. The channel's bottom is supported, eliminating deformation due to its own weight. While the top could theoretically sag, this effect is minimal for two reasons. First, the hoop stress from internal pressure ($\mathrm{kPa}$ range) is orders of magnitude greater than gravitational stress, with a typical ratio of ($10^3 - 10^4$). Second, the channel is embedded in a larger Ecoflex body, providing additional support that suppresses sagging. This is consistent with experimental observations: even unsupported Ecoflex tubes of $2\ \mathrm{mm}$ diameter maintain a circular shape. Pressurization further stabilizes the cross-section.

In summary, gravity effects are negligible in both the fluid and structural domains and are omitted from the model.

### FSI Solver details

The transient simulation employs an implicit second-order Backward Differentiation Formula (BDF) time-stepping scheme with an adaptive physics-controlled (free) step size. The fluid-solid interaction problem is solved using a segregated solver with Anderson acceleration due to its computational efficiency. Relative and absolute tolerance are set to $10^{-4}$ and $10^{-5}$ respectively, with a maximum number of iterations per time step set to 20 to balance computational efficiency and accuracy (<1% variation in displacement and velocity fields). To improve stability, a damping factor of 0.7 is applied to the fluid-solid interaction update. Iterations are solved with direct solvers for the fluid velocity and pressure, the fluid-structure interaction (PARDISO), and the solid displacement field (MUMPS).

## S9. Data acquisition for the 2D sensor

Following the 1D sensor, we tested the 2D sensor with all three media. We measured the pressure at two ends of the 2D channel while indenting the sensor at 16 locations $\tilde{s} = (2i - 1)/32$ (where $i \in [1,16]$), and at 16 indentation depths $\Delta z \in [0\text{ mm}, 4\text{ mm}]$. As shown in the Fig S13, the sensor operating with oil100 exhibits a much larger working range compared to those with air and water. This is because the faster flowback in air and water leads to similar left and right pressures. Furthermore, the results indicate that the water-filled sensor achieves a higher working range than the air sensor, likely due to flow damping caused by the intricate Hilbert-curve channels. Finally, while air and water show no significant difference in the rise time between left and right pressures, the oil sensor displays a clear difference. This distinct rise-time behavior in oil could serve as an additional feature for our learning and estimation.

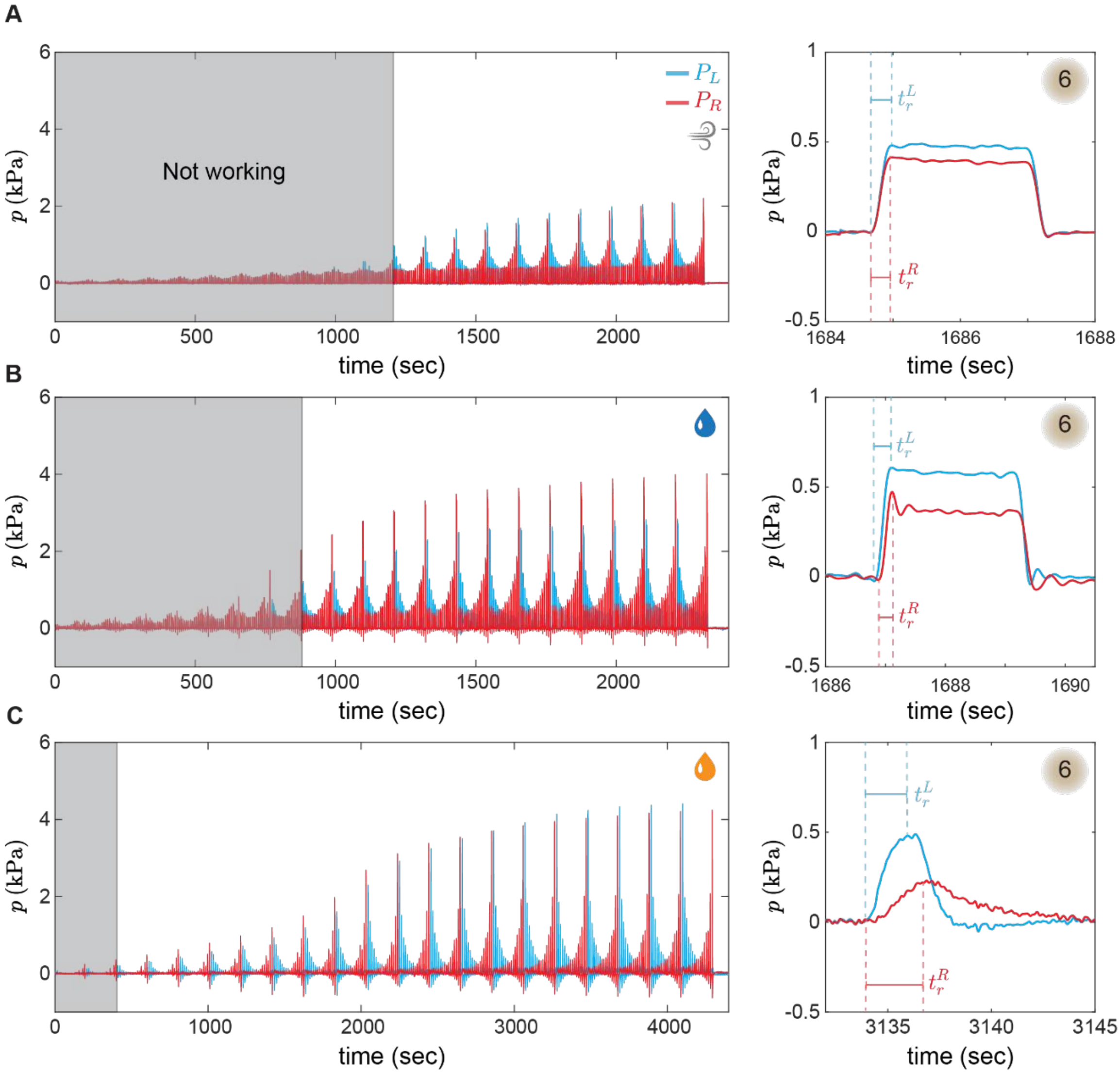

**Figure S13.** Preliminary 2D results for all 3 media for indentation at 21 different depths and 16 points per depth (left) with the zoomed-in results for $\Delta z = 3\ mm$ and the $\tilde{s} = \frac{11}{32}$ (right). (**A**) Air-filled channels show a wide non-working range due to flowback and incomplete sealing, with no observable rise time. (**B**) Water-filled channels exhibit similar behavior, with limited functionality and absence of rise time. (**C**) Oil-filled channels display a larger working range and a distinct rise-time feature, advantageous for 2D sensing.

## S10. Repeatability test for 2D sensor

To evaluate the sensor's long-term stability, we conducted a preliminary repeatability study in which the 2D sensor's pressure response was monitored over 2 weeks across all indentation depths. Fig. S14 shows a representative example at the indentation depth of $\Delta z = 3\ \mathrm{mm}$. The top row shows the raw pressure data without baseline correction, where a gradual drift can be observed over the two-week period. As discussed in the manuscript, this drift originates from slow diffusion of the silicone oil within the elastomer. The bottom row shows the same data after applying baseline pressure correction; therefore, the slow baseline drift is not visible in the figure. After baseline subtraction, the sensor exhibits a stable and reproducible pressure response over time, demonstrating good long-term repeatability.

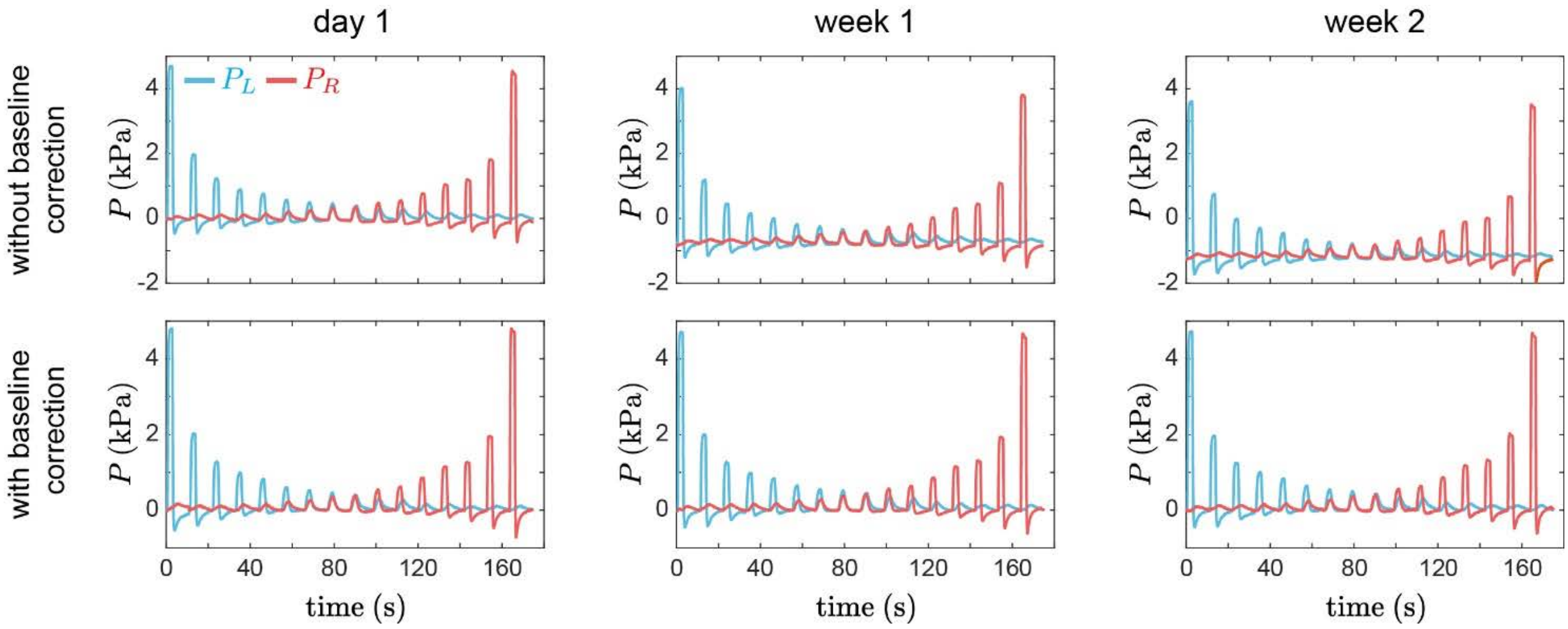

**Figure S14:** The pressure response of the 2D sensor at 16 points along the Hilbert curve over a 2-week period for $\Delta z = 3$ mm.

## S11. FCM–ANFIS modeling approach

In this study, we developed a model to estimate the position and magnitude of the applied force on our soft tactile sensor using a combined fuzzy C-means (FCM) clustering and adaptive neuro-fuzzy inference system (ANFIS) approach.

### Fuzzy C-Means (FCM) Clustering

We first employed a soft clustering approach, which is an unsupervised learning method used to group similar data points while allowing overlap between clusters. Unlike hard clustering, where each data point belongs strictly to one cluster, in soft clustering each data point is assigned a degree of membership to multiple clusters. The FCM algorithm, introduced by Bezdek[2,3], partitioned a dataset ($X = \{x_1, x_1, \dots, x_N\}$) into $c$ clusters by minimizing the following objective function:

$$J_m = \sum_{i=1}^{N}\sum_{j=1}^{C} u_{ij}^{m}\left\|x_i - c_j\right\|^2, \quad 1 \le m \le \infty \qquad (S7.1)$$

Where $N$ is the number of data samples, $c$ is the number of clusters, $u_{ij} \in [0,1]$ is the membership degree of data point $x_i$ to cluster $j$ with the center of $c_j$, and the parameter $m$ is the fuzzification coefficient, which controls the level of cluster fuzziness and the degree of overlap among clusters. The operator $\|*\|$ represents the Euclidean norm, used to measure the distance between a data sample and a cluster center.

By iteratively updating the cluster centers and membership values, the objective function ($J_m$) is minimized using the following update rules:

$$u_{ij} = \frac{1}{\sum_{k=1}^{C}\left(\frac{\left\|x_i - c_j\right\|}{\left\|x_i - c_k\right\|}\right)^{\frac{2}{m-1}}}, \qquad c_j = \frac{\sum_{i=1}^{N} u_{ij}^{m} x_i}{\sum_{i=1}^{N} u_{ij}^{m}} \qquad (S7.2)$$

These updates continue until convergence, defined by:

$$\max\left|u_{ij}^{t+1} - u_{ij}^{t}\right| < \in \qquad (S7.3)$$

Where ∈ is a small threshold.

The resulting clustering information is then used to construct a fuzzy interface system (FIS) for each cluster. Each cluster corresponds to a single fuzzy rule. We FCM implementation provided in the Fuzzy Logic Toolbox of MATLAB, specifically through the genfis3 function, to generate a fuzzy inference system based on the clustered input-output data. The optimal number of clusters was determined by cluster analysis through evaluating models with varying cluster counts and comparing root mean square error (RMSE) and mean error values in both training and testing phases.

### Adaptive Neuro-Fuzzy Inference System (ANFIS)

After clustering, the resulting FIS structure was refined using an adaptive neuro-fuzzy inference system (ANFIS), which integrates an artificial neural network (ANN) with Takagi-Sugeno fuzzy inference. This combination leverages the interpretability of fuzzy logic and the adaptive learning capabilities of neural networks.

The ANFIS model maps the nonlinear relationship between the input pressure data ($[P_L,\ P_R]$) and the output position and force ($[X, F]$). The ANFIS model consists of five layers, as described below.

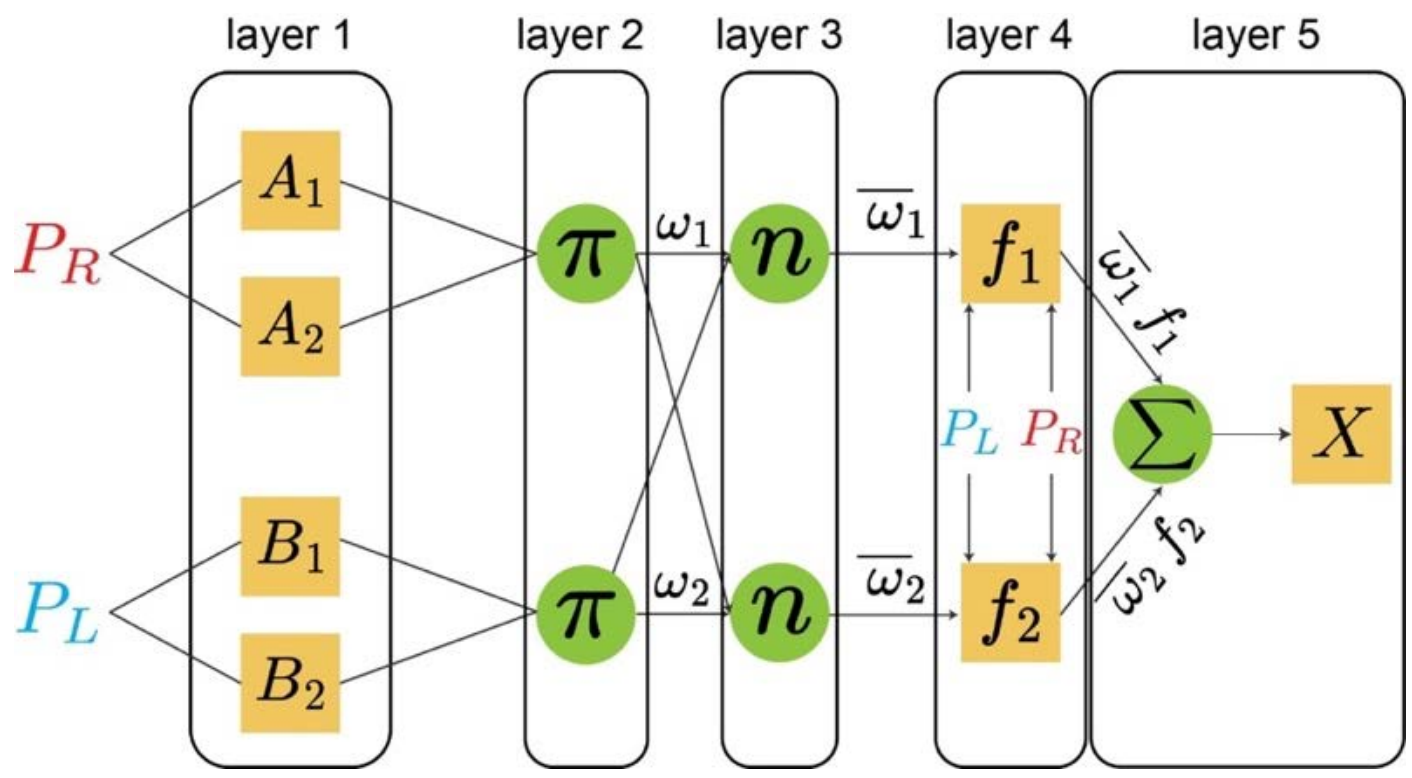

**Figure S15:** The FCM and ANFIS model schematic showing all 5 layers of the model from fuzzification to the output estimation

**Layer 1: Fuzzification (adaptive)**

Each input variable is passed through fuzzy membership functions (MFs) initialized based on the FCM clusters. These MFs (e.g., $A_1, A_2$ for $P_L$ and $B_1, B_2$ for $P_R$) quantify the degree of membership and are represented by parameterized functions such as Gaussian or sigmoid.

**Layer 2: Rule Layer (fixed)**

The degree of activation of each fuzzy rule is computed by applying T-norm operators (typically the product) to the fuzzified inputs.

$$o_{2,i} = w_i = \mu_{A_i}(P_l)\mu_{B_i}(P_r) \qquad (S7.4)$$

**Layer 3: Normalization (fixed)**

The degree of activation from all rules are normalized in this later to ensure that they contribute proportionally to the final output.

$$o_{3,i} = \overline{w}_i = \frac{w_i}{\sum w_i} \qquad (S7.5)$$

**Layer 4: Defuzzification (adaptive)**

Each fuzzy rule is associated with a first-order Sugeno output function of the form:

$$o_{4,i} = \overline{w}_i f_i = \overline{w}_i(a_i P_l + b_i P_r + c_i) \qquad (S7.6)$$

Where the parameters $a_i$, $b_i$, and $c_i$ are estimated using least squares, enabling the model to capture nonlinear relationships.

**Layer 5: Output (fixed)**

The final output (predicted position or force) is computed as:

$$o_{5,i} = \sum \overline{w}_i f_i = \frac{\sum w_i w f_i}{\sum w_i} \qquad (S7.7)$$

This hybrid framework combines the self-learning capability of neural networks with the interpretability of fuzzy systems, allowing adaptive refinement of both the fuzzy input representation and the input–output mapping. The Fig. S15 represents the ANFIS architecture, where adaptive nodes are represented as squares and fixed nodes as circles.

## S12. Learning results

**Model comparison:** To evaluate the accuracy of the FCM, compared its performance with four conventional regression methods: multi-layer perceptron (MLP), support vector regression (SVR), linear regression (LR), and polynomial regression (PR). These methods were previously applied in our earlier work[4]. The comparison was performed using the 1D sensor data for different channel diameters and media. As shown in Figs. S16 and S17, the FCM model consistently outperforms the other methods for both position and force estimation. While MLP provides high accuracy in position estimation and, in some cases, performs comparably or slightly better than FCM, it shows significantly lower accuracy in force estimation due to the complex and nonlinear nature of the pressure–force relationship in the system. Among the remaining methods, SVR demonstrates moderate accuracy but exhibits sensitivity to viscosity and channel diameter, whereas LR shows noticeably weaker performance, particularly in capturing nonlinear behavior.

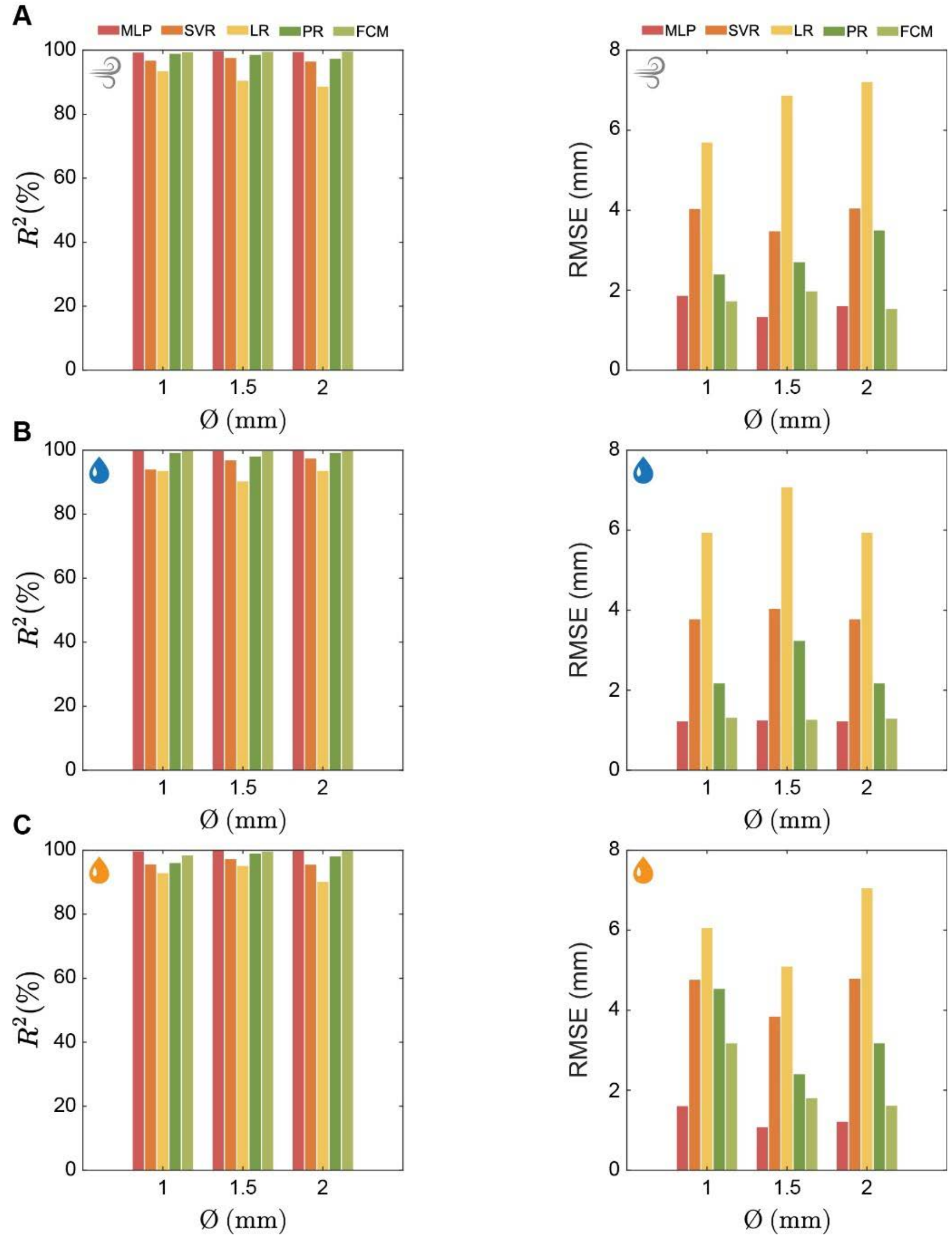

**Figure S16:** The position estimation comparison between FCM with MLP, SVR, LR, and PR for (**A**) air-filled channel (**B**) water-filled channel (**C**) oil-filled channel and for 3 different channel diameters.

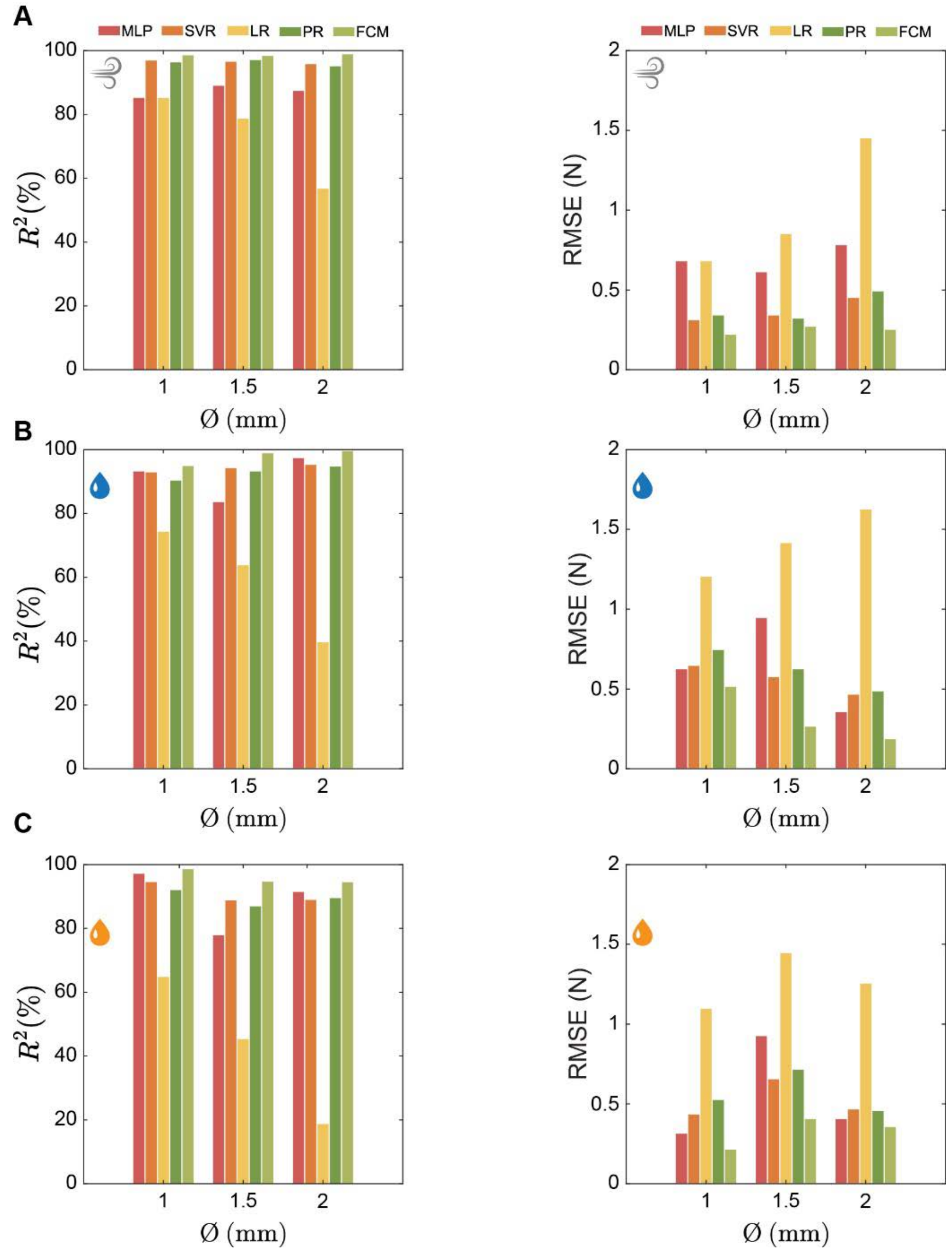

**Figure S17:** The force estimation comparison between FCM with MLP, SVR, LR, and PR for (**A**) air-filled channel (**B**) water-filled channel (**C**) oil-filled channel and for 3 different channel diameters.

### FCM for 1D sensor

Following the comparison of FCM with other regression methods, we present here the full results of the FCM model for all three media and three channel diameters, where the estimation points are also color-coded by indentation depth. As presented in Fig S18, the FCM achieves highly accurate position estimation across all conditions, with $R^2$ exceeding 99% and root mean square errors generally below 2 $\mathrm{mm}$. A slightly higher error is observed for the oil-filled channel with a 1 $\mathrm{mm}$ diameter, particularly at low indentation depths, which may result from a suboptimal selection of the working range.

Similarly, as illustrated in Fig. S19, the FCM provides reliable force estimation, with air- and water-filled channels exhibiting better performance than the oil-filled channel. This difference likely arises from the more linear pressure–force relationship in air and water, which facilitates the model's ability to learn the trend, while the oil-filled channel exhibits a more nonlinear response. Additionally, due to the reduced flowback effect in the viscous oil medium, its working range starts at lower indentation depths and smaller forces compared to the other two media.

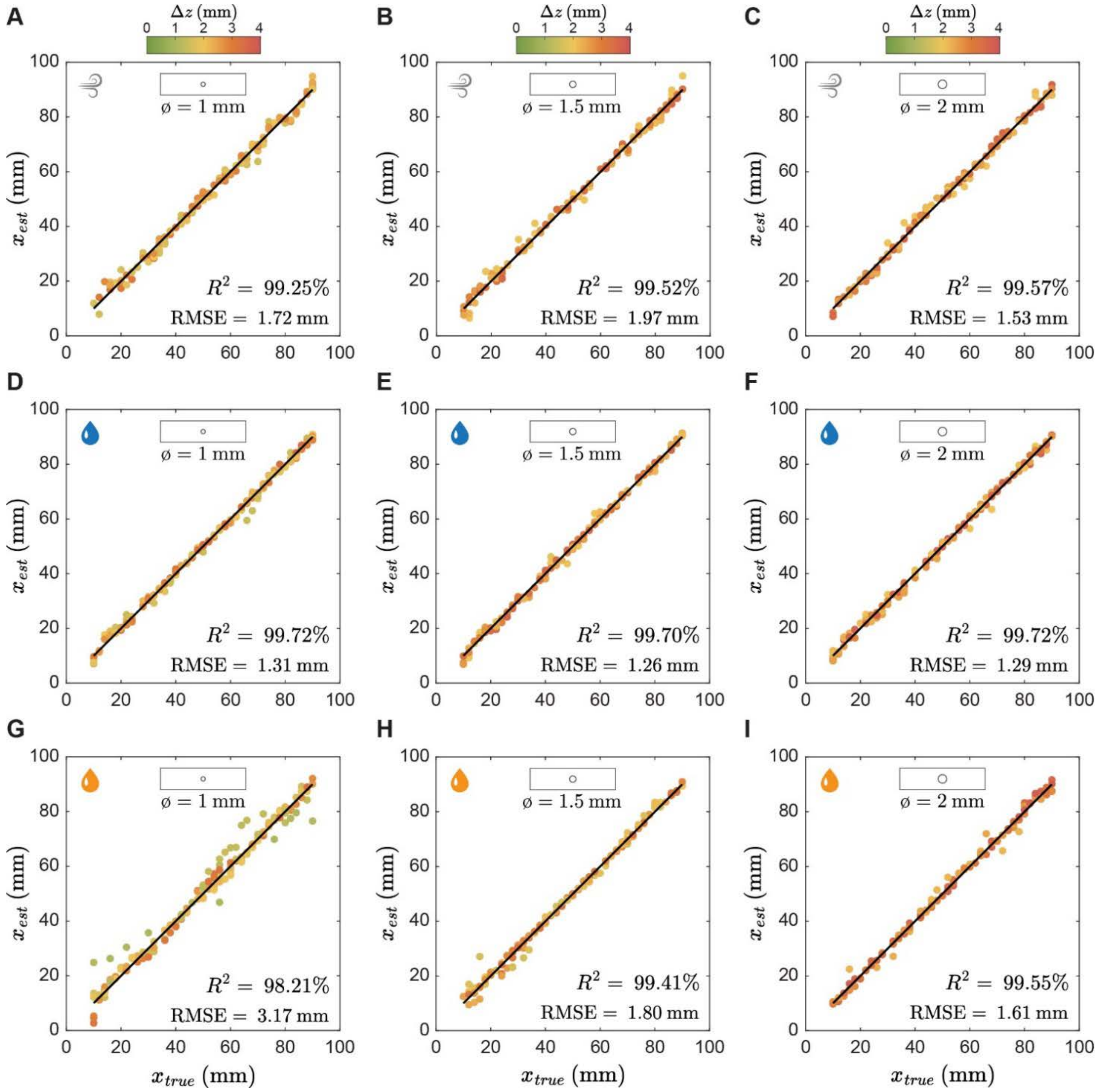

**Figure S18:** Position estimation results of the FCM model for three channel diameters and three media; (**A-C**) air-filled (**D-F**) water filled (**G-I**) oil-filled channel. The results presenting high accuracy ($R^2$ > 99%) across all cases, with errors generally below 2 $mm$.

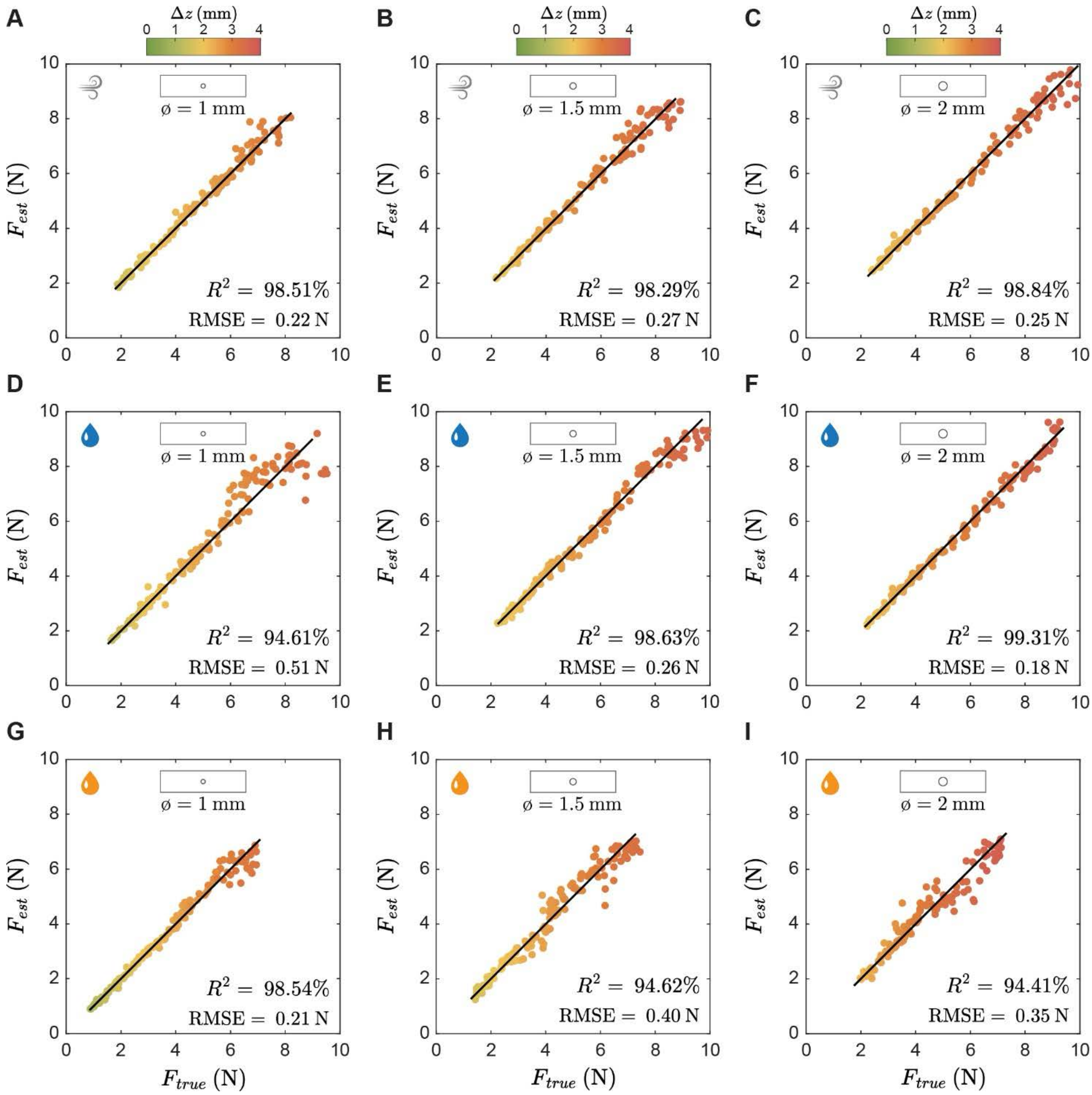

**Figure S19:** Force estimation results of the FCM model for three channel diameters and three media; (**A-C**) air-filled (**D-F**) water filled (**G-I**) oil-filled channel, where air- and water-filled channels exhibit more linear and accurate responses compared to the oil-filled channels.

**Cluster Analysis**

One of the key parameters in the FCM model is the number of clusters used during the initial soft clustering of the data. Generally, increasing the number of clusters enhances model accuracy but also raises computational cost. As shown in Fig. S20, we conducted a cluster analysis for both 1D and 2D cases. The results indicate that while increasing the cluster number initially improves estimation accuracy, the improvement gradually saturates beyond a certain point, accompanied by higher computation time and potential overfitting. This trade-off highlights the need to select an optimal cluster number that balances accuracy and efficiency.

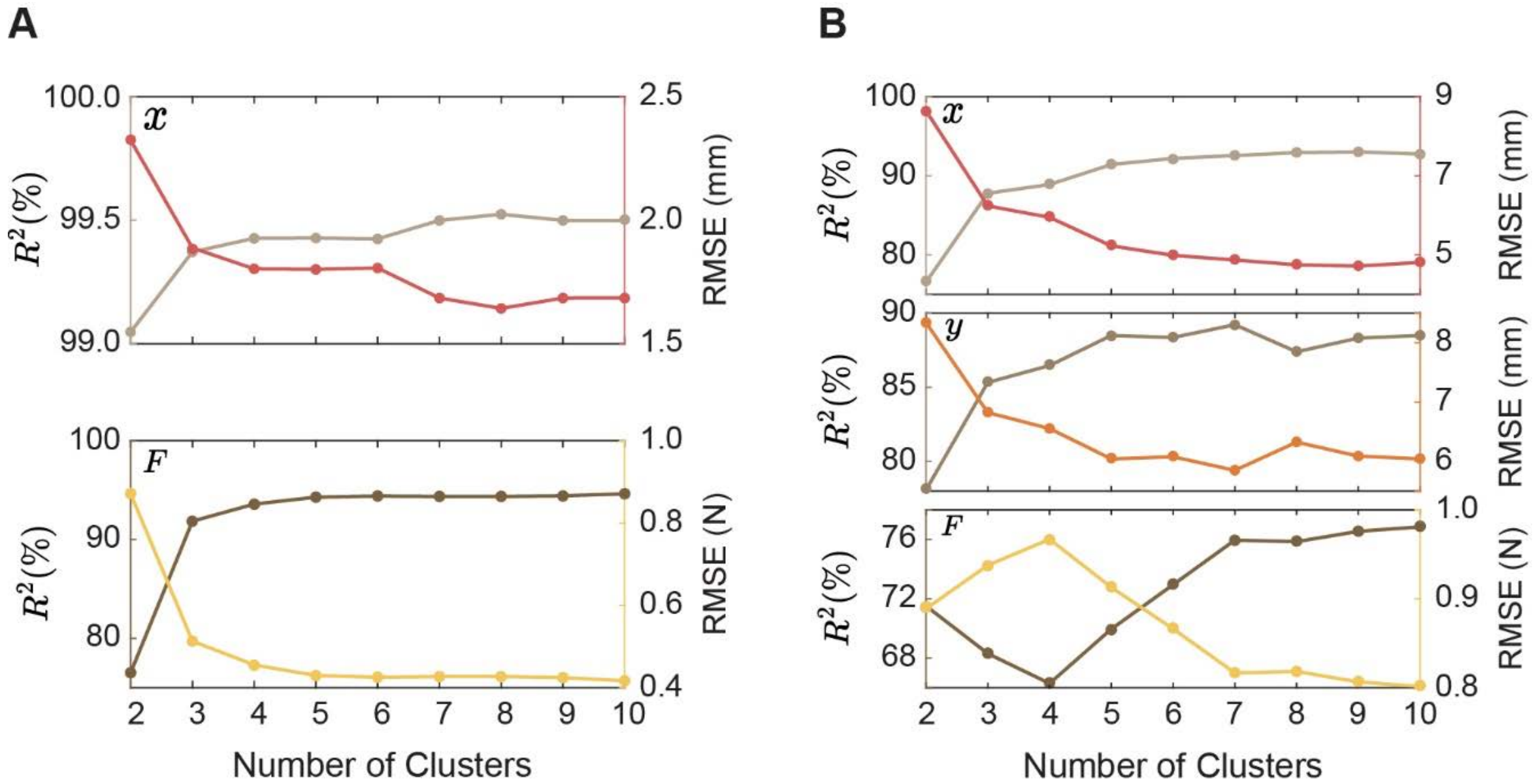

**Figure S20:** Cluster analysis for both 1D and 2D models to determine the optimal number of clusters that achieves high accuracy while minimizing computation time.

## S13. Additional results for the pressure response of the 2D sensor

In this section, we present additional results that provide deeper insight into the response of the 2D sensors, with the aim of better understanding the sources of error in touch prediction in specific regions. First, we take a closer look at the extracted features fed into the learning algorithm. Figure S21 shows the pressure peaks recorded by two pressure transducers for both on-channel and off-channel positions at three different indentation depths. The left and right pressure maps are nearly symmetric, and a clear pressure gradient is observed along the Hilbert curve, with more pronounced decay at lower pressures. Consequently, the pressure signals are considerably weaker in the second (top left) and third (top right) quarters of the 2D sensor.

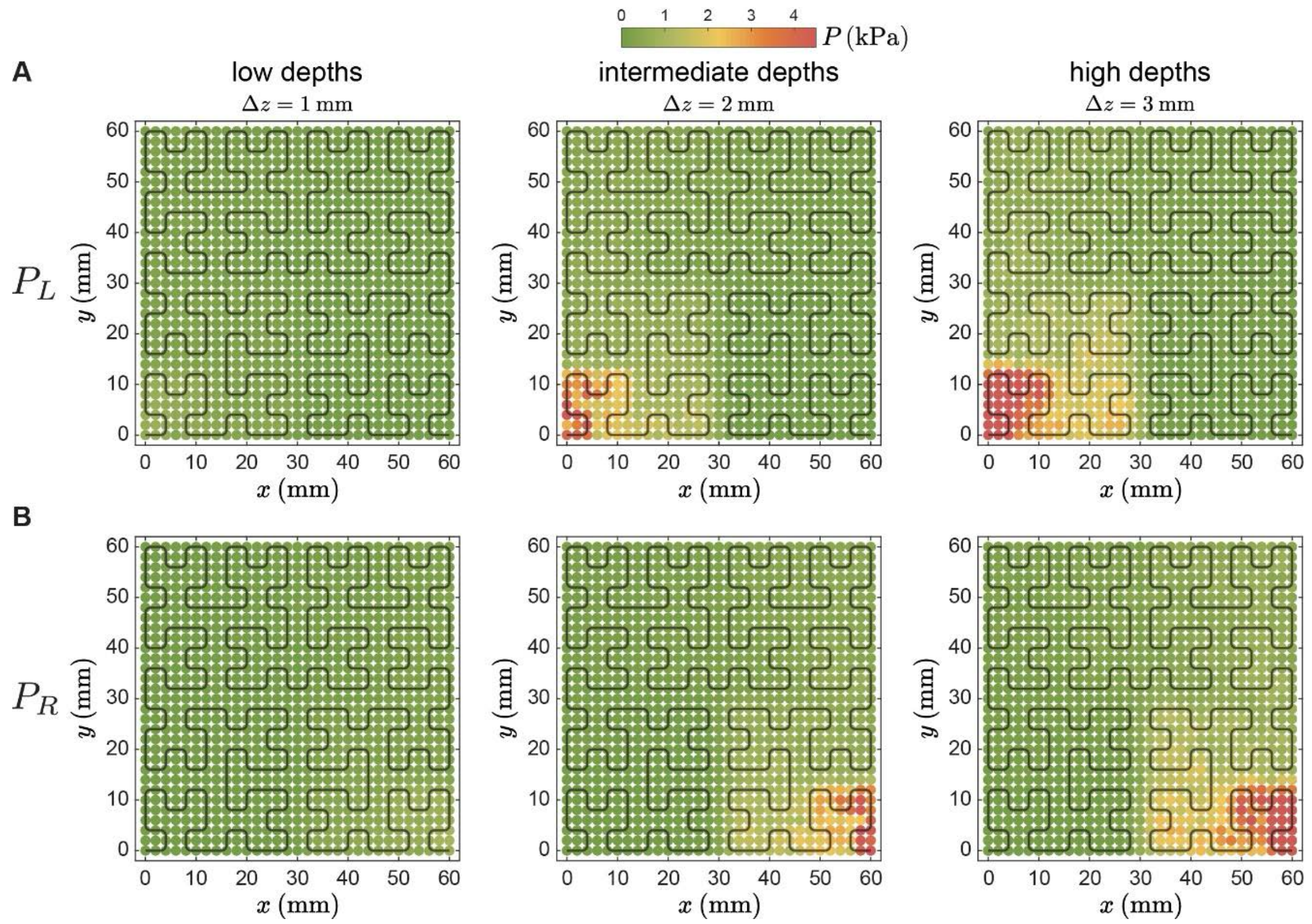

**Figure S21**. Maximum pressure maps of the 2D sensor recorded by pressure transducers at two ends for three different indentation depths. (A) $P_L$, (B) $P_R$.

Fig. S22 shows the rise time of the pressure signals measured by the left and right pressure transducers during indentations at three different depths. As expected, the $t_r^L$ and $t_r^R$ exhibit a symmetric pattern. In contrast to the pressure magnitude, which decays rapidly, the spatial gradient of the rise time is more clearly visible along the Hilbert curve. This observation is notable, as it highlights the role of viscous effects in transmitting tactile information over long distances. However, variations in rise time across different indentation depths are less pronounced. Together, these results suggest that combining pressure peak magnitudes with rise-time information could enable improved prediction of both touch position and indentation intensity.

After examining the feature space, we investigate the average errors at different indentation depth ranges: Low depths ($\Delta z = 0.6 - 1.4$ mm), medium depths ($\Delta z = 1.6 - 2.4$ mm), high depths ($\Delta z = 2.6 - 3.6$ mm). Figures S23A–C show the position errors in the X and Y coordinates, as well as the Euclidean norm of the error, respectively. Across all metrics, errors decrease with increasing indentation depth, as stronger pressure signals are generated and flowback within the channel becomes more constrained. Overall, errors in the X direction are consistently smaller than in Y, likely due to the chosen outlet locations. A similar trend is observed for force prediction errors, which also decrease at higher indentation depths as shown in Fig. S24.

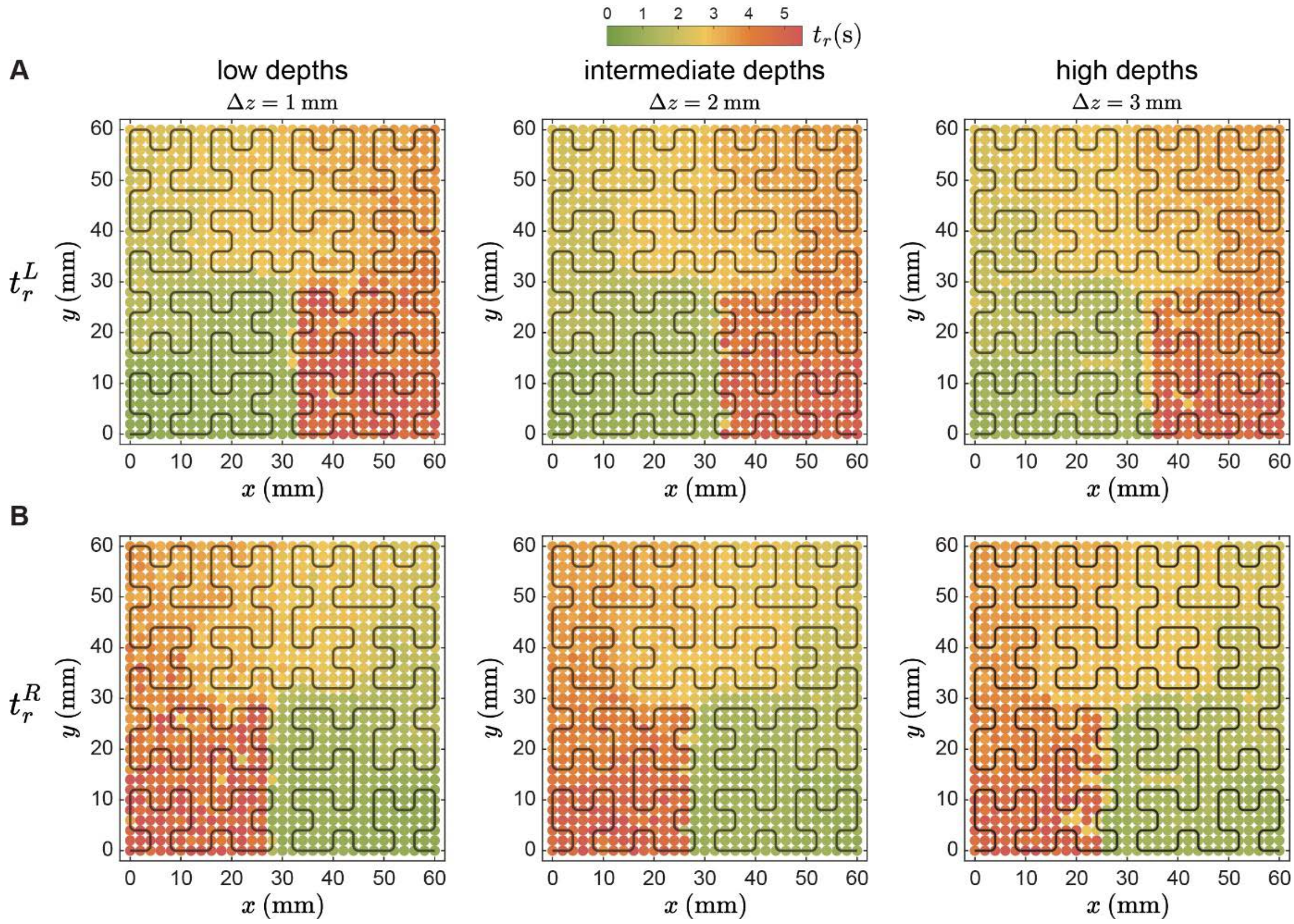

**Figure S22.** Rise time in pressure signals probed by left and right pressure transducers for the 2D sensor. (A) $t_r^L$, (B) $t_r^R$.

We observed that errors in specific regions of the 2D sensor are notably higher than in other regions. This effect is particularly pronounced in areas where the Hilbert curve length between two adjacent points is significantly larger than in other regions. In Fig. S25, we report pressure histories along a straight line for five regions with different Hilbert curve topologies subjected to a large indentation depth ($\Delta z = 3$ mm). However, for certain points, (e.g., see the last three subplots in Fig. S25 corresponding to the red and brown regions), changes in the pressure response are effectively indistinguishable. These points correspond to larger prediction errors. Nevertheless, establishing a clear relationship between the characteristics of the pressure histories and the prediction error is not straightforward, given the complexity and inherent nonlinearity of the system.

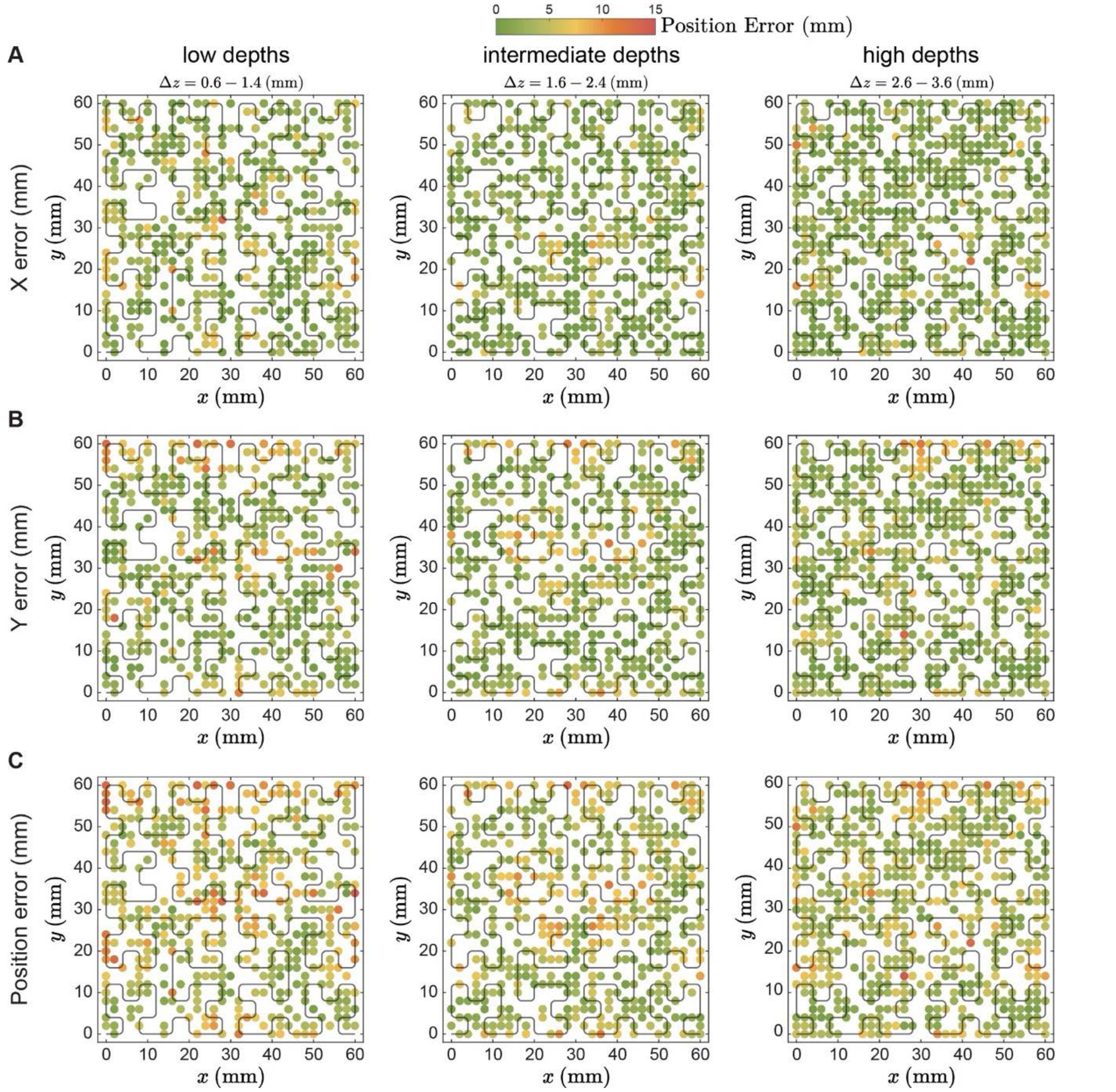

**Figure S23:** Normalized error for the predicted touch position averaged for different indentation depth ranges. (A) X, (B) Y, and (C) Euclidean position error.

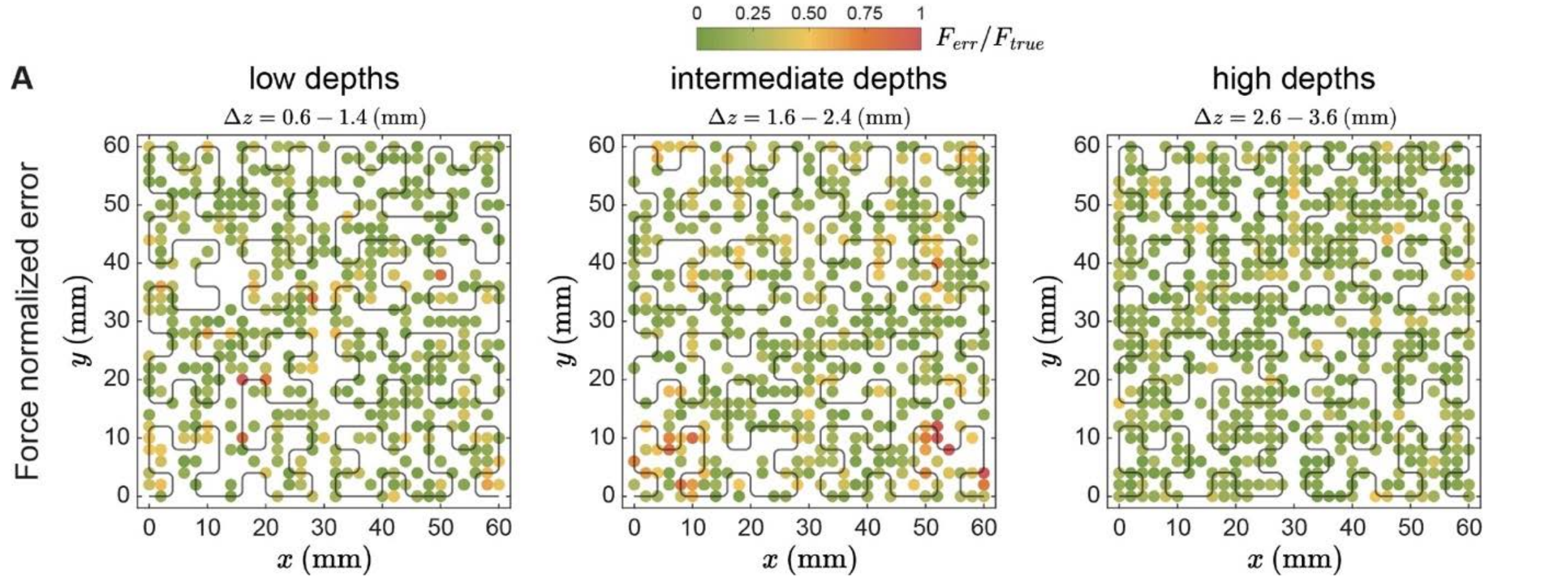

**Figure S24.** Normalized error for the predicted touch force averaged for different indentation depth ranges.

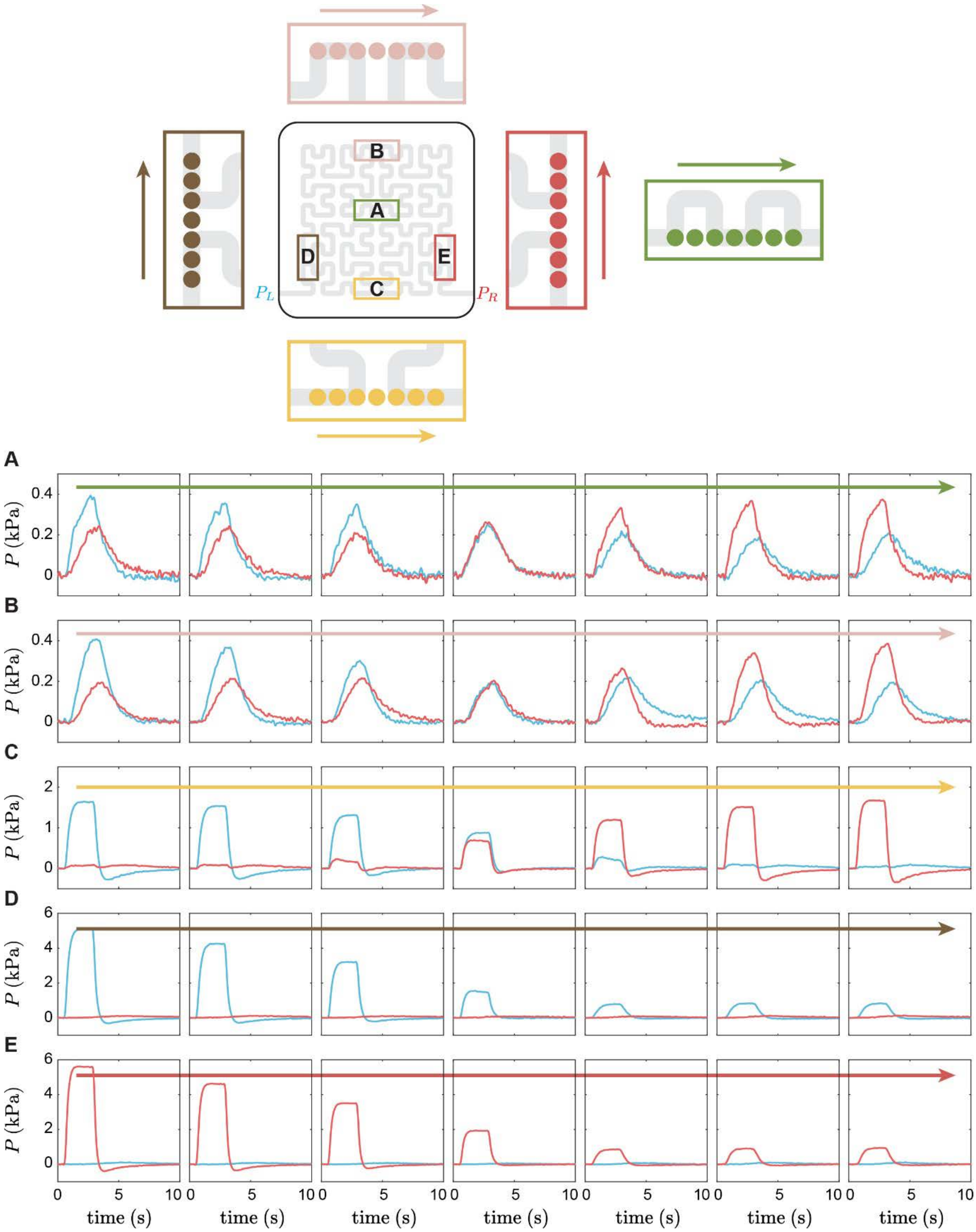

**Figure S25.** Pressure responses of the 2D sensor at selected regions (A, B, C, D, E) of the Hilbert curve.

## S14. Demonstration

### Underwater Touch

The 1D response for underwater tactile sensing is presented in Fig. 6A, with a comparative analysis of raw and baseline-corrected data provided in Fig. S26. As expected, increasing the water column height exerts greater hydrostatic pressure on the sensor, resulting in a predictable shift in the baseline signal. However, by applying baseline correction, we successfully eliminated this depth-dependent offset, yielding nearly identical response profiles across all underwater conditions.

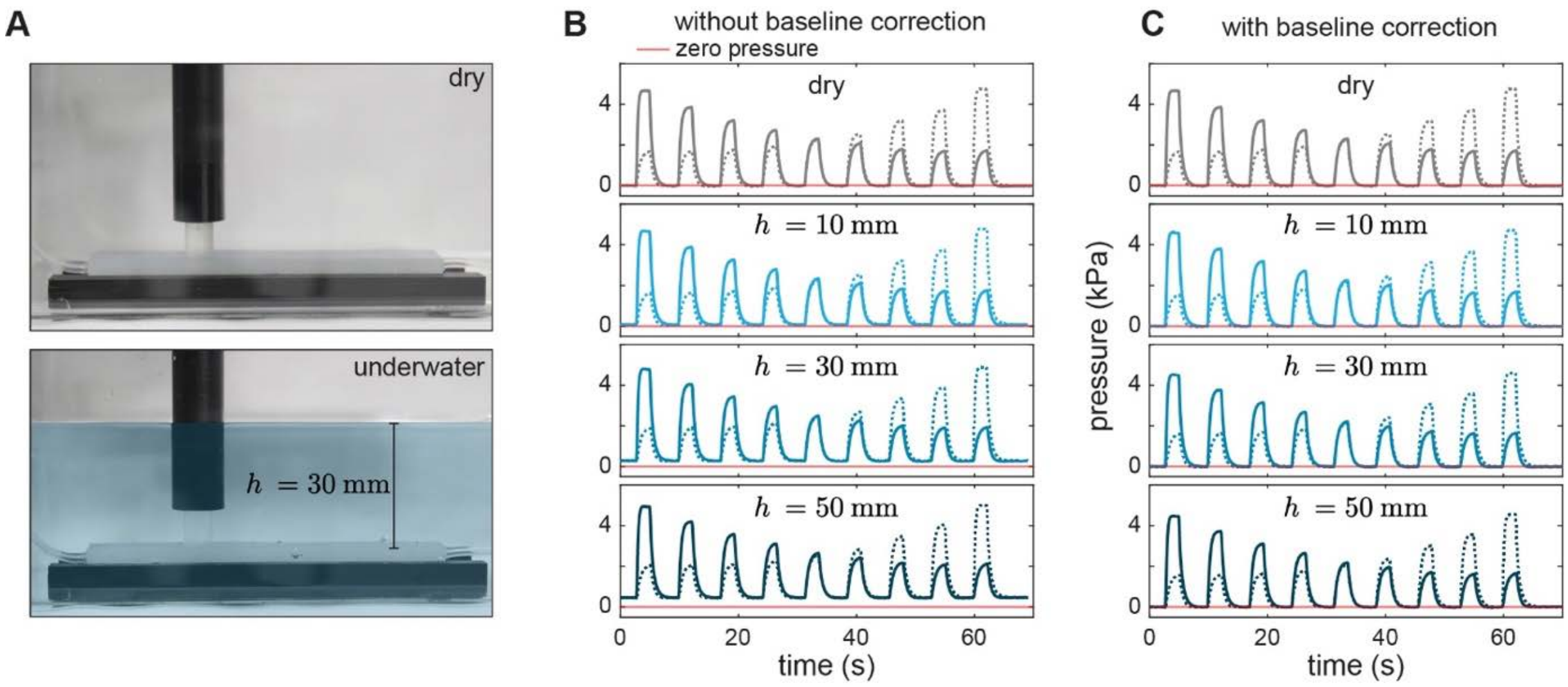

**Figure S26:** Underwater performance of the 2D tactile sensor at varying water depths (10, 30, and 50 $mm$), compared with the reference case in dry case. (**B**) The pressure response without removal of the baseline pressure. (**C**) The pressure response with baseline removal.

Moreover, we performed the underwater experiments for 2D sensor as well by indenting the sensor at 16 locations $\tilde{s} = (2i - 1)/32$ (where $i \in [1,16]$), at the $\Delta z = 2$ mm. As shown in Fig. S27A, experiments were performed under dry conditions and at different water levels. Without baseline removal (Fig. S27B), the measured pressure exhibits a systematic offset that increases with water level due to hydrostatic pressure. After baseline removal (Fig. S27C), the pressure responses under dry and submerged conditions become nearly identical across all tested water levels. The pressure difference between dry and underwater measurements remains below 5 Pa, even at a water height of 100 mm. These results demonstrate that baseline removal is essential to compensate for hydrostatic pressure–induced drift and to preserve the sensor's functionality underwater.

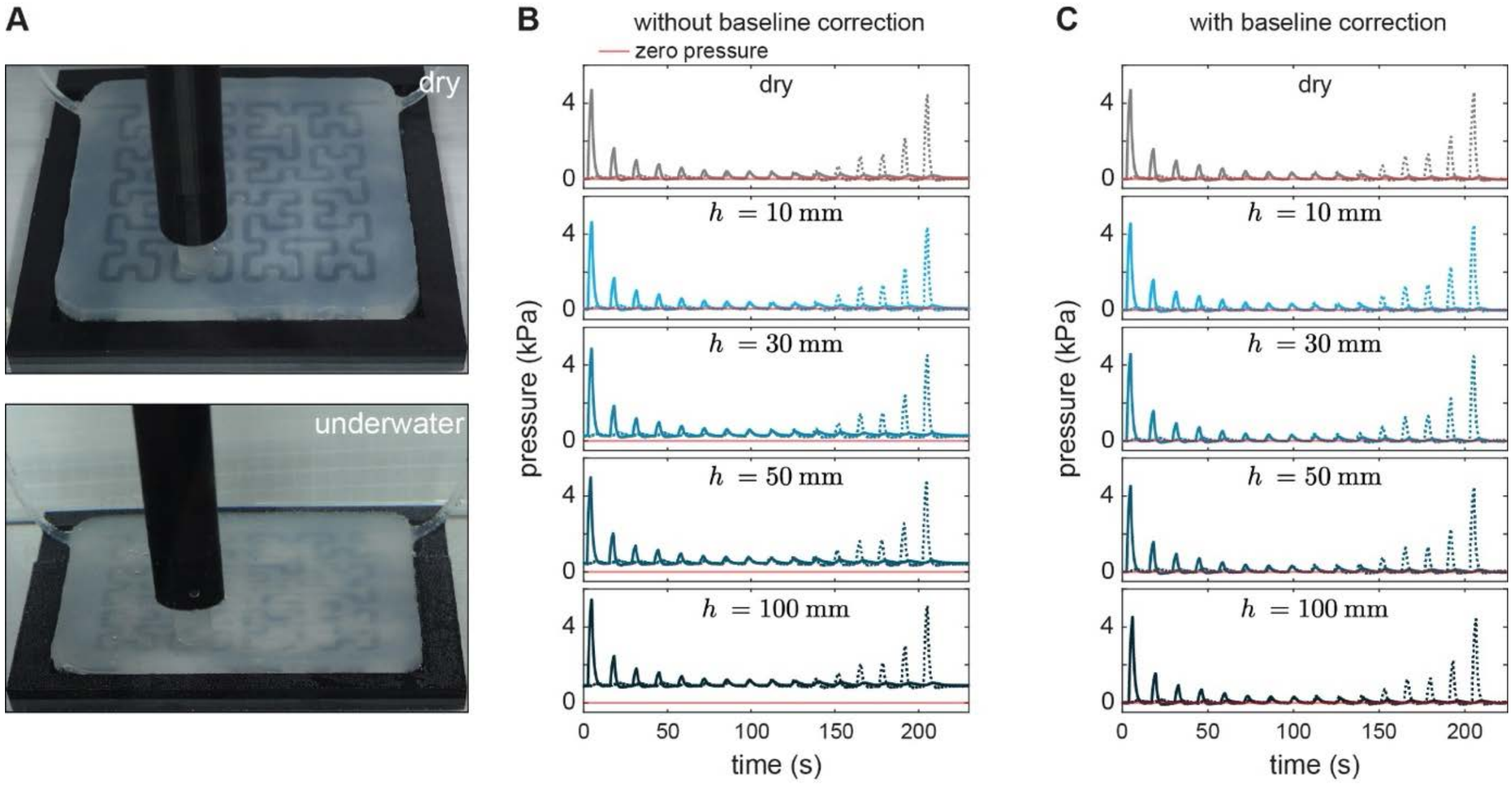

**Figure S27:** Underwater performance of the 2D tactile sensor at varying water depths (10, 30, 50, and 100 $mm$), compared with the reference case in dry case. (**B**) The pressure response without removal of the baseline pressure. (**C**) The pressure response with baseline removal.

## Magnetic Touch

In addition to the 1D sensor under the magnetic indentation, we performed a similar experiment with plastic indenter with the same shape of the magnetic indenter for the 2D sensor. In this experiment, we indented the sensor at 16 locations $\tilde{s} = (2i - 1)/32$ (where $i \in [1,16]$), at the $\Delta z = 2$ mm. As presented in Fig S28, the pressure signals in both cases are identical, showcasing the performance of the 2D sensor under magnetic interference as well.

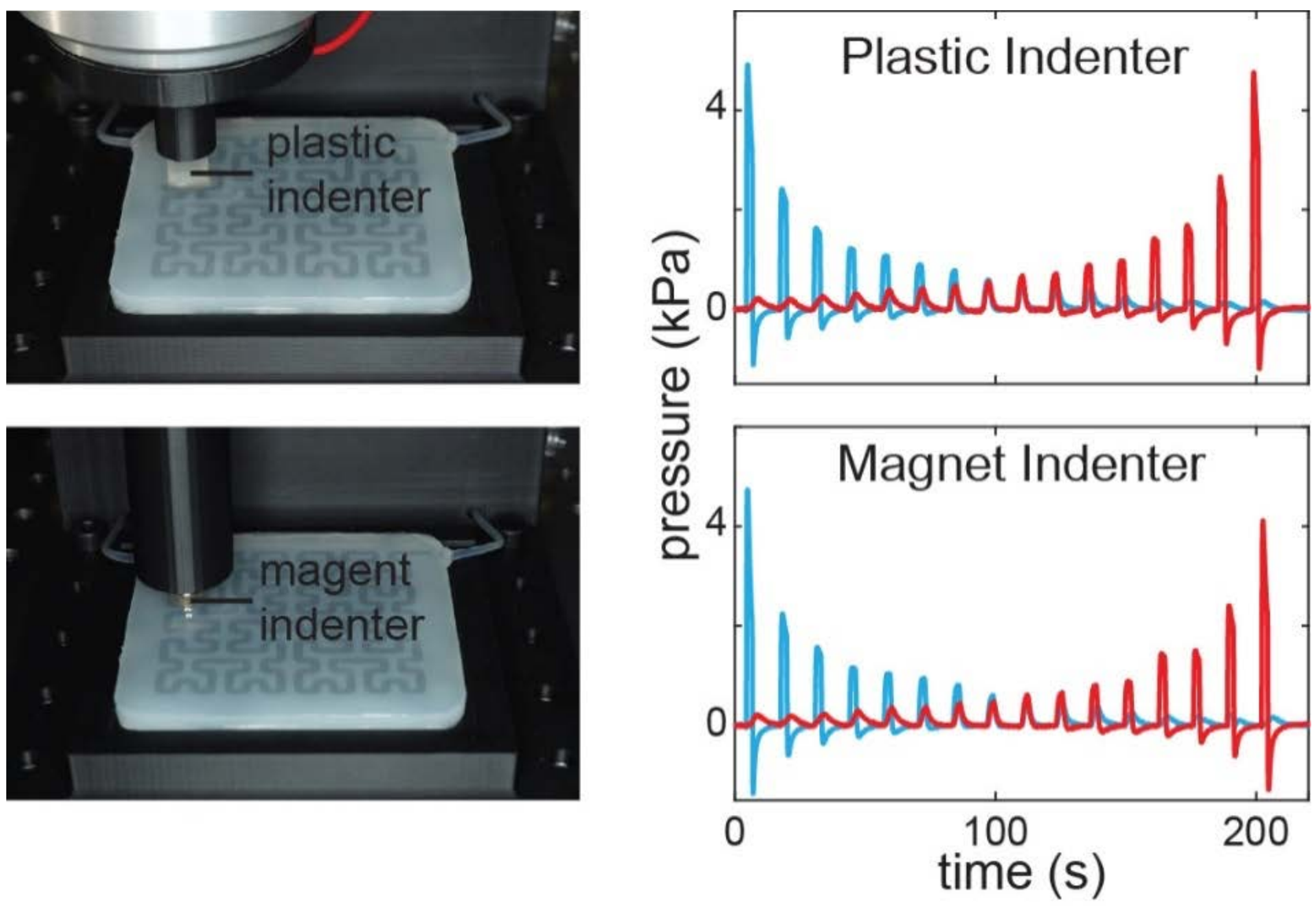

**Figure S28:** The 2D sensor pressure response with plastic indenter (top) and the magnetic indenter (bottom), showing similar results and the sensors capability of working under strong magnetic interference.

### Indenter Shape

In Fig S29, we present the indenter shapes used in the experiments reported in Fig. 6D.

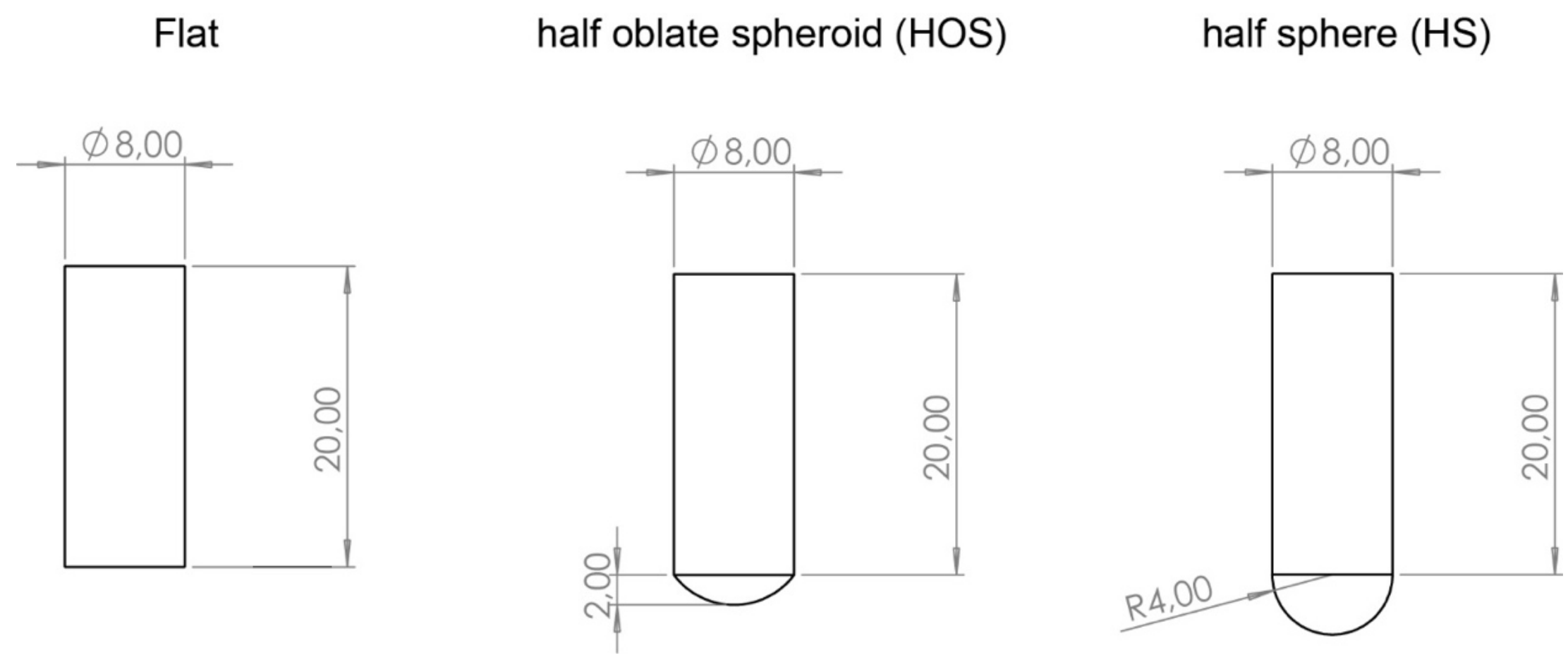

**Figure S29:** The drawing of the different indenter shapes used in the paper. All experiments except for the indenter shapes are conducted with HOS indenters.

## SUPPLEMENTARY REFERENCES